\documentclass[aps,twocolumn,secnumarabic,balancelastpage,amsmath,amssymb,nofootinbib,floatfix,superscriptaddress]{revtex4-2}
\usepackage{graphicx}      % tools for importing graphics
\usepackage{siunitx}
\usepackage{adjustbox}
\usepackage{xcolor}
\usepackage{braket}
\usepackage{float}
\usepackage{booktabs,caption}
\usepackage{array}
\usepackage[flushleft]{threeparttable}
\usepackage{textcomp}
\usepackage{gensymb}%\degree must load package textcomp before
\usepackage[colorlinks=true]{hyperref}  % this package should be added after 
\newcommand{\modif}[1]{\textcolor{black}{#1}}

\begin{document}
\raggedbottom

\title{Suppression of ferromagnetism governed by a critical lattice parameter in CeTiGe$_3$ with hydrostatic pressure or V substitution}
\author{Hanshang Jin}
\affiliation{Department of Physics and Astronomy, University of California, Davis, California 95616, USA}
\author{Weizhao Cai}
\affiliation{Department of Physics and Astronomy, University of Utah, Salt Lake City, Utah 84112, USA}
\author{Jared Coles}
\affiliation{Department of Physics and Astronomy, University of Utah, Salt Lake City, Utah 84112, USA}
\author{Jackson R. Badger}
\affiliation{Department of Chemistry, University of California, Davis, California 95616, USA}
\author{Peter Klavins}
\affiliation{Department of Physics and Astronomy, University of California, Davis, California 95616, USA}
\author{Shanti Deemyad}
\affiliation{Department of Physics and Astronomy, University of Utah, Salt Lake City, Utah 84112, USA}
\author{Valentin Taufour}
\affiliation{Department of Physics and Astronomy, University of California, Davis, California 95616, USA}

\begin{abstract}
We combine structural and magnetic measurements to compare the different magnetic phase diagrams between the pressure and substitution studies in CeTiGe$_3$. We report on the structural, magnetic, and electrical transport properties of single crystals of CeTi$_{1-x}$V$_x$Ge$_3$ ($x = 0$, $0.1$, $0.2$, $0.3$, $0.4$, $0.9$, and $1$), and of polycrystalline samples ($x = 0.5$, $0.6$, $0.7$, $0.8$), as well as structural properties of CeTiGe$_3$ under pressure up to $9$\,GPa. The ferromagnetic ordering in CeTiGe$_3$ is suppressed with V doping in CeTi$_{1-x}$V$_x$Ge$_3$, and suggests a possible ferromagnetic quantum critical point near $x = 0.45$. We perform a detailed crystalline electric field (CEF) analysis, and the magnetic susceptibility data in pure CeTiGe$_3$ and CeVGe$_3$ can be well explained by the CEF model. The proposed CEF energy levels suggest that there is a gradual change of the ground state from $\ket{\pm 5/2}$ in CeTiGe$_3$ to $\ket{\pm 1/2}$ in CeVGe$_3$, and a suppression of CEF splitting energies near the quantum critical region. When hydrostatic pressure is used instead of chemical substitution, the quantum critical point is avoided by the appearance of magnetic phases above around $4.1$\,GPa. In the substitution study, the ferromagnetic and antiferromagnetic regions are well separated, whereas they touch in the pressure study. We observe a different trend in the temperature dependence of the resistivity maximum in both studies, suggesting that the CEF splitting energy is suppressed by V substitution but enhanced by pressure. We also observe different responses in lattice constants between the two studies, highlighting the fact that substitution effects cannot be reduced to chemical pressure effects only. Nevertheless, when the magnetic phase diagrams of both hydrostatic pressure and substitution are compared, we find a common critical lattice constant $c = 5.78$\,\si{\angstrom} where the ferromagnetic ordering is suppressed in both studies.
\end{abstract}
\maketitle

%%%%%%%%%%%%%%%%%%%%%%%%%%%%%%%%%%%%%%%%%%%%%%%%%%%%%%%%%%%%%%%%%%

\section{Introduction}

Quantum phase transitions have received a lot of attention because they are related to the origin of many exotic quantum phenomena in condensed matter physics~\cite{Brando2016RMP,Stewart2001RMP,RevModPhys.79.1015}. A variety of metallic ferromagnets and antiferromagnets have been tuned towards a quantum phase transition where their magnetic transition temperature is suppressed to zero by the application of pressure, chemical doping, or magnetic field. In these cases, unconventional properties are observed such as non-Fermi liquid behavior, and unconventional superconductivity~\cite{unconvsc}. While the origin of unconventional superconductivity remains a challenge to our theoretical understanding, the quantum phase transition itself is often difficult to predict as well. This difficulty is the result of subtle relationships between the electronic, structural, and magnetic properties, which remain hidden due to limited experimental information and the lack of sufficient tuning parameters to control these properties. In addition, materials with quantum phase transitions often exhibit multiple correlations in competition. A typical example is the competition between ferromagnetism and antiferromagnetism, which has been recently observed in many families of unconventional superconductors~\cite{Sefat2009PRB,Hu2010JPCC,Filsinger2017PRB,Xu2012JPCS,Sefat2009PRBBaCo2As2,Pandey2013PRB,Jayasekara2013PRL,Guterding2017PRL,Wiecki2015PRB,Wiecki2015PRL,Jesche2017PSSB,Lausberg2012PRL,Kopp2007PNAS,Kurashima2018PRL,Ishikawa2001JPCM,Ishikawa2002JPCS,Ran2019Science,Sundar2019PRB,Duan32020PRL}. Another example is provided by materials that have a quantum tricritical point (QTCP) at which two order parameters vanish continuously~\cite{Kaluarachchi2018PRB,NbFe2}.

In CeTiGe$_3$, a QTCP has been observed in the temperature-pressure-magnetic field phase diagram near $2.8$\,T, and $5.4$\,GPa~\cite{Kaluarachchi2018PRB}. In the absence of magnetic field, CeTiGe$_3$ undergoes a pressure induced first-order ferromagnetic (FM) to antiferromagnetic (AFM) transition, with a quantum phase transition at $4.1$\,GPa.

Substitutions of titanium with various elements such as V~\cite{Kittler:2013bc}, Ni~\cite{Khan:vx}, and Cr~\cite{Das:wy}, have been studied in search of a possible FM quantum critical point (QCP). The parent compound CeVGe$_3$ is known to be antiferromagnetic below $T_\textrm{N} = 5.5$\,K~\cite{Bie:uh, Inamdar:2014io} and has the same crystal structure as CeTiGe$_3$~\cite{MANFRINETTI2005444}. As a result, it makes sense that V substitution would suppress the FM state. This also gives us a unique opportunity to investigate the nature of the FM-order suppression by comparing chemical substitution and hydrostatic pressure.

In previous studies of polycrystalline samples of CeTi$_{1-x}$V$_x$Ge$_3$, a possible ferromagnetic quantum critical point at $x \approx 0.35$~\cite{Kittler:2013bc,Fritsch2015} was found, and evidence for a competition between ferro- and antiferromagnetic correlations was found in NMR measurements~\cite{Majumder:2018vt}. To further study the suppression of ferromagnetism in CeTiGe$_3$, we investigate single crystal samples of CeTi$_{1-x}$V$_x$Ge$_3$ and study the phase diagram within the full substitution range, as well as structural properties of CeTiGe$_3$ under pressure up to $9$\,GPa. We observe a suppression of ferromagnetism with increasing V content, with a suggested QCP at around $x = 0.45$. With the help of the single crystals across a large region of substitution, we are able to perform a detailed CEF analysis on the magnetic susceptibility data along both axes. The curvature behavior in inverse magnetic susceptibility data of the pure CeTiGe$_3$ and CeVGe$_3$ can be well explained by the CEF model. The proposed CEF energy levels suggest that there is a gradual change of the ground state from $\ket{\pm 5/2}$ in CeTiGe$_3$ to $\ket{\pm 1/2}$ in CeVGe$_3$, and a suppression of CEF splitting energies near the quantum critical region. The CeTi$_{1-x}$V$_x$Ge$_3$ magnetic phase diagram shows a clear separation between the FM and AFM regions. The shape of the maximum in temperature dependent resistivity in the pressure and substitution studies suggest that the CEF splitting energy and the Kondo coherence energy are of similar magnitude. However, the different temperature trend in resistvity maximum suggests that the CEF splitting energy is suppressed by V substitution, but enhanced by pressure. By comparing the changes in lattice constants in both substitution and pressure works, we observed a different response in lattice constants between the hydrostatic pressure study and the substitution study. Nevertheless, we find a common critical lattice constant $c$ of $5.78$\,\si{\angstrom} where the FM ordering is suppressed in both studies. This provides an example of a simple structural parameter playing the role of a descriptor for the ferromagnetic instability. 
%%%%%%%%%%%%%%%%%%%%%%%%%%%%%%%%
%%%%%%%%%%%%%%%%%%%%%%%%%%%%%%%%

\section{Methods}
\subsection{V substitution study}
Single crystals of CeTi$_{1-x}$V$_x$Ge$_3$ with $x = 0, 0.1, 0.2, 0.3 ,0.4, 0.9, 1$, were synthesized via self-flux solution growth with similar, but modified, temperature and stoichiometry profiles compared to the previous report~\cite{Inamdar:2014io}. The starting materials  [Ce pieces (Ames Lab), Ti granules (4N), V pieces (etched with nitric acid), Ge lumps (6N)] were initially arc-melted to ensure a homogeneous mixture. While a majority of the initial compositions have a stoichiometric ratio of Ce$_4$Ti$_{1-x}$V$_x$Ge$_{19}$, some were modified to avoid unwanted phases as discussed below. The arc-melted mixture was placed in a $2$\,mL Canfield Crucible Set~\cite{Canfield:uq}, and sealed in a fused silica ampoule in a partial-pressure of argon. The sealed ampoule was placed in a furnace where it was held at \SI{1200}{\celsius} for 10 hours, cooled to \SI{1050}{\celsius} in 2.5 hours, held at \SI{1050}{\celsius} for one hour, and slowly cooled to \SI{860}{\celsius} over 140 hours. At \SI{860}{\celsius}, the ampoule was removed from the furnace and quickly centrifuged to separate the single crystals from the molten flux. 

In the titanium-rich regime, we found an increasing amount of the cubic impurity phase, (Ce$_{0.85}$Ti$_{0.15}$)Ge$_3$O$_{0.5}$, as vanadium doping increased, showing that Ti and O can stabilize CeGe$_3$~\cite{JIN2021158354}, which is otherwise only obtained using high-pressure synthesis technique~\cite{Fukuoka2004CL}. To avoid this impurity phase, we adjusted the stoichiometric ratio to $\mathrm{Ce}:(\mathrm{Ti} + \mathrm{V}):\mathrm{Ge} = 15:6:79$, which successfully removed the cubic impurity phase up to $x = 0.4$.  Single crystals of CeTi$_{1-x}$V$_x$Ge$_3$ with $x = 0, 0.1, 0.2, 0.3 ,0.4, 0.9, 1$, were obtained. For the vanadium-rich side, we used the ratio of $\mathrm{Ce}:(\mathrm{Ti} + \mathrm{V}):\mathrm{Ge} = 15:4:81$ to optimize the synthesis of $x = 0.9$ samples.

While we were not able to grow single crystals between $x = 0.5$ and $0.8$ due to unknown liquidus surfaces that involves four different elements, we successfully grow polycrystalline samples in this range. Polycrystalline samples of CeTi$_{1-x}$V$_x$Ge$_3$ with $x = 0.5, 0.6, 0.7, 0.8$, were grown following the method in Ref.~\cite{Kittler:2013bc}. The elements with the ratio of $1 : 1-x : x : 3$ were first arc-melted several times and then annealed at 950$^{\circ}$C for one week.

\begin{figure}[!htp]
\includegraphics[width=\linewidth]{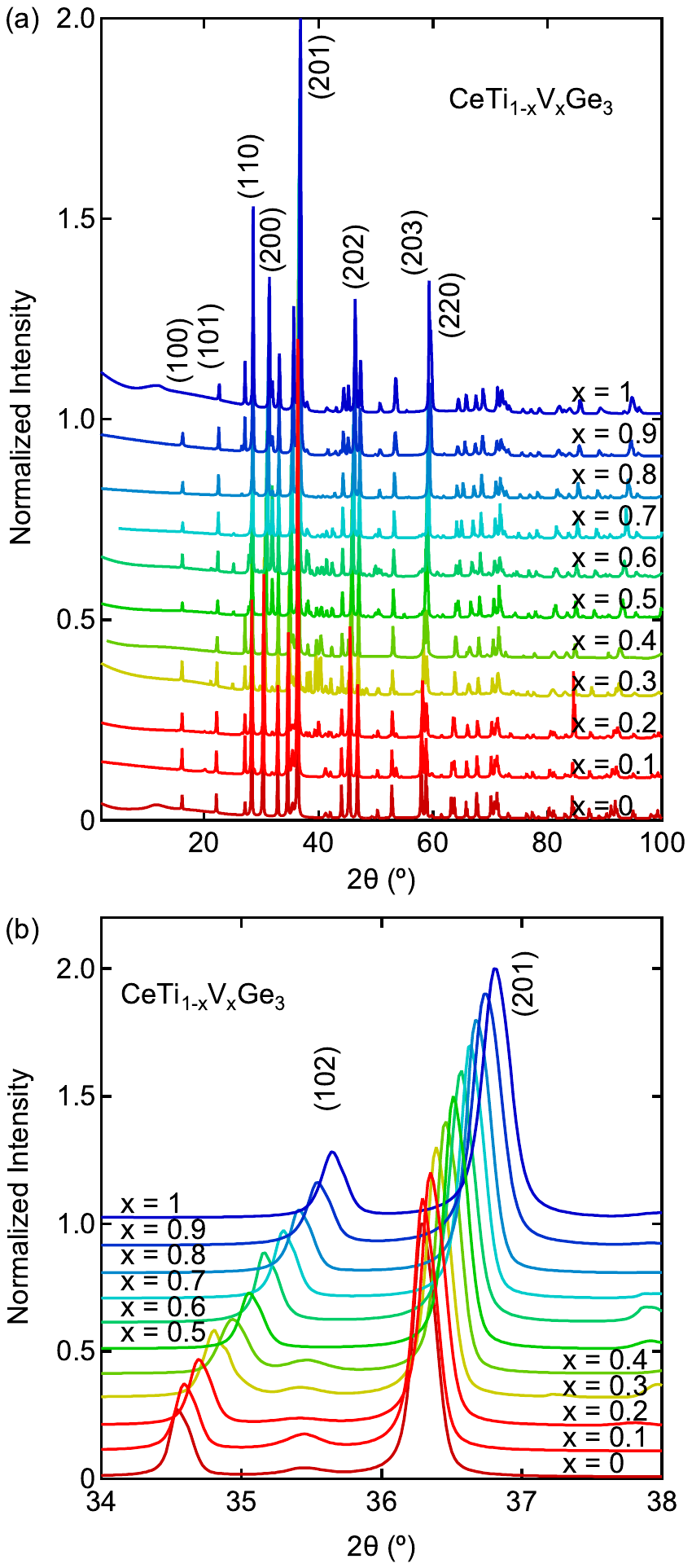}
\caption{X-ray diffraction pattern of CeTi$_{1-x}$V$_x$Ge$_3$ with a consistent offset of 0.1 normalized intensity. The red end is Ti-rich side and the blue end is V-rich side. (a) The overall pattern and (b) the consistent shifts of the main XRD peak of CeTi$_{1-x}$V$_x$Ge$_3$ indicates that the structure type remains the same along the substitution.}
\label{fig:xrd_full}
\end{figure}

The phase identification of the samples was carried out by powder x-ray diffraction (PXRD) on a Rigaku Miniflex 600 diffractometer with Cu K$\alpha$ ($\lambda$ = 1.54178\,\si{\angstrom}) radiation at room temperature as shown in Fig.~\ref{fig:xrd_full}. The powder patterns contributed by the main phases can be indexed by a hexagonal BaBiO$_3$-type structure. We note that there are some peaks from other impurities near the $2\theta = 40$\degree~ region in Fig.~\ref{fig:xrd_full}(a). Impurities are often a combination of CeGe$_{2-x}$, Ti$_{6-x}$V$_x$Ge$_5$, and TiV. With the exception of CeGe$_{2-x}$ which is magnetic~\cite{Bud_ko_2014}, the other potential impurities do not affect our results on magnetic phase diagrams. The major peaks of CeTi$_{1-x}$V$_x$Ge$_3$ shift consistently as shown in Fig.~\ref{fig:xrd_full}(b), indicating that the substitution is successful and the structure type remains the same along the substitution.

\begin{figure}[!htp]
\includegraphics[width=\linewidth]{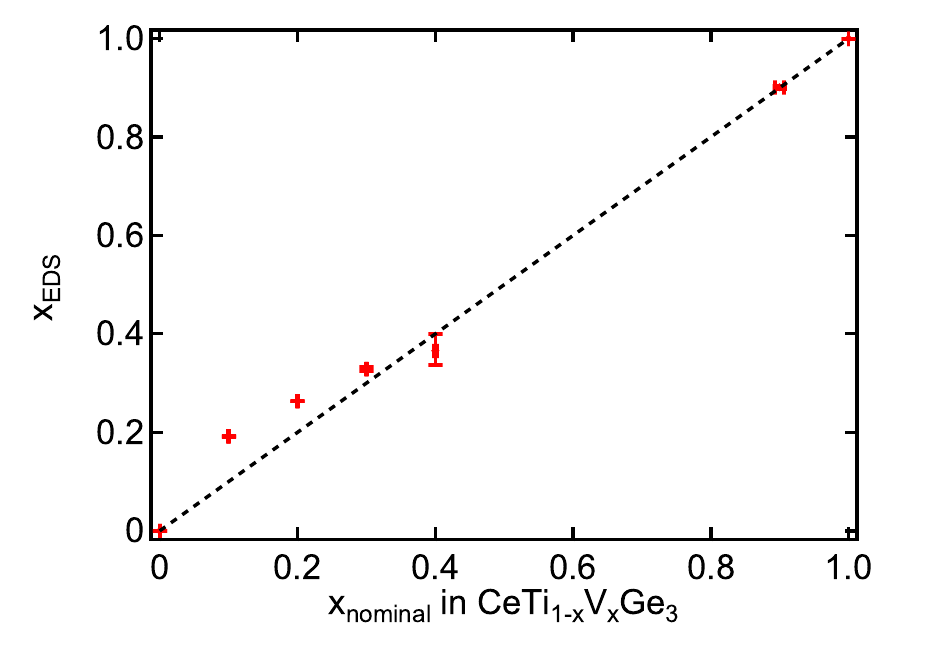}
\caption{The V content of CeTi$_{1-x}$V$_x$Ge$_3$ single crystals obtained from EDS, compared with the nominal value (the dashed line).}
\label{fig:eds_nominal}
\end{figure}

Scanning electron microscope (SEM) images and elemental composition were obtained for all doped samples on a Thermo Fisher Quattro S equipped with a Schottky field emission gun. Point scans and elemental mappings were collected to determine the atomic composition across several crystals with electron beam of 20\,kV, and three spectra were collected for each crystal. The averaged atomic percentage of the component elements of selected samples were analyzed, and the percentage of vanadium doping (V\%) was calculated by the ratio of V/(Ti + V). The results, compared with the nominal doping level, are shown in Fig.~\ref{fig:eds_nominal}. The error bars of the nominal value are estimated from weighing errors of the initial composition of the batch. The error bars of the electron diffraction spectroscopy (EDS) results show the range of the three scans on different spots of the crystal, to represent the homogeneity of the sample.

Due to the energy resolution of the instrument and the overlap of vanadium K series and cerium L series, the vanadium spectrum peak is buried under the tail of the cerium peak for Ti-rich samples. Therefore, the deviation at lower doping samples is most likely because of overlapping of peaks, rather than from a true difference in substitution level. Given that major XRD peaks have a consistent shift, we will use the nominal values for the rest of this paper, and use the EDS error bars of the doping level to represent the homogeneity of the sample.

Based on the EDS results, all samples show a good degree of homogeneity, where the doping levels on different spots of the same crystal are within a difference of $\pm 0.004$. The only exception is the $x = 0.4$ sample, which shows a $\pm 0.03$ difference in the doping level across the crystal.

Magnetic properties were characterized by a SQUID magnetometer (Quantum Design MPMS XL) in magnetic fields up to $7$\,T, and in the temperature range of $2$\,K to $300$\,K. Resistivity measurements were carried out by a Quantum Design Physical Property Measurement System from $1.8$\,K to $300$\,K. The AC resistivity ($f=17$\,Hz, $I = 1$\,mA) was measured by the standard four-probe method using Pt wires (0.002 inch diameter) applied with silver-filled epoxy (EPO-TEK\textsuperscript{®} H20E).

\subsection{High pressure single crystal x-ray diffraction}
Pressure was generated by the Boehler-Almax plate diamond anvil cells (DACs) with $500$\,$\mu$m culet size and an opening angle of about $60^{\circ}$. To ensure the hydrostatic environments up to 9.0\,GPa, helium was used as the pressure transmitting medium. All high-pressure single-crystal x-ray diffraction data of CeTiGe$_3$ were collected at 13-BM-C, GSECARS of the Advanced Photon Source, Argonne National Laboratory with the x-ray wavelength of 0.4341\,\si{\angstrom}. Diffraction data were collected using a PILATUS3 1M (Dectris) detector. The exposure time was set as $1$\,s$/^{\circ}$, and each diffraction image covered 1 degree in the $\phi$ axis. The collected x-ray images were reduced using the APEX3 package and related lattice parameters were obtained~\cite{bruker2016apex3}. The crystal structures of CeTiGe$_3$ can be solved by the direct method using SHELXS-97~\cite{Sheldrick:sc5010} and refined with SHELXL interfaced by Olex2-1.2~\cite{Dolomanov:kk5042}. All the Ce, Ti, and Ge atoms were refined anisotropically. 
%%%%%%%%%%%%%%%%%%%%%%%%%%%%
%%%%%%%%%%%%%%%%%%%%%%%%%%%%
%%%%%%%%%%%%%%%%%%%%%%%%%%%%
\section{Results and Discussion}
\subsection{Different responses in lattice parameters}
\begin{figure}[!htp]
\includegraphics[width=\linewidth]{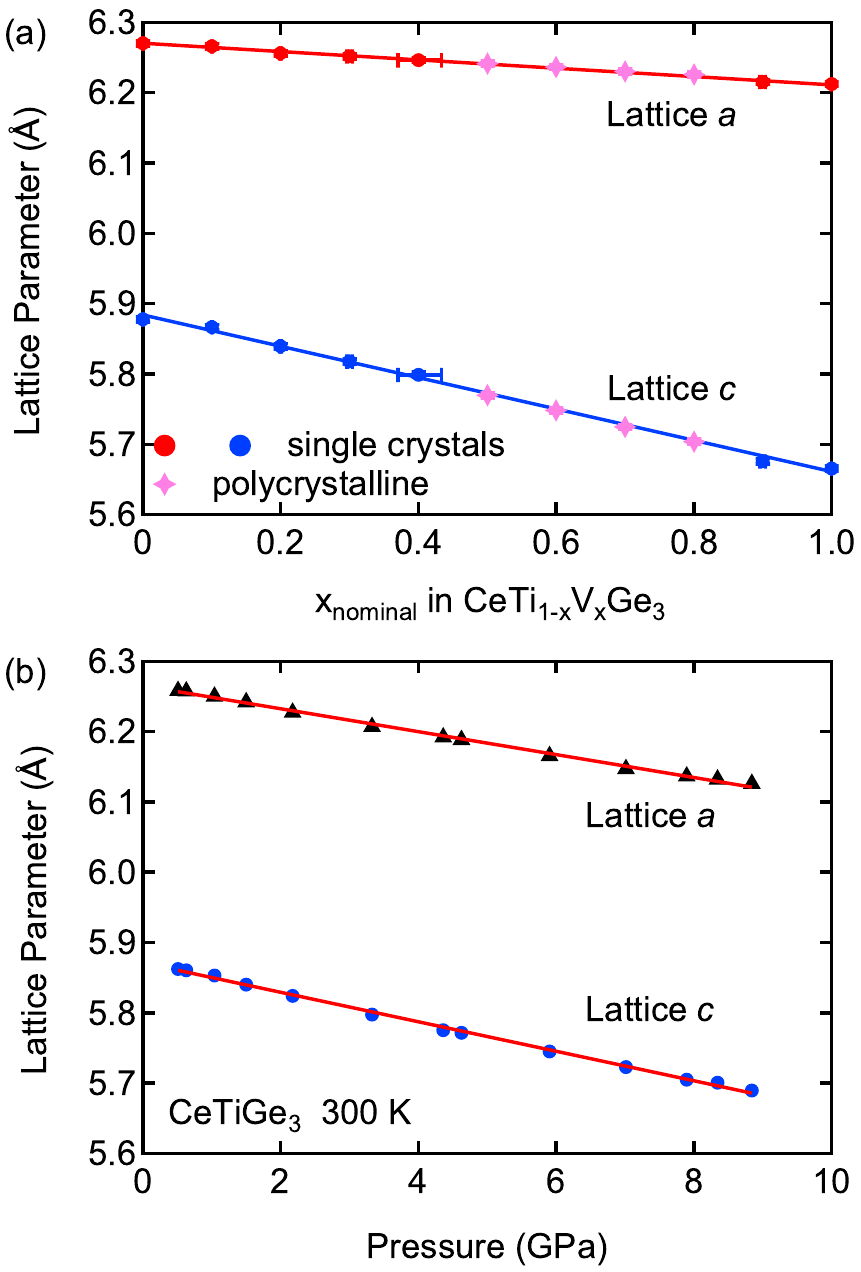}
\caption{(a) Lattice parameters of the CeTi$_{1-x}$V$_x$Ge$_3$ in both the single crystals and the polycrystalline samples obey the Vegard's law. (b) Lattice parameters of the CeTiGe$_3$ as a function of pressure.}
\label{fig:Lattice_parameter}
\end{figure}

Rietveld refinements for all the CeTi$_{1-x}$V$_x$Ge$_3$ samples were performed using the program GSAS II. The lattice parameters are shown in Fig.~\ref{fig:Lattice_parameter}(a). The lattice parameters $a$ and $c$ follow the Vegard's law, which also agrees with the previous report~\cite{Kittler:2013bc}. This further supports that the substitution is successful. The error bars for the lattice parameters are estimated based on multiple measurements on pure Si, which is about 0.06\%.

The lattice parameters of CeTiGe$_3$ under pressure, shown in Fig.~\ref{fig:Lattice_parameter}(b), were obtained from our high pressure single crystal x-ray diffraction measurements. We did not observe any pressure-induced phase transition up to $9.0$\,GPa at room temperature. 

\begin{figure}[!htb]
\center
\includegraphics[width=\linewidth]{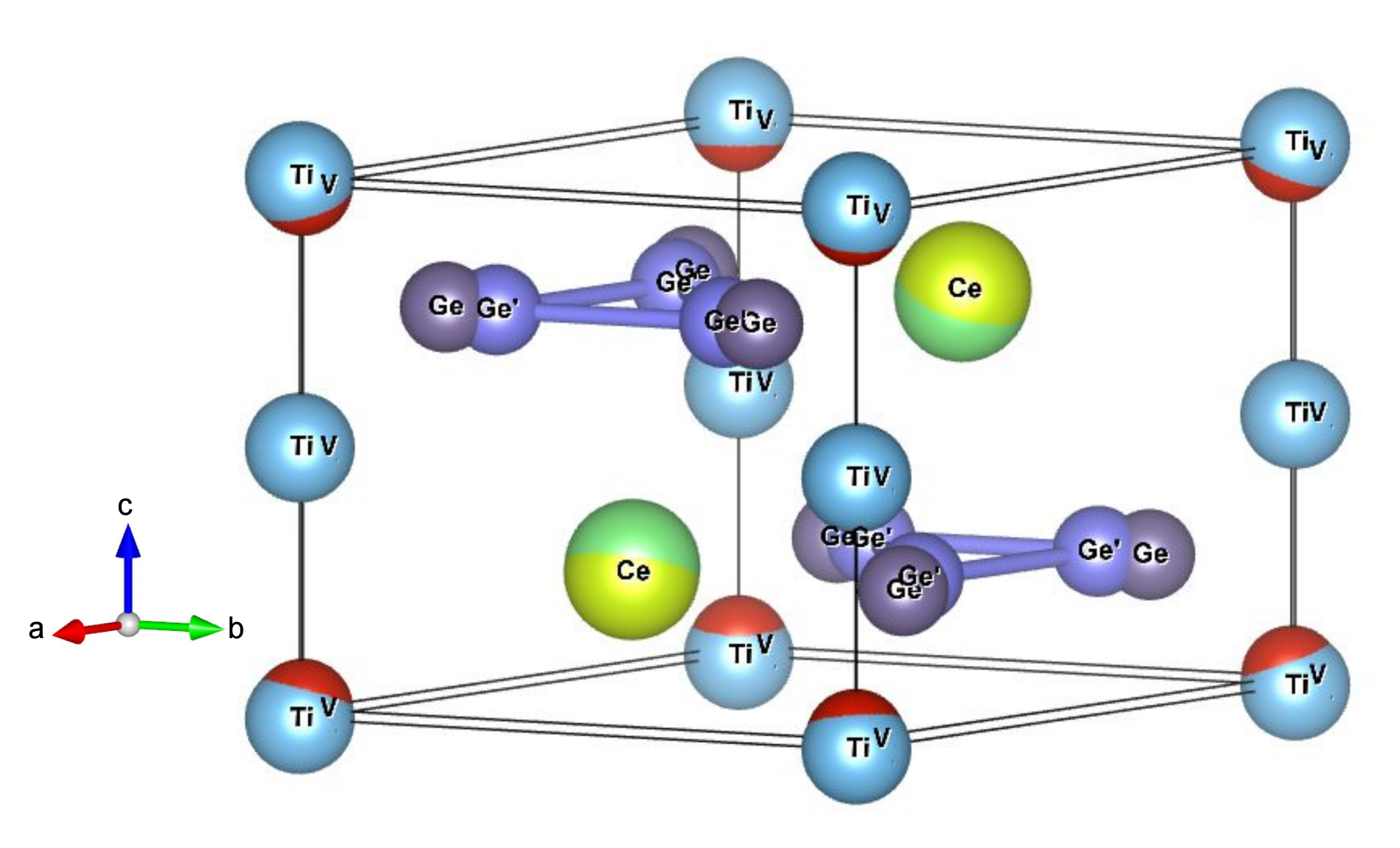}
\caption{Overlapped unit cells of CeTiGe$_3$ and CeVGe$_3$. The blue atom is Ti, the red atom is V, the yellow and green atoms are the Ce atom from CeTiGe$_3$ and CeVGe$_3$ respectively. The Ge atoms from CeVGe$_3$ were labeled with a prime notation.}
\label{fig:CeTiVGe3_lattice}
\end{figure}

As visible in Fig.~\ref{fig:Lattice_parameter}, the relative rate at which the lattice constants $a$ and $c$ change is different in substitution versus pressure. In the pressure work, the lattice constants $a$ and $c$ change about $0.26$\,\%/GPa and $0.36$\,\%/GPa, respectively. In the substitution work, the lattice constants $a$ and $c$ change about $0.0095$\,\%/V\% and $0.038$\,\%/V\%. Therefore, for the lattice constant $a$, the pressure effect is about $0.0364$\,GPa/V\%, but the effect is about $0.106$\,GPa/V\% on lattice constant $c$. This result shows that the substitution of vanadium has much stronger chemical pressure effect along the $c$ axis than along the $a$ axis. Figure~\ref{fig:CeTiVGe3_lattice} shows the overlapped unit cells of CeTiGe$_3$ and CeVGe$_3$. We can see that the Ti/V atoms are primarily shifted along the $c$ direction. This observation explains why the chemical substitution has a stronger pressuring effect along the $c$ axis. On the other hand, the Ge atoms get squeezed closer to each other in the $ab$ plane in response to the chemical pressure.

\subsection{The magnetic phase diagram and the CEF evolution with V substitution}

\begin{figure}[!htp]
\center
\includegraphics[width=\linewidth]{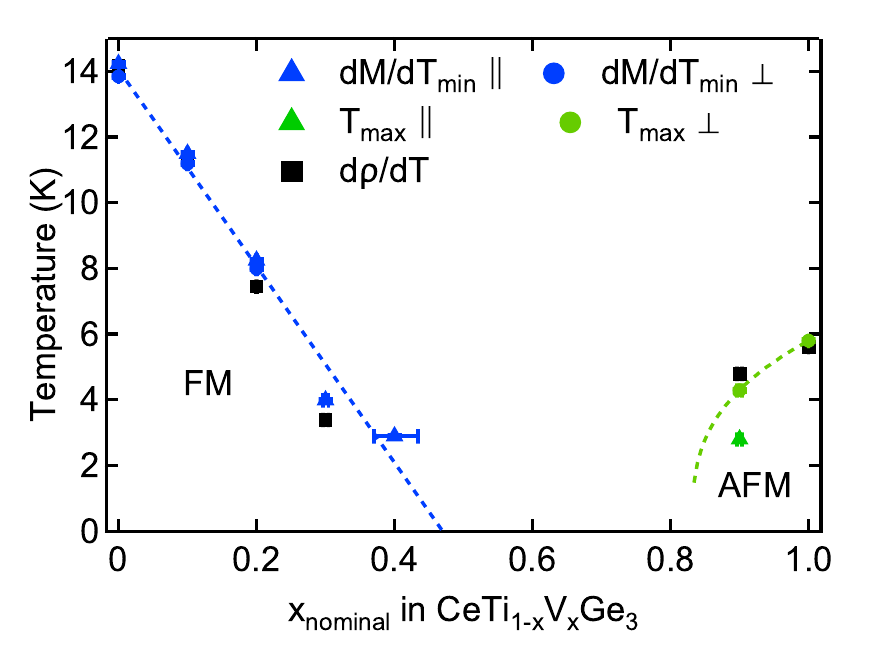}
\caption{The phase diagram of CeTi$_{1-x}$V$_x$Ge$_3$ obtained by magnetization and resistivity measurements.}
\label{fig:phase_diagram}
\end{figure}

The detailed phase diagram of CeTi$_{1-x}$V$_{x}$Ge$_3$ shown in Fig.~\ref{fig:phase_diagram} was obtained by combining magnetization and resistivity data. A striking result from this phase diagram is that the FM and AFM regions are separated: there is no ordering in the region separating the FM and AFM ground state, as opposed to the pressure phase diagram \modif{where both regions are connected around $4.1$\,GPa}~\cite{Kaluarachchi2018PRB}.

\begin{figure}[!htp]
\includegraphics[width=\linewidth]{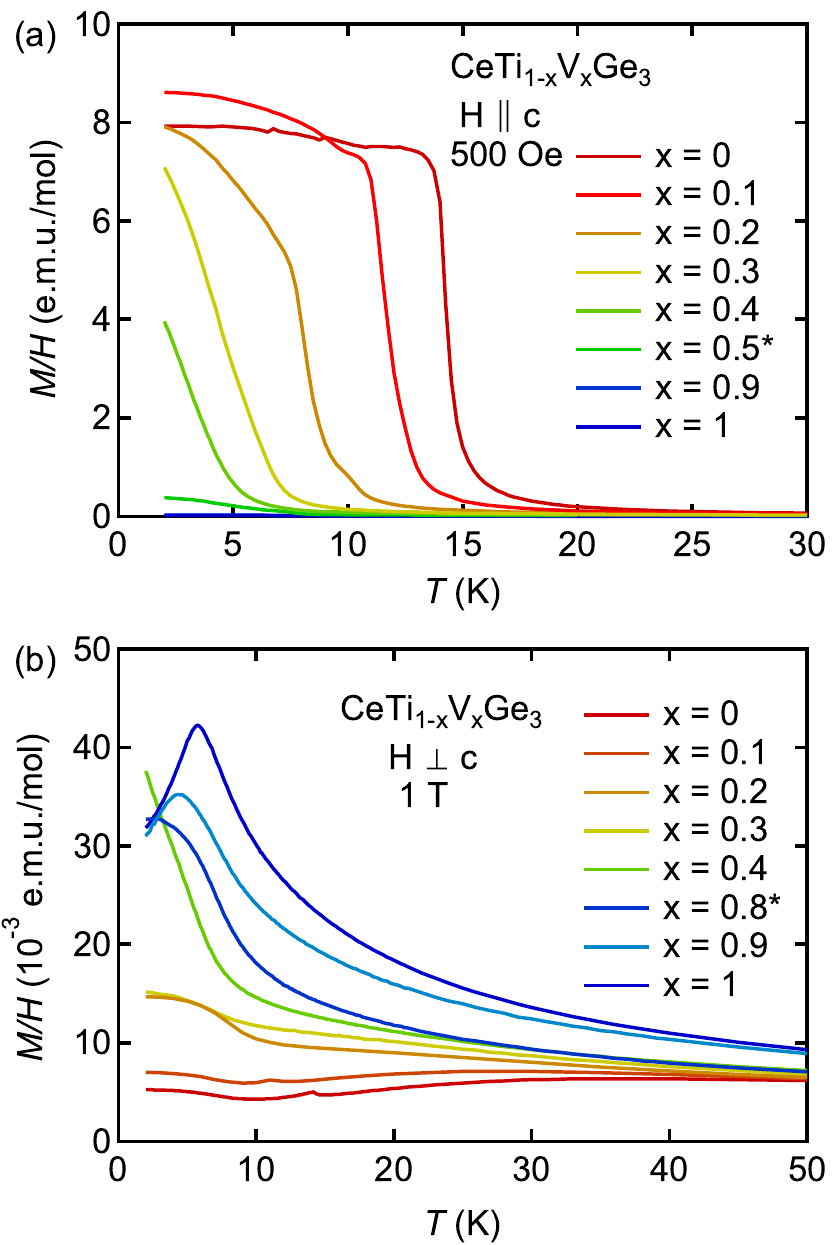}
\caption{Magnetic susceptibility vs temperature of CeTi$_{1-x}$V$_x$Ge$_3$ in (a) parallel to the $c$ axis, and in (b) perpendicular to the $c$ axis ($ab$ plane). Samples that are labeled with the star(*) symbol are polycrystalline.}
\label{fig:MvsT}
\end{figure}

Magnetic susceptibility as a function of temperature along the $c$ axis with applied field of $0.05$\,T below $30$\,K, and along the $ab$ plane with applied field of $1$\,T below $50$\,K, are shown in Fig.~\ref{fig:MvsT}. Only the field-cooled data are shown for simplicity. For the Ti-rich region (FM region), the Curie temperature $T_\textrm{C}$ is determined by the peak of the derivative of the magnetization along the easy axis (parallel to $c$ axis). There is a reduction of $T_\textrm{C}$ with an increase in substitution, which suggests a possible QCP at $x_c\approx0.45$, which is higher than the results of polycrystals for which $x_c\approx0.35$~\cite{Kittler:2013bc}.

For the V-rich region (AFM region), the N\'eel temperature is determined by the maximum of the magnetization along the easy plane (perpendicular to $c$ axis), denoted as $T_{max\bot}$. A similar reduction of $T_{max\bot}$ with increasing Ti substitution is observed. Due to the lack of single crystals in $x = 0.5$ to $0.8$ region, however, we are limited to data from polycrystalline samples and there is no obvious maximum above $2$\,K in the $x = 0.8$ sample as shown in Fig.~\ref{fig:MvsT}(b). 
In the polycrystalline samples, due to the presence of a small amount of CeGe$_{1.75}$, the derivative of the magnetization data shows that there is always a small contribution from CeGe$_{1.75}$ near $7$\,K. Since there are no other anomalies, we conclude that there is no magnetic ordering of CeTi$_{1-x}$V$_{x}$Ge$_3$ from $x = 0.5$ to $x= 0.8$ above $2$\,K. Of course, lower temperature measurements on single crystals and a method to avoid the CeGe$_{1.75}$ impurity would be desirable to better study the suppression of the antiferromagnetic order.
We note that for the CeTi$_{0.1}$V$_{0.9}$Ge$_3$ single crystal sample, the $T_{max\parallel}$ along the $c$ axis (hard axis) is smaller than $T_{max\bot}$ along the easy plane.

\begin{figure}[!htp]
\includegraphics[width=\linewidth]{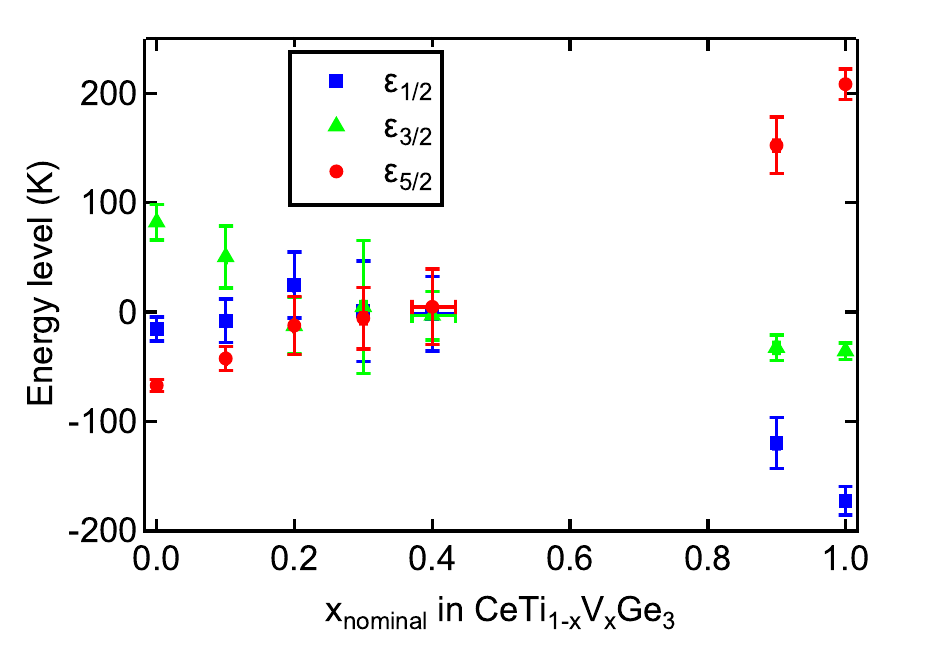}
\caption{The CEF energy levels $\epsilon _{1/2}$, $\epsilon _{3/2}$, and $\epsilon _{5/2}$ with respective uncertainties derived from the magnetic susceptibility of the CeTi$_{1-x}$V$_x$Ge$_3$ single crystals.}
\label{fig:energylevel}
\end{figure}

Further analysis of the magnetization data is shown in the \hyperref[appendix:mag]{Appendix} section. In particular, we find that the Curie-Weiss temperature along the easy axis becomes negative from $x = 0.2$, and in the CEF analysis, the molecular field contribution $\lambda$ remains negative in most samples.  This negative $\theta_\mathrm{CW}$ or the negative molecular field contribution $\lambda$ in the ferromagnetic region is very likely due to the Kondo interaction~\cite{MANFRINETTI2005444,Gruner_1974,PhysRevLett.35.1101}. We also observe that the magnetic moment in the polycrystalline samples ($0.5 \leq x \leq 0.8$ region) remains near the theoretical value for Ce$^{3+}$ despite the absence of ordering above $2$\,K. The magnetic susceptibility data in the pure CeTiGe$_3$ and CeVGe$_3$ can be well explained by the CEF model. Figure~\ref{fig:energylevel} shows the evolution of the CEF energy levels derived from the magnetic susceptibility data (see \hyperref[appendix:CEF]{Appendix B}). The proposed CEF energy levels suggest that there is a gradual change of the ground state from $\ket{\pm 5/2}$ in CeTiGe$_3$ to $\ket{\pm 1/2}$ in CeVGe$_3$, which is compatible with the change of anisotropy from axial to planar around $x\approx0.6-0.8$ in $M$ vs $H$ measurement. The proposed CEF energy levels along the Ti/V substitution also suggest a suppression of CEF splitting energies near the quantum critical region. The detailed discussion of the CEF model, the CEF fittings to the inverse magnetic susceptibility data, the evolution of the CEF parameters, and of the energy levels can be found in~\hyperref[appendix:CEF]{Appendix B}.

As one of the rare examples among Kondo lattice ferromagnets that actually orders along the CEF easy axis~\cite{KLFM}, we indeed observe a large ratio of $M_{\parallel c}^{7\,T}/ M_{\perp c}^{7\,T}$ being 55.4 in the pure CeTiGe$_3$ sample, and the value drops to 4.4 in the $x=0.4$ sample. Although the value is considered small compared to that of pure CeTiGe$_3$ and CeRu$_2$Al$_2$B, we do not observe the crossing of the planar and axial magnetic susceptibility curves that was observed for the hard axis ordering in other Kondo-lattice ferromagnets~\cite{KLFM}.

%%%%%%%%%%%%%%
%%%%%%%%%%%%%%
%%%%%%%%%%%%%%
%%%%%%%%%%%%%%

%%%%%%%%%%%%%%%%%%%%%%%%%%%%%%%%%%%%%%%%%%%%%%%%%%%%%%%%%%%%%%%%%%%%%%%%%%

%%%%%%%%%%%%%%%%%%%%%%%%%%%%%%%%
%%%%%%%%%%%%%%%%

\subsection{Resistivity behavior near QCP}
\begin{figure}[!htp]
\centering
\includegraphics[width=\linewidth]{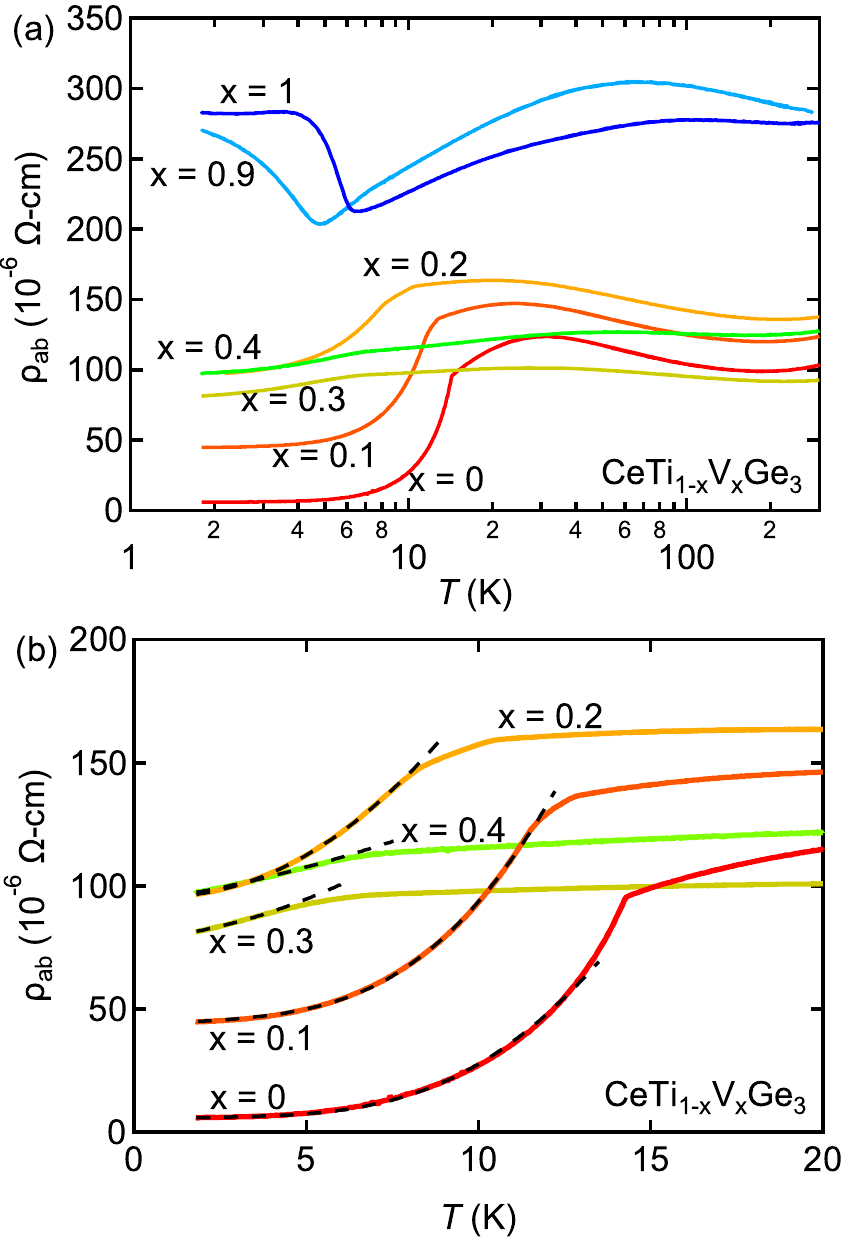}
\caption{(a) Temperature dependence of the electrical resistivity along the $ab$ plane $\rho_{ab}(T)$ for all of the single crystals ($x = 0, 0.1, 0.2, 0.3, 0.4, 0.9, 1$) in log$T$ scale. (b) The low-temperature $\rho_{ab} (T)$ data for samples in the FM region ($x \leq 0.4$). The black-dashed lines are the fitting of Eq.~(\ref{resistFM}) or a power law, $\rho (T) = \rho_{0} + A^\prime T^n$
}
\label{fig:resistivity}
\end{figure}

Figure~\ref{fig:resistivity} shows the temperature dependence of the electrical resistivity along the $ab$ plane $\rho_{ab}(T)$ for the single crystals. We determined the transition temperature based on the peak in the temperature derivative of the electrical resistivity. The results are shown in the phase diagram in Fig.~\ref{fig:phase_diagram}. The transition temperature based on the resistivity measurements agrees with the magnetization measurements, except that there is no clear peak in the derivative in the $x = 0.4$ sample. This is consistent with the larger inhomogeneity of the $x = 0.4$ single crystals.

To further investigate the influence of spin fluctuation, we analyze the resistivity data in our system using the expression~\cite{Sidorov:2003et,Khan:vx}:
\begin{equation} \label{resistFM}
\rho (T) = \rho_{0} + AT^2 + BT\Delta^{-1}(1 + 2T\Delta^{-1})e^{-\Delta/T}
\end{equation}
where $\rho_{o}$ is the residual resistivity, $A$ is the coefficient for electron-electron scattering, $B$ is the coefficient corresponding to electron - magnon scattering, and $\Delta$ is the energy gap in the magnon excitation~\cite{f:wd}. Fits using Eq.~(\ref{resistFM}) are shown as dashed lines in in Fig.~\ref{fig:resistivity}(b) for samples $x=0, 0.1, 0.2$. The low temperature data for $x = 0.3$ and $0.4$ does not fit well to either Fermi-liquid behavior, $\rho (T) = \rho_{0} + AT^2$, nor Eq.~(\ref{resistFM}). Therefore, we try to fit the low temperature $\rho (T)$ using a power-law: $\rho (T) = \rho_{0} + A^\prime T^n$, and we observed that the resistivity exponent decreases, exhibiting a non-Fermi-liquid behavior that is compatible with previous studies in metallic ferromagnetic systems~\cite{Yamamoto15704,Smith2008}. The fitting curves are shown as dashed lines in Fig.~\ref{fig:resistivity}(b). The fitting parameters are summarized in Table~\ref{resis_data}.

\begin{table}[!htp]
\caption{Fitting parameters to Eq.~(\ref{resistFM}) or a power law obtained from the low temperature resistivity data with the current along $ab$ plane in the FM region of CeTi$_{1-x}$V$_x$Ge$_3$ with $x\leq0.4$, and the temperature of the maximum in resistivity $T_{\textrm{max,}\rho}$ in the last column}
\begin{tabular}[t]{p{0.05\linewidth}p{0.15\linewidth}p{0.26\linewidth}p{0.23\linewidth}p{0.07\linewidth}|l}
\toprule
$x$ &$\rho_{0}$ & $A$ & $B$ & $\Delta$ & $T_{\textrm{max,}\rho}$\\
 & ($\mu\Omega$ cm) & ($\mu\Omega$ cm K$^{-2}$) & ($\mu\Omega$ cm K$^{-1}$) & (K) & (K)\\
\midrule
0&5.6&0.056&0.60&29&31.4\\
0.1&45&0.20&0.95&27&24.1\\
0.2&95&0.53&0.24&7.6&20.8\\
\bottomrule
\toprule
$x$ &$\rho_{0}$ & $A'$ & $n$ &  & $T_{\textrm{max,}\rho}$\\
 & ($\mu\Omega$ cm) & ($\mu\Omega$ cm K$^{-2}$)&&&(K)\\
\midrule
0.3&79&0.94&1.75& &27.0\\
0.4&94&1.4&1.40& &56.7\\
\bottomrule
\end{tabular}
\label{resis_data}
\end{table}

We find that for $x \leq 0.2$, the residual resistivity $\rho_{0}$ and the electron-electron scattering coefficient $A$ consistently increases with increasing vanadium content. The energy gap $\Delta$ decreases with increasing V content with a sharp drop at $x = 0.2$, above which the model no longer describes the temperature dependence. For $x = 0.3$ and $0.4$, the power-law behavior gives a slight decrease in residual resistivity compared to the $x = 0.2$ sample, and continues the increasing trend of the coefficient $A$. Overall, these results are consistent with the departure from Fermi liquid behavior near the possible quantum critical point around $x\approx0.45$, even though the reduced homogeneity of the $x=0.4$ single crystal and the lack of single crystals at $x=0.5$ restrict further analysis of the resistivity.

The last column of the Table~\ref{resis_data} shows the temperature of the maximum in resistivity, which is associated with the Kondo effect. The temperature decreases for $x\leq 0.2$ and then increases for $x\geq0.3$ indicating \modif{a complicated evolution of the interplay} between the Kondo effect, and the CEF effect as a function of the Ti/V substitution.

\modif{In the V-rich side, the low-T resistivity above 2\,K cannot be fitted to the expression with the AFM magnon contribution~\cite{CeAgSb2}. We note that the residual resistivity at 2\,K is much higher than even that of the polycrystalline samples, and this could be due to the presence of the CeGe$_{1.75}$ impurity phase in the V-rich samples. This observation is consistent with different samples in different batches, and with the previous study that also reported the presence of the CeGe$_{1.75}$ phase and a similar residual resistivity of around 300\,$\mu\ohm$ in their sample~\cite{Inamdar:2014io}. We also observe the upturn of the resistivity below $T_N$, but the resistivity remains relatively constant as the temperature decreases, unlike in the previous report where the resistivity drops~\cite{Inamdar:2014io}. These features might be the result of the opening of an energy gap from a spin density wave (SDW)~\cite{Inamdar:2014io}. A similar jump due to a SDW state has also been observed in URu$_2$Si$_2$~\cite{URu2Si2}, BaFe$_{2-x}$Co$_x$As$_2$~\cite{BaFeCoAs2}, and Mo$_3$Sb$_7$~\cite{Mo3Sb7}. In this case, the height of the resistivity jump may reflect the magnitude of the energy gap, and thus $T_{\mathrm{\scriptscriptstyle{SDW}}}$ and the energy gap are both suppressed by the substitution by Ti. Given that compounds with SDW have very interesting properties under pressure such as superconductivity, it will be interesting to study how pressure affects the properties of CeVGe$_3$. We did not observe superconductivity or other anomaly above 1.8\,K when the SDW is suppressed by Ti substitution in polycrystalline samples.}

\begin{figure*}[!htb]
\center
\includegraphics[width= \textwidth]{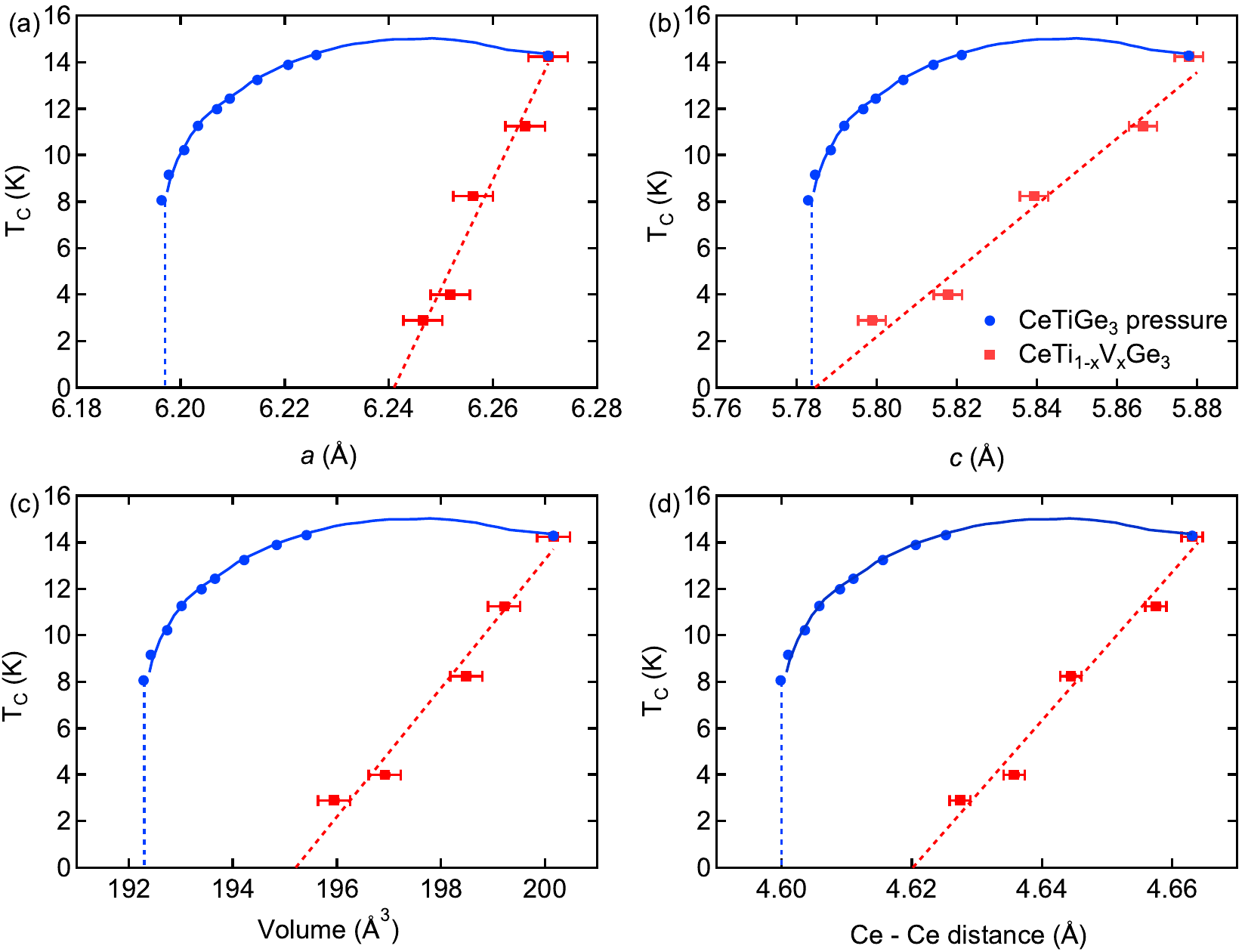}
\caption{The comparison of the Curie temperature of CeTiGe$_3$ under pressure and under V substitution (CeTi$_{1-x}$V$_x$Ge$_3$) as a function of lattice constant (a) $a$, (b) $c$, (c) unit-cell volume, and (d) Ce-Ce distance. The critical temperature under hydrostatic pressure is taken from Ref.~\cite{Kaluarachchi2018PRB}.}
\label{fig:4Tc}
\end{figure*}

\subsection{Comparison with the pressure phase diagram}
The phase diagram of CeTiGe$_3$ under pressure has already been reported~\cite{Kaluarachchi2018PRB}. With the addition of our structural measurements under pressure, we are now able to compare the pressure and chemical doping effects. At first glance, unlike the first order transition from FM to AFM observed in the pressure paper, \modif{ V substitution results in a FM QCP with a clean separation of the FM and AFM regions. The shape of the phase diagrams in both works are in line with the expectations that the quantum ferromagnetic transition in clean metals is a first-order transition with tricritical wings under an external magnetic
field~\cite{Kaluarachchi2018PRB}, and that the strong disorder leads to a continuous second-order QCP~\cite{Fritsch2015,disorderFMQPT,Brando2016RMP}.}

The Curie temperatures of CeTiGe$_3$ under pressure and of CeTi$_{1-x}$V$_x$Ge$_3$ are plotted together in Fig.~\ref{fig:4Tc}, as a function of lattice parameters $a$, $c$, lattice volume, and Ce-Ce distance. The critical temperatures of the pressure study were taken from Ref.~\cite{Kaluarachchi2018PRB}. The lattice parameters of CeTiGe$_3$ under pressure, shown in Fig.~\ref{fig:Lattice_parameter}(b), as well as the critical temperatures and lattice parameters of CeTi$_{1-x}$V$_x$Ge$_3$ were obtained from this work. To construct the phase diagram in the pressure work, we first identify the transition temperatures at given pressures from Ref.~\cite{Kaluarachchi2018PRB}, and convert the pressures to lattice $a$ or $c$ based on the linear relationship shown in Fig.~\ref{fig:Lattice_parameter}(b). With simple calculations, we can also plot the phase diagram as a function of lattice volume and Ce-Ce distance.

By comparing the FM regions in both works, we observe that under the parameter of lattice constant $c$, the suggested QCP in the substitution work is very close to the point where the FM order disappears in the pressure work, which is about $c = 5.78$\,\si{\angstrom}.

\begin{figure}[!htb]
\centering
\includegraphics[width=\linewidth]{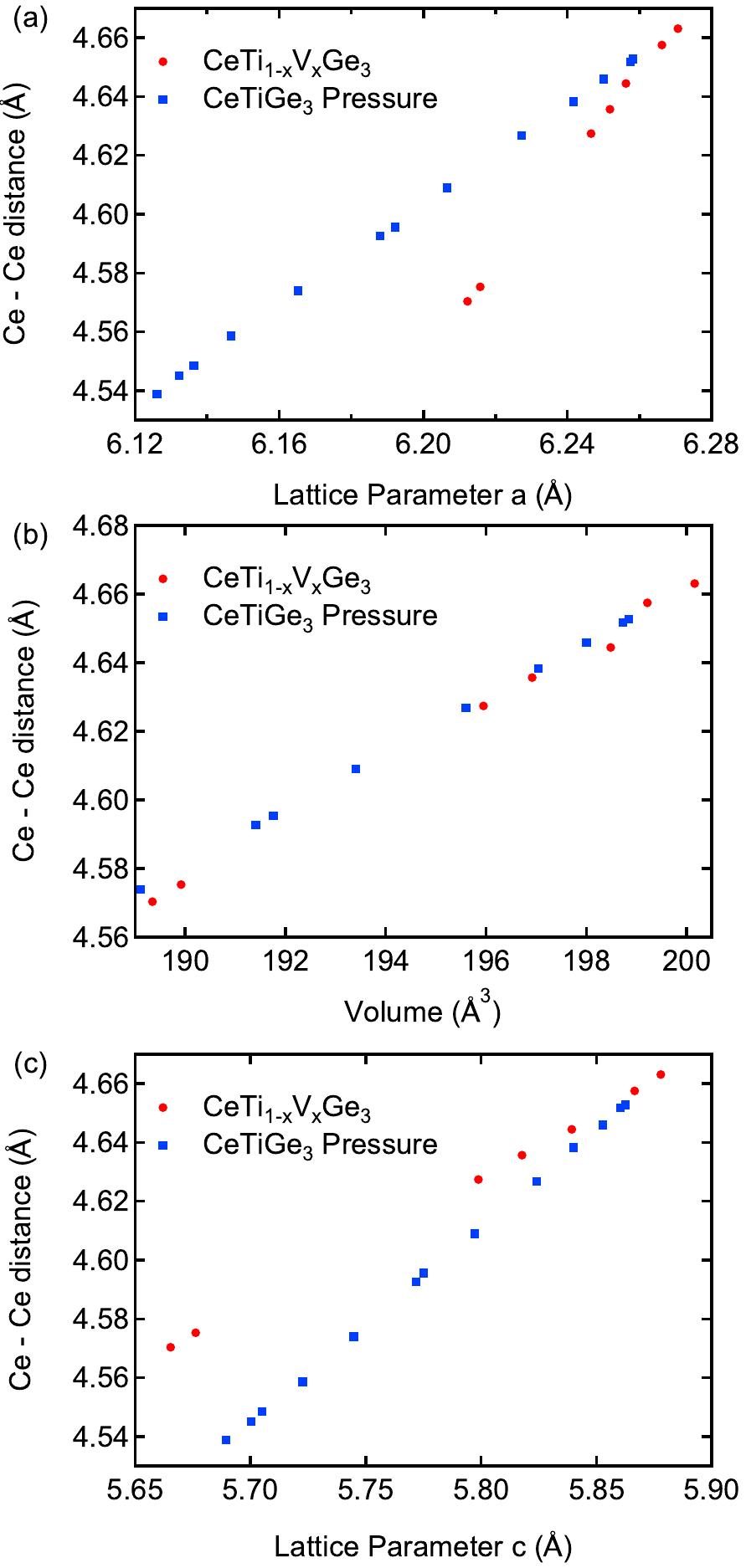}
\caption{The Ce-Ce distance of both substitution and hydrostatic pressure work as a function of (a) lattice parameter $a$, (b) lattice parameter $c$, and (c) lattice volume.}
\label{fig:CeCe}
\end{figure}

Given that CeTiGe$_3$ is a ferromagnetic Kondo-lattice system, the ferromagnetic order originates from the Ruderman-Kittel-Kasuya-Yoshida (RKKY) interaction, which is mainly affected by the Fermi wavevector and the nuclei distance, i.e., the Ce-Ce distance~\cite{PhysRev.96.99,10.1143/PTP.16.45}. The magnetic ordering temperature often can be described by the Doniach phase diagram as a result of a competition between the Kondo effect and the RKKY interaction~\cite{DONIACH1977231,PhysRevB.56.11820}. \modif{At first glance, the faster suppression
of the ferromagnetism in the chemical doping study could
be explained by the stronger Kondo effect from the excess
screening electrons provided by vanadium. However, the non-monotonic evolution of the temperature of the maximum in resistivity $T_{max,\rho}$ suggests a complicated evolution of the interplay between the Kondo effect and CEF effect, and thus the appearance of the FM QCP is a consequence of this complex competition between the RKKY interaction, the Kondo effect and the CEF effect. The single and broad resistivity maximum across the whole substitution study or under pressure~\cite{Kaluarachchi2018PRB} suggests that the CEF splitting energy $\Delta_\mathrm{CEF}$ and the Kondo coherence energy scale $k_BT_{coh}$ are of similar magnitude~\cite{KondoCEF_Jackson}. However, the increasing trend of $T_{max,\rho}$ under pressure suggests that the CEF splitting energy could be increasing under pressure, whereas V substitution causes it to decrease. This difference might explain the different phase diagrams obtained in the pressure and substitution studies.}

While it is difficult to determine the Fermi wavevector, one can easily obtain the Ce-Ce distance based on the lattice parameters. In Fig.~\ref{fig:CeCe}, we plot the Ce-Ce distance as a function of lattice parameter $a$, $c$ and lattice volume. We observe the anisotropic lattice changes in the substitution work, which is manifested as different slopes compared to the pressure one in Figs.~\ref{fig:CeCe}(a) and \ref{fig:CeCe}(b). We can put this in relation to the results discussed earlier that the chemical pressure effect is stronger in the $c$-direction. By comparing the different responses of the lattice parameters in both the hydrostatic pressure and chemical substitution works, we see that the hydrostatic pressure effect is relatively isotropic while the chemical pressure effect due to V substitution is anisotropic. But surprisingly, in Fig.~\ref{fig:CeCe}(c), the Ce-Ce distance as a function of lattice volume, which can also be interpreted as a function of lattice pressure, is similarly changed by chemical/hydrostatic pressure. This suggests that the anisotropic lattice change induced by chemical substitution is not solely responsible for the difference in magnetic ordering.

At the same time, the changes in lattice constants will inevitably lead to a change of the density of states. In particular for the chemical substitution with V, the anisotropic changes in lattice constants will result in a different change of the density of states. In addition, V substitution also provides extra electrons, and therefore an additional shift in the Fermi level. \modif{A substitution series of CeTi(Ge$_{1-x}$Si$_{x}$)$_{3}$ would be ideal since it will also mimic the pressure effect without directly altering the band filling. However, to the best of our knowledge, there are no reports of partial substitution of Ge by Si. The end compound CeTiSi$_3$ has not been reported either, so it is likely this
substitution is not feasible.}

We also already mentioned that the Ge atoms get squeezed closer in the $ab$ plane in response to the chemical pressure. This will inevitably change the surrounding of Ce atoms, and could possibly be responsible for the change of magnetic anisotropy, and the faster suppression of the ferromagnetism due to larger screening effects from the extra electrons. It is quite remarkable that, despite all these effects, the same critical lattice parameter $c$ is observed in the pressure and substitution studies. 

We hope our paper will stimulate calculations of the change in electronic structure under pressure and from the chemical substitution to shed light on the qualitative difference in magnetic phase diagrams. In particular, the $q$ dependence of the RKKY interaction $J$, which depends on the details of the Fermi surface should depend strongly on the Ti/V substitution since the Ce $f$ electrons are mostly localized.

%%%%%%%%%%%%%%%%

\section{Conclusions}

In summary, we have successfully synthesized single crystals of CeTi$_{1-x}$V$_x$Ge$_3$ ($x = 0, 0.1, 0.2, 0.3, 0.4, 0.9$ and $1$) by the flux method, and polycrystalline samples from $x = 0.5$ to $x = 0.8$. Magnetic properties have been studied for all the samples, and electrical resistivity measurements were performed on the single crystal samples.  We observed a suppression of ferromagnetism with increasing content of V, with a suggested QCP at around $x = 0.45$. No magnetic ordering was observed in the polycrystalline samples from $x = 0.5$ to $x = 0.8$ above $2$\,K. \modif{We perform a detailed CEF analysis on the magnetization data in all the single crystals, and the magnetic susceptibility data in the pure CeTiGe$_3$ and CeVGe$_3$ can be well explained by the CEF model. The proposed CEF energy levels suggest that there is a gradual change of the ground state from $\ket{\pm 5/2}$ in CeTiGe$_3$ to $\ket{\pm 1/2}$ in CeVGe$_3$, and a suppression of CEF splitting energies near the quantum critical region.}

\modif{Compared to the magnetic phase diagram from the pressure study of CeTiGe$_3$, the CeTi$_{1-x}$V$_x$Ge$_3$ magnetic phase diagram shows a clear separation between the FM and AFM regions.}

\modif{By comparing the trend and the shape of the maximum in temperature dependent resistivity, we observe that the CEF splitting energy and the Kondo coherence energy are of similar magnitude in both works, but the different temperature trend in resistivity maximum suggests that CEF splitting energy is suppressed by V substitution, but enhanced by pressure. By comparing the changes in lattice constants in both substitution and pressure works, we observed a different response in lattice constants between the hydrostatic pressure study and the substitution study. The anisotropic changes in lattice constants that leads to stronger squeezing effect on Ge atoms in the $ab$ plane and possibly to the change of CEF ground state, plus the extra electrons provided by the V substitution are most likely the main factors for such differences in the magnetic phase diagram between the pressure and the substitution studies.}

\modif{Nevertheless, by comparing the transition temperature as a function of lattice constants, we find a common critical lattice constant $c$ of $5.78$\,\si{\angstrom} where the FM ordering is suppressed in both studies.}

%%%%%%%%%%%%%%%%%%%%%%%%%%%%%%%
%%%%%%%%%%%%%%%%%%%%%%%%%%%%%%%%%%%%%%%%%%%%%%%%%%%%%%%%%%%%%%%%%%%%%%%%%%%%%

%%%%%%%%%%%%%%%%%%%%%%%%%%%%%%%%%%%%%%%%%%%%%%
\section{Acknowledgement}
This work is financially supported by the Physics Department, University of California, Davis, U.S.A. The authors are grateful to Jing-Tai Zhao from Guilin University of Electronic Technology, Manuel Brando from Max Planck Institute for Chemical Physics of Solids, Rahim Ullah and Sergey Savrasov from University of California, Davis for useful discussions. The authors thank Andrew Thron for discussions and training on the EDS measurement in AMCaT Laboratory in UC Davis. We acknowledge support from the Physics Liquid Helium Laboratory fund. We thank Dr. Sergy Tkachev for the helium gas loading, and Dr.  Dongzhou Zhang   for experimental support of x-ray measurements. The high-pressure single crystal x-ray diffraction data were collected at 13-BM-C of GeoSoilEnviroCARS (The University of Chicago, Sector 13) , Advanced Photon Source (APS), Argonne National Laboratory. GeoSoilEnviroCARS is supported by the National Science Foundation-Earth Sciences (EAR-1634415), and Department of Energy-GeoSciences (DE-FG02-94ER14466). Use of the COMPRES-GSECARS gas loading system and PX2 was supported by COMPRES under NSF Cooperative Agreement EAR-1661511, and by GSECARS through NSF Grant No. EAR-1634415 and DOE Grant
No. DE-FG02-94ER14466. Work at Argonne (sample preparation, characterization, and crystal growth) is supported by the U.S. DOE, Office of Basic Energy Science, Materials Science and Engineering Division. Use of the Advanced Photon Source at Argonne National Laboratory was supported by the U.S. Department of Energy, Office of Science, Office of Basic Energy Sciences, under Contract No. DE-AC02-06CH11357. This work at University of Utah was supported by the U.S. Department of Energy, Office of Science, Fusion Energy Sciences under Award No.
DE-SC0020340 (S.D.). J.C. thanks Undergraduate Research Opportunities Program of University of Utah for providing travel funds to Argonne National laboratory.

%%%%%%%%%%%%%%%%%%%%%%%%%%%%%%%%%%%%%%%%%%%%%%%%%%%%%%%%%%%%%%%%%%%%%%%%%%%%%

\appendix
\section{Magnetic properties of CeTi$_{1-x}$V$_x$Ge$_3$} \label{appendix:mag}
\begin{figure*}[!htb]
\includegraphics[width=0.9\linewidth]{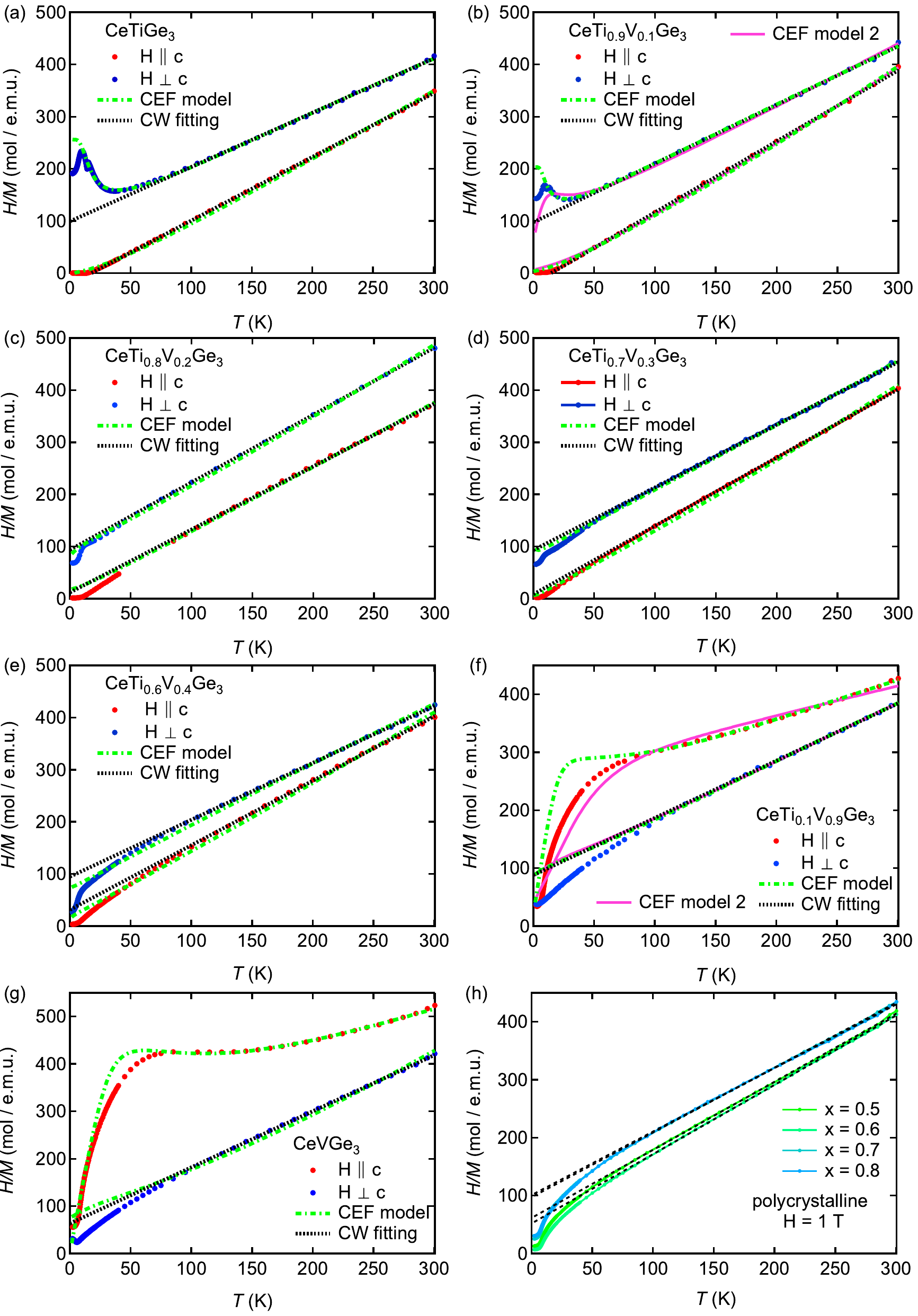}
\caption{Temperature dependence of the inverse magnetic susceptibility of CeTi$_{1-x}$V$_x$Ge$_3$ single crystals [(a)-(g)] along both axes, and all polycrystalline sample in (h). The green-dashed lines show the results from the fitting with the CEF model described in the text, and the black-dashed lines are the results from Curie-Weiss fittings above 100K. The pink lines in (b) and (f) are from the fitting with a different procedure (CEF model 2) described at the end of the section.}
\label{fig:HoverM_all}
\end{figure*}

\begin{figure}[!htb]
\includegraphics[width=0.95\linewidth]{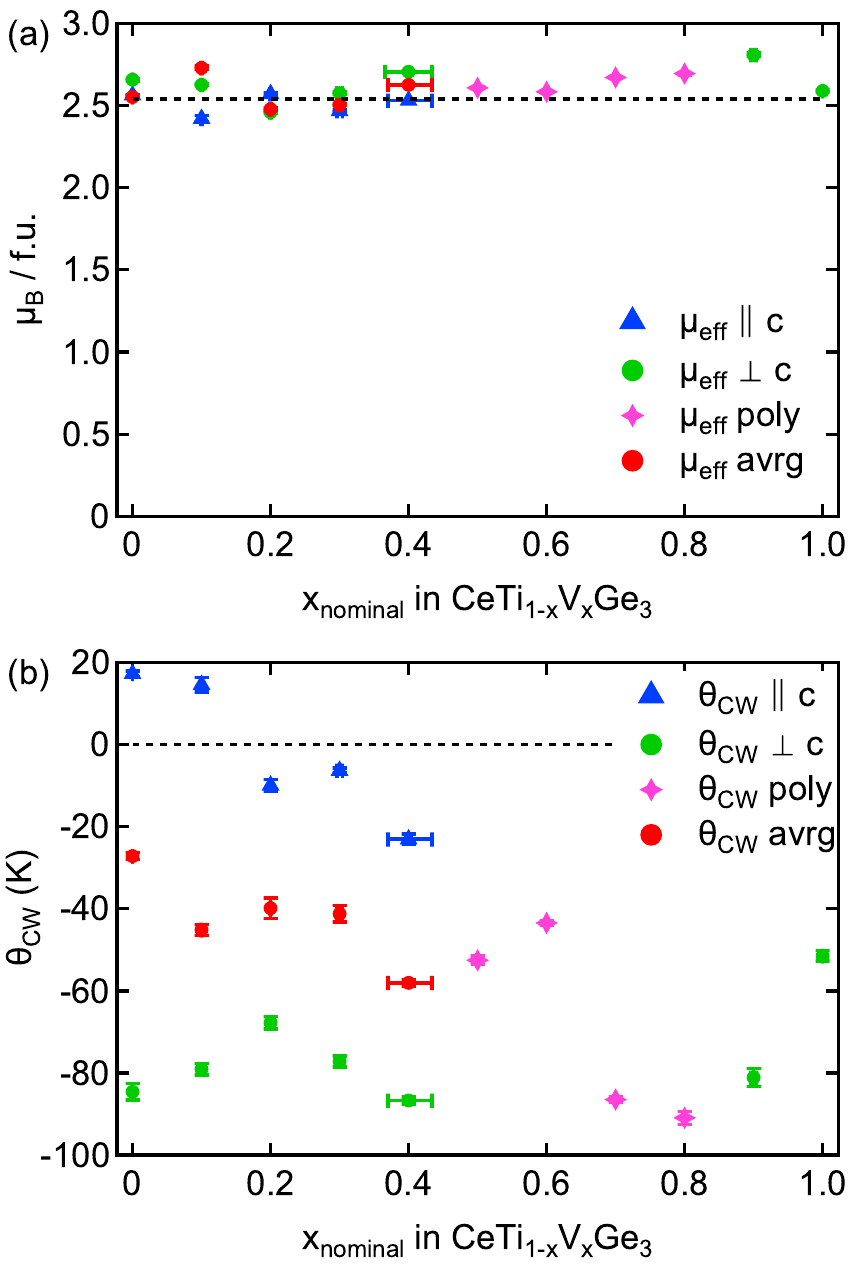}
\caption{(a) The effective paramagnetic moment, $\mu_{\mathrm{\scriptscriptstyle{eff}}}$, and (b) the Curie-Weiss temperature, $\theta_{\mathrm{CW}}$, of CeTi$_{1-x}$V$_x$Ge$_3$ in single crystals along $c$ axis (blue triangle) and $ab$ plane (red circle), their powder average (green circle), and the polycrystalline samples (pink star). The absence of Curie-Weiss fitting results for the AFM region samples along the $c$ axis is due to the bad Curie-Weiss fitting.}
\label{fig:CWT}
\end{figure}

The high temperature susceptibility data for both axes in single crystal samples and in polycrystalline samples from 100\,K to 300\,K were fit to the Curie-Weiss law in Fig.~\ref{fig:HoverM_all}, and the results are shown in Fig.~\ref{fig:CWT}. 

The effective paramagnetic moments $\mu_{\mathrm{\scriptscriptstyle{eff}}}$ of all the samples across the substitution are near the theoretical Ce$^{3+}$  value (2.54$\mu_\mathrm{B}$). The Curie-Weiss temperatures, $\theta_\mathrm{CW}$, are shown in the Fig.~\ref{fig:CWT}(b). In the FM region, the $\theta_{\mathrm{CW}\parallel}$, which is along the easy axis for this region, decreases as substitution is increased and becomes negative at $x = 0.2$, even though the ordering is still ferromagnetic. \modif{Given that the previous study with elastic neutron-scattering experiments confirms ferromagnetic and exclude ferrimagnetic
order of CeTiGe$_3$~\cite{Kittler:2013bc}, this negative $\theta_\mathrm{CW}$ is very likely due to the Kondo interaction~\cite{MANFRINETTI2005444}}. Due to the strong anisotropic magnetic behavior of the system, $\theta_{\mathrm{CW}\bot}$ is much lower than $\theta_{\mathrm{CW}\parallel}$ for $x\leq0.4$. We note that the powder average $\theta_{p}$, where $\chi_{\mathrm{\scriptscriptstyle{avg}}}$ = (2$\chi_{\mathrm{\scriptscriptstyle{ab}}}$ + $\chi_{c}$)/3, remains negative throughout the entire FM region ($x\leq0.4$). This is consistent with the finding of AFM fluctuations across the whole range of CeTi$_{1-x}$V$_x$Ge$_3$ in an NMR study on polycrystalline samples~\cite{Majumder:2018vt}.

In the AFM region, for our single crystal samples in x = 0.9 and x = 1, the susceptibility along the $c$ axis (hard axis) strangely does not fit to the simple Curie-Weiss law, whereas the Curie-Weiss behavior was reported in Inamda's study~\cite{Inamdar:2014io}. \modif{This is due to the larger CEF effect in the V-rich samples, and the high-$T$ behavior can be explained by the CEF model. The detailed CEF analysis is discussed in the next section.}

\begin{figure}[!htb]
\includegraphics[width=0.95\linewidth]{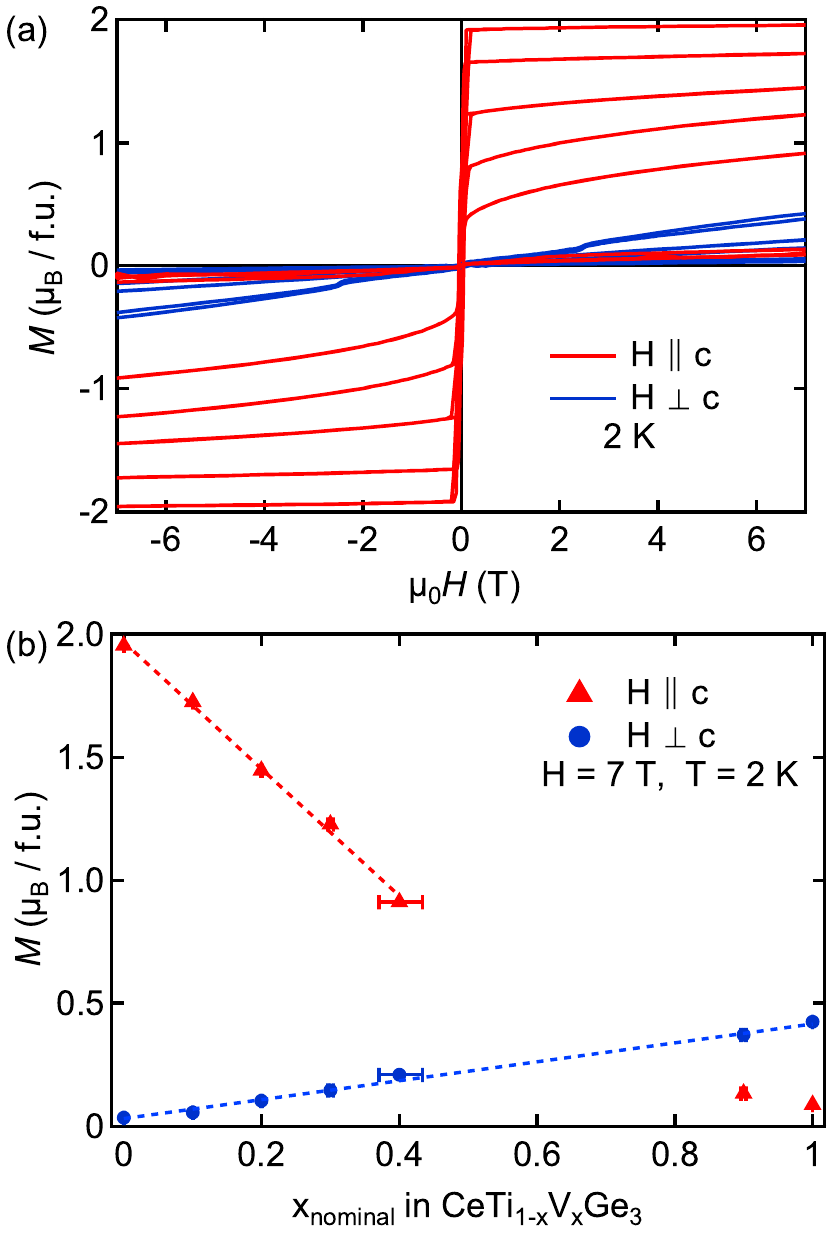}
\caption{(a) Magnetization versus magnetic field of all the single crystals CeTi$_{1-x}$V$_x$Ge$_3$ parallel and perpendicular to the $c$ axis. (b) Magnetization of single crystals CeTi$_{1-x}$V$_x$Ge$_3$ at $7$\,T and $2$\,K parallel and perpendicular to the $c$ axis. We can see a change of anisotropy from easy axis to easy plane as a function of substitution.}
\label{fig:MvsH}
\end{figure}

Figure~\ref{fig:MvsH} is a summary of $M$ vs $H$ data of CeTi$_{1-x}$V$_x$Ge$_3$ along both axes. It shows that $M_{7T, 2K}$, the magnetization of single crystal CeTi$_{1-x}$V$_x$Ge$_3$ measured at $H = 7\,T, T = 2\,K$, in the FM region, were linearly decreasing along the $c$ axis and increasing perpendicular to the $c$ axis. And at some point in the polycrystalline region, there will be a change of anisotropy, where the easy axis changes from the $c$ axis in the FM region to the $ab$ plane in the AFM region. \modif{This reflects the change of CEF ground states, which will be discussed in the next section.} It is also interesting that it seems like the increasing rate of $M^{7\,T}$ perpendicular to the $c$ axis is consistent not only in the FM region but across the entire substitution. In the AFM region, we observed the metamagnetic transition in both $x = 0.9$ and $x = 1$ sample.

\modif{A previous study observed that Ce- and Yb-based Kondo lattice ferromagnets order mainly along the magnetically CEF hard axis, and that this is due to the presence of the Kondo effect with a Kondo temperature close to $T_\mathrm{C}$~\cite{KLFM}. However, CeTiGe$_3$ is one of the very few example in which the moments order along the CEF easy axis because of it's large CEF anisotropy. The magnetization data along both axes at $7$\,T, and $2$\,K shown in Fig.~\ref{fig:MvsH}(b) reflect the anisotropy of the effective $g$ factor, and the ratio of these magnetization values, $M_{\parallel c}^{7\,T}/ M_{\perp c}^{7\,T}$, will also reflect the CEF anisotropy. The calculated values are listed in Table~\ref{CEFanisotropy}. We indeed observe a large magnetization anisotropy of 55.4 in the pure CeTiGe$_3$ sample, and the value drops to 4.4 in the $x=0.4$ sample. Although the value is considered small compared to that of pure CeTiGe$_3$ and CeRu$_2$Al$_2$B~\cite{KLFM}, we do not observe the crossing of the planar and axial magnetic susceptibility curves that was observed for the hard axis ordering in other Kondo-lattice ferromagnets~\cite{KLFM}.}

\modif{As will be shown later, the ground state of CeTiGe$_3$ is a pure $\ket{\pm 5/2}$ state, and that of CeVGe$_3$ is a pure $\ket{\pm 1/2}$ state. The theoretical electron spin resonance (ESR) $g$ factors can be calculated based on the CEF ground-state wavefunctions, where $g_{i}(\mathrm{ESR}) = 2g_{J}\braket{J_{i}}=2M_{i}^{\mathrm{sat}}/\mu_\mathrm{B}$. We find that, for CeTiGe$_3$, the $g_z$ and $g_{xy}$ are 4.28 and 0 respectively, and for CeVGe$_3$, the $g_z$ and $g_{xy}$ are 0.86 and 2.57 respectively. The fact that the $\ket{\pm 5/2}$ ground state has expected $g_{xy} = 0$ is consistent with the large anisotropy in magnetization in CeTiGe$_3$. To compare with the experimental data, instead of using the saturation moment, we use magnetization value at $7$\,T, and $2$\,K to estimate the effective $g$ factors, and the values are listed in Table~\ref{CEFanisotropy}. It matches well with the expected $g$ factors of pure CeTiGe$_3$, but not with that of pure CeVGe$_3$ since the magnetization is far from saturation.}

\begin{table}[!htbp]
\caption{$2M_{i}^{7\,T}/\mu_B$, and the ratio of magnetization along the $c$ axis vs the $ab$ plane at $\mathrm{H} = 7$\,T, $\mathrm{T} = 2$\,K of CeTi$_{1-x}$V$_x$Ge$_3$.}
\renewcommand{\arraystretch}{1.2}
\begin{tabular}{c|lllllll}\hline
 $x$ & 0  & 0.1 & 0.2  & 0.3 & 0.4& 0.9 & 1.0 \\ \hline
 $2M_{\parallel c}^{7\,T}/\mu_\mathrm{B}$ & 3.91 & 3.45 & 2.89 & 2.45 & 1.83 & 0.27 & 0.18 \\ \hline
$2M_{\perp c}^{7\,T}/\mu_\mathrm{B}$ & 0.07 & 0.11 & 0.21 & 0.29 & 0.42 & 0.75 & 0.85 \\ \hline
$M_{\parallel c}^{7\,T}/ M_{\perp c}^{7\,T}$ & 55.4 & 30.2 & 13.8 & 8.4 & 4.4 & 0.36 & 0.21 \\ \hline
\end{tabular}
\label{CEFanisotropy}
\end{table}

\section{Crystalline Electric Field Analysis of CeTi$_{1-x}$V$_x$Ge$_3$ magnetic susceptibility}
\label{appendix:CEF}

It is known that CeTiGe$_3$ and CeVGe$_3$ have strong magnetic anisotropic behavior due to the strong crystalline electric field effect (CEF) and it is crucial to do the CEF analysis to understand the magnetic properties of Ce-based compounds~\cite{CePd2As2}. For example, the temperature dependence of the magnetic susceptibility below room temperature in CeVGe$_3$ and CeTi$_{0.1}$V$_{0.9}$Ge$_3$ deviates from a Curie-Weiss behavior and we will see that such deviation can be explained by the CEF effect. To learn more about the magnetic anisotropy in CeTi$_{1-x}$V$_{x}$Ge$_3$ system, we perform a detailed CEF analysis based on our magnetic susceptibility data. For a Ce atom in a hexagonal site symmetry, the CEF Hamiltonian can be written as, 
\begin{equation} \label{eq:H}
H_{\mathrm{CEF}}=B_{2}^{0}O_{2}^{0}+B_{4}^{0}O_{4}^{0},
\end{equation}
where $B_{m}^{n}$ and $O_{m}^{n}$ are the CEF parameters and the Stevens operators, respectively~\cite{Stevens_1952,HUTCHINGS1964227}. The theoretical expression of the magnetic susceptibility with the Van Vleck contribution is given by

\begin{equation} \label{eq:chi}
\begin{aligned}
\chi =\frac{N_{A}g_{J}^{2}\mu _\mathrm{{B}}^{2}\mu_{0} }{Z}&\Big[\sum_{n}^{}\beta|\braket{J_{i,n}}|^2e^{-\beta E_n}\\ + 2&\sum_{m\neq n}|\Braket{m|J_{i,n}|n}|^2(\frac{e^{-\beta E_{m}}-e^{-\beta E_{n}}}{E_{n}-E_{m}})\Big]
\end{aligned}
\end{equation}
where $\beta=1/k_BT$, $Z=\sum_{n}e^{-\beta E_{n}}$, $i=x,z$, and $n,m=0,1,2$ for the three doublets of Ce$^{3+}$ with $J=5/2$. A clear and detailed derivation of the theoretical magnetic susceptibility expression for the trigonal point symmetry of the Ce
atoms can be found in Ref.~\cite{PhysRevB.98.195120}. Following similar calculation, the final paramagnetic and Van Vleck susceptibilities in both axes in the hexagonal point symmetry are given by
\begin{equation} \label{eq:chipara}
\chi_{B\parallel z}^{total} =\frac{N_{A}g_{J}^{2}\mu _\mathrm{{B}}^{2}\mu_{0}\beta}{Z}\Big[\frac{25}{4}+\frac{9}{4}(e^{-\beta E_{1}})+\frac{1}{4}(e^{-\beta E_{2}})\Big]
\end{equation}

\begin{equation}
\label{eq:chiperp}
\begin{split}
\chi_{B\perp  z}^{total} =\frac{N_{A}g_{J}^{2}\mu _\mathrm{{B}}^{2}\mu_{0}}{k_{B}Z}
\Big[\frac{9e^{-\beta E_{2}}}{4T} + 4(\frac{e^{-\beta E_{1}}-e^{-\beta E_{2}}}{E_{2}-E_{1}})\\+\frac{5}{2}(\frac{1-e^{-\beta E_{1}}}{E_{1}})\Big]
\end{split}
\end{equation}
where $E_{1} = \epsilon _{3/2} -\epsilon _{5/2}$ and $E_{2} = \epsilon _{1/2} -\epsilon _{5/2}$ represent the splitting energies. Each energy level can be expressed by CEF parameters as $\epsilon _{1/2} = -8B_{2}^{0}+120B_{4}^{0}$,
 $\epsilon _{3/2} = -2B_{2}^{0}-180B_{4}^{0}$,
 and $\epsilon _{5/2} = 10B_{2}^{0}+60B_{4}^{0}$. 
The zero energy represents the case where there is no CEF effect. Compared to the trigonal point symmetry, the sixfold hexagonal point symmetry does not have any mixing of the $\ket{\pm 1/2}$ and the
$\ket{\pm 5/2}$ states, thus the above expression can also be obtained by setting the $B_{4}^{3}$ and mixing angle $\alpha$ in the trigonal symmetry expression (Eq. (3) in Ref.~\cite{PhysRevB.98.195120}) to zero.
Also, for the expression shown above, although it assumes that we have a ground state of $\ket{\pm 5/2}$ state, and followed by $\ket{\pm 3/2}$ and $\ket{\pm 1/2}$, it is mathematically equivalent to other derived expressions assuming a different ordering of the states.

Before doing the fitting, we can estimate the first CEF parameter $B_{2}^{0}$ from the paramagnetic Curie-Weiss temperatures $\theta_{CW}^{\parallel}$ and $\theta_{CW}^{\perp}$ by the following expression~\cite{Bowden_1971,WANG1971383}, which gives the strength of the magnetocrystalline anisotropy:
\begin{equation} \label{B20CW}
B_{2}^{0} = (\theta _{CW}^{\perp}-\theta _{CW}^{\parallel})\frac{10k_{B}}{3(2J-1)(2J+3)}
\end{equation}

\begin{figure}[!htp]
\includegraphics[width=\linewidth]{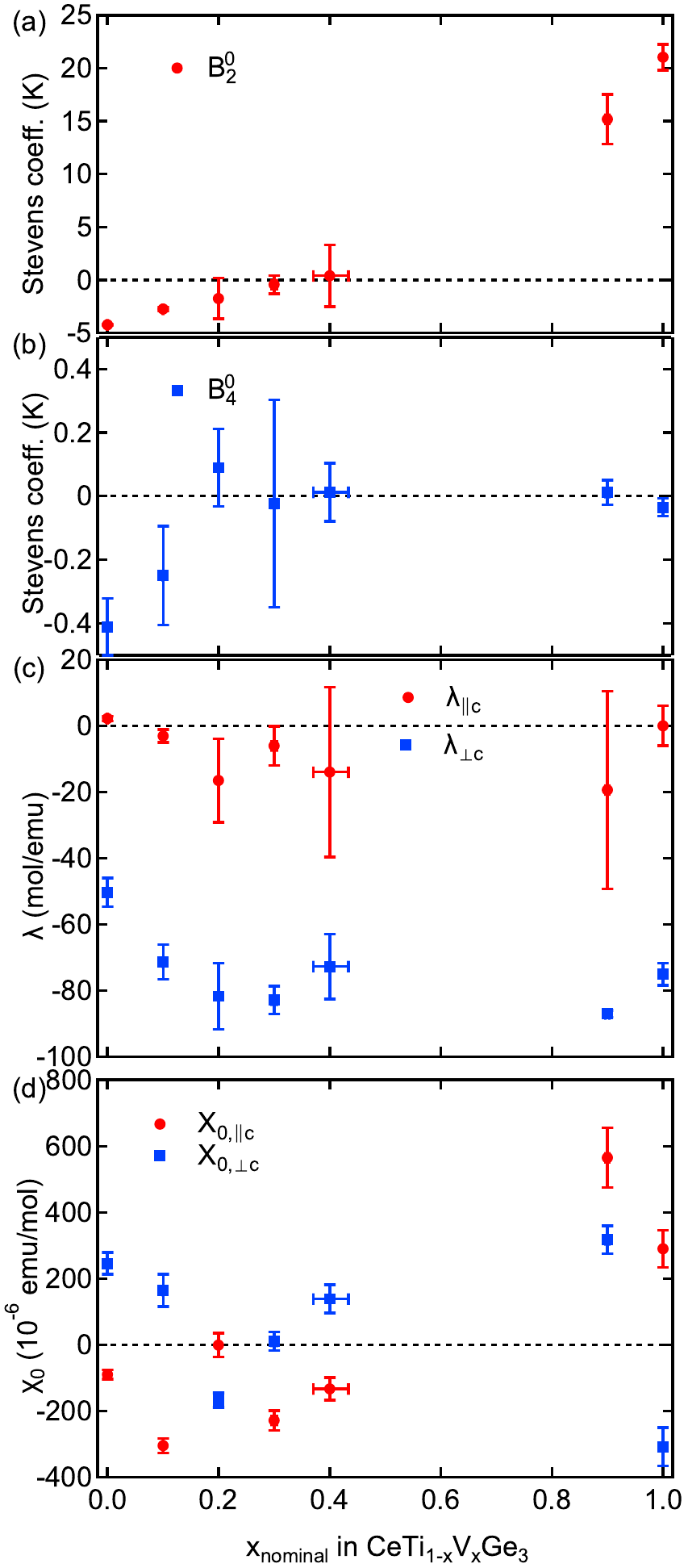}
\caption{The CEF parameters (a)~$B_{2}^{0}$, (b)~$B_{4}^{0}$, (c)~molecular field contribution $\lambda$, and (d)~the residual magnetic susceptibility $\chi_{0}$ from the fitting with the CEF model described in text.}
\label{fig:cefcoeff}
\end{figure}

The estimated parameters are listed in the second column of Table~\ref{fittings}, and it is expected that the values are moving from negative favoring $c$ axis to positive favoring the $ab$ plane as $x$ increases. Using these estimated values as the starting point, the inverse magnetic susceptibility including the molecular field contribution $\lambda_{i}$ and the residual susceptibility $\chi_{0}^{i}$ is calculated as~\cite{Ce2Pd2Pb}:
\begin{equation} \label{eq:inverchi}
\chi_{i}^{-1}=\left ( \frac{\chi_{i}^\mathrm{CEF}(T)}{1-\lambda_{i} \chi_{i}^\mathrm{CEF}(T)} + \chi_{0}^{i} \right )^{-1}
\end{equation}
The inverse magnetic susceptibility is fit to the experimental data for both orientations simultaneously, and the fit curves are shown in Fig.~\ref{fig:HoverM_all}. The CEF fittings are done for the high-temperature region (above $100$\,K), and the results with their uncertainties are shown in Fig.~\ref{fig:cefcoeff} and in Table~\ref{fittings}.

In the Ti-rich region, the uncertainty for the samples $x=0$ and $x=0.1$ are much smaller because the CEF model is able to capture the upturning feature in the 20\,K to 50\,K region perpendicular to the $c$ axis. However, for the samples $x\geq0.2$, the simple CEF model is no longer able to describe the lower-temperature region (below 100\,K) well, and this indicates a more complicated competition between the Kondo effect, RKKY interaction and the CEF effect. However, the high-temperature data are not very sensitive to the changes in those fitting parameters. This means that there are several sets of the fitting parameters that can give good fittings in the high-temperature region, and thus it will generate relatively large uncertainties. Given that we know $B_{2}^{0}$ is changing sign from negative to positive reflecting the change of anisotropy, we added a lower limit to $B_{2}^{0}$ (chosen as the value obtained from the lower $x$ curve) in order to constrain the fitting. In the V-rich region, we can see that the CEF model gives a relatively good fitting over the curvature on the susceptibility above 80\,K parallel to the $c$ axis, which cannot be easily fit to the Curie-Weiss law. Similar curvature behavior was also found in other antiferromagnetic Ce compounds~\cite{PhysRevB.98.195120,CePd2As2,Ce2Pd2Pb,CeAgSb2}, which can be well explained by the CEF scheme.

\begin{table*}[!htp]
\caption{Fitting parameters to the inverse magnetic susceptibility data based on the Eq.~(\ref{B20CW}) ($B_{2,\mathrm{CW}}^{0}$), Eq.~(\ref{eq:chipara})($B_{2}^{0}$, $B_{4}^{0}$, $\lambda_{\parallel}$, and $\chi_{0}^{\parallel}$) and Eq.~(\ref{eq:chiperp})($B_{2}^{0}$, $B_{4}^{0}$, $\lambda_{\perp}$, and $\chi_{0}^{\perp}$).}
\begin{adjustbox}{width=\textwidth}
\begin{tabular}{c|c|c|c|c|c|c|c}\hline
$x$    & $B_{2,CW}^{0}$ (K)                                & $B_{2}^{0}$ (K)                                 & $B_{4}^{0}$ (K)              & $\lambda_{\parallel}$ (mol/e.m.u)       & $\lambda_{\perp}$ (mol/e.m.u) & $\chi_{0}^{\parallel}$ (e.m.u/mol)              & $\chi_{0}^{\perp}$ (e.m.u/mol)            \\ \hline
0   & $-$11.7 & $-$4.24 $\pm$ 0.039 & $-$0.410 $\pm$ 0.089  & 2.18 $\pm$ 0.688       & $-$50.3 $\pm$ 4.30  & $-$9.02$\times10^{-5}$ $\pm$ 1.32$\times10^{-5}$ & 2.45$\times10^{-4}$ $\pm$ 3.26$\times10^{-5}$  \\
0.1 & $-$10.3  & $-$2.75 $\pm$ 0.174  & $-$0.250 $\pm$ 0.155   & $-$3.07 $\pm$ 1.90       & $-$71.4 $\pm$ 5.28  & $-$3.04$\times10^{-4}$ $\pm$ 2.18$\times10^{-5}$ & 1.64$\times10^{-4}$ $\pm$ 4.88$\times10^{-5}$  \\
0.2 & $-$6.49                        & $-1.75$ $\pm$ 1.91                        & 0.0895 $\pm$ 0.123   & $-$16.5$\pm$ 12.6        & $-81.7$ $\pm$ 10.0  & $-$6.79$\times10^{-7}$ $\pm$ 3.62$\times10^{-5}$ & $-$1.67$\times10^{-4}$ $\pm$ 2.25$\times10^{-5}$ \\
0.3 & $-$7.38                      & $-0.438$ $\pm$ 0.87                      & $-$0.0226 $\pm$ 0.327  & $-$6.07 $\pm$ 5.88       & $-82.9$ $\pm$ 4.23  & $-2.29$$\times10^{-4}$ $\pm$ 2.95$\times10^{-5}$ & 1.01$\times10^{-5}$ $\pm$ 2.90$\times10^{-5}$  \\
0.4 & $-$6.63                        & 0.388 $\pm$ 2.92                        & 0.0131 $\pm$ 0.0913  & $-$13.9 $\pm$ 25.6       & $-72.7$ $\pm$ 9.80  & $-$1.34$\times10^{-4}$ $\pm$ 3.37$\times10^{-5}$ & 1.385$\times10^{-4}$ $\pm$ 4.26$\times10^{-5}$ \\
0.9 &                           & 15.2 $\pm$ 2.34                         & 0.0117 $\pm$ 0.0384  & $-$19.3 $\pm$ 29.9       & $-$87.0 $\pm$ 1.04  & 5.65$\times10^{-4}$ $\pm$ 9.04$\times10^{-5}$  & 3.18$\times10^{-4}$ $\pm$ 4.26$\times10^{-5}$  \\
1.0 & 25.4                         & 21.0 $\pm$ 1.22                         & $-$0.035 $\pm$ 0.0276 & 0 $\pm$ 6.01 & $-$75.1 $\pm$ 3.39  & 2.90$\times10^{-4}$ $\pm$ 5.60$\times10^{-5}$  & $-$3.08$\times10^{-4}$ $\pm$ 5.79$\times10^{-5}$ \\ \hline
\end{tabular}
\end{adjustbox}
\label{fittings}
\end{table*}

\begin{table*}[!htp]
\caption{The CEF energy levels $\epsilon _{1/2}$, $\epsilon _{3/2}$, $\epsilon _{5/2}$, and the corresponding splitting energies $\Delta_1$ and $\Delta_2$ with respective uncertainties derived from the magnetic susceptibility of the CeTi$_{1-x}$V$_x$Ge$_3$ single crystals.}
\begin{tabular}{c|c|c|c|c|c}
\hline
  $x$  & $\epsilon _{1/2}$ (K) & $\epsilon _{3/2}$ (K) & $\epsilon _{5/2}$ (K) & $\Delta_1$ (K) & $\Delta_2$ (K) \\ \hline

0   & $-15.3$  $\pm$ 11.0             & 82.3  $\pm$ 16.1              & $-67.0  \pm 5.73$            &  51.6  $\pm$ 6.0   &  149.3  $\pm$ 21.8 \\ \hline  
0.1 & $-8.0$  $\pm$ 20.3             & 50.5  $\pm$ 28.3              & $-42.5$  $\pm$ 11.1            &  34.5  $\pm$ 12.5  &  93.0  $\pm$ 39.4  \\ \hline 
0.2 & 24.8  $\pm$ 30.0              & $-12.6$  $\pm$ 25.9             & $-12.2$  $\pm$ 26.4            &  0.4  $\pm$ 52.4   &  37.4  $\pm$ 48.3  \\ \hline 
0.3 & 0.8  $\pm$ 46.2             & 4.9  $\pm$ 60.6              & $-5.7$  $\pm$ 28.3            &  6.5  $\pm$ 35.3   &  10.7  $\pm$ 88.9  \\  \hline
0.4 & $-1.5$  $\pm$ 34.3             & $-3.1$  $\pm$ 22.3             & 4.7  $\pm$ 34.7             &  1.6  $\pm$ 44.9   &  7.8  $\pm$ 56.9   \\  \hline
0.9 & $-120 $ $\pm$ 23.4              & $-32.5$  $\pm$ 11.6             & 152  $\pm$ 25.7              &  87.6  $\pm$ 25.6  &  272.5  $\pm$ 44.5 \\ \hline 
1.0 & $-172$  $\pm$ 13.1              & $-35.7  \pm 7.41 $            & 208  $\pm$ 13.8              &  136.7  $\pm$ 15.6 &  380.7  $\pm$ 23.6 \\ \hline 
\end{tabular}
\label{eigen}
\end{table*}

Overall, as shown in Fig.~\ref{fig:cefcoeff}, we see a relatively smooth change in CEF parameter $B_{2}^{0}$, and the values agree reasonably with the $B_{2,\mathrm{CW}}^{0}$ values from the Curie-Weiss temperatures. The $B_{4}^{0}$ increases from around $-0.4$\,K to near zero and stays around zero. The molecular field contribution $\lambda$ represents the exchange interaction, and it remains negative in the $ab$ plane across the entire doping series. Along the $c$ axis in the Ti-rich region, $\lambda_{\parallel}$ starts with a small positive value at the pure Ti sample and turns negative while it still has a ferromagnetic ordering at the low temperature. Similar negative molecular field was also found in ferromagnetic hexagonal Ce compound, CeRh$_6$Ge$_4$, in both axes~\cite{CeRh6Ge4}, and such negative $\lambda$ can be resulted from the Kondo effect~\cite{Gruner_1974,PhysRevLett.35.1101}. The residual magnetic susceptibility does not have a clear trend but remains small compared to the measured susceptibility values, and this is probably due to some impurity phases, quality of the samples, and different magnetic backgrounds when mounting the samples.

The calculated energy levels $\epsilon _{1/2}$, $\epsilon _{3/2}$, and $\epsilon _{5/2}$ and the corresponding splitting energy $\Delta_1$ and $\Delta_2$ are shown in Table~\ref{eigen}. For the pure samples, we found out that the ground state of CeTiGe$_3$ is a $\ket{\pm 5/2}$ state, and the expected saturation magnetization parallel and perpendicular to the $c$ axis are:
\begin{equation}
\begin{split}
M^{sat}_{\parallel c} & = \Braket{\pm 5/2|J_z|\pm 5/2} g_{J}\mu_\mathrm{B} = 2.14\,\mu_\mathrm{B}/\mathrm{Ce}^{3+},\\
M^{sat}_{\perp c} & = \Braket{\pm 5/2|J_x|\pm 5/2} g_{J}\mu_\mathrm{B} = 0\,\mu_\mathrm{B}/\mathrm{Ce}^{3+}.\\
\end{split}
\label{eqTi}
\end{equation}
This matches with the experimental data where the measured saturation magnetization along the $c$ axis is $M^{sat}_{\parallel c} = 1.94$\,$\mu_\mathrm{B}/\mathrm{Ce}^{3+}$, and a small magnetization along the $ab$ plane of $M_{\perp c} = 0.04$\,$\mu_\mathrm{B}/\mathrm{Ce}^{3+}$ at $7$\,T. Such small difference is likely due to the Kondo screening effect. 

For CeVGe$_3$, the proposed ground state is a $\ket{\pm 1/2}$ state with expected saturation magnetization parallel and perpendicular to the $c$ axis being:
\begin{equation}
\begin{split}
M^{sat}_{\parallel c} & = \Braket{\pm 1/2|J_z|\mp 1/2} g_{J}\mu_\mathrm{B} = 0.43\,\mu_\mathrm{B}/\mathrm{Ce}^{3+},\\
M^{sat}_{\perp c} & = \Braket{\pm 1/2|J_x|\mp 1/2} g_{J}\mu_\mathrm{B} = 1.29\,\mu_\mathrm{B}/\mathrm{Ce}^{3+}.\\
\end{split}
\label{eqV}
\end{equation}
Our experimental data also shows that the basal plane is the easy plane of magnetization, with $M_{\parallel c}=0.088$\,$\mu_\mathrm{B}/\mathrm{Ce}^{3+}$ and $M_{\perp c}=0.42$\,$\mu_\mathrm{B}/\mathrm{Ce}^{3+}$ at $7$\,T. If the magnetization keeps increasing linearly under the field, we expect that the saturation fields will be around $H^{sat}_{\parallel c} = 44$\,T and $H^{sat}_{\perp c} = 23$\,T.

In our calculations, CeTiGe$_3$ has level splittings of $\Delta_1 = 51.6$\,K and $\Delta_2 = 149.3$\,K, and CeVGe$_3$ has level splittings of $\Delta_1 = 136.7$\,K and $\Delta_2 = 380.7$\,K. From the fittings for the Schottky heat capacity done in Ref.~\cite{Inamdar:2014io}, the inferred level spacings for CeTiGe$_3$ are $\Delta_1 = 54$\,K and $\Delta_2 = 221$\,K, and for CeVGe$_3$ are $\Delta_1 = 245$\,K and $\Delta_2 = 585$\,K. We have a fairly good agreement on the level spacings for CeTiGe$_3$, but a large difference in CeVGe$_3$. This might be due to noise in the heat capacity measurements of CeVGe$_3$, or to the fact that our CEF model does not fit well in the $20 - 80$\,K region. An inelastic neutron scattering measurement will be helpful to give us further insights on the CEF splittings.

The evolution of the proposed energy levels with respective uncertainties are plotted in Fig.~\ref{fig:energylevel}. On the Ti-rich side, we can see that along the substitution, the splitting energies $\Delta_1$ and $\Delta_2$ keep decreasing, and three energy levels become very close to each other around $x\geq0.3$. This is compatible with the observation of exceptionally high values in 4$f$-derived entropy in polycrystalline CeTi$_{1-x}$V$_{x}$Ge$_3$ samples, indicating that the splitting energies are indeed small across the Ti-rich region and taking account of the higher CEF levels is necessary~\cite{Kittler:2013bc}. Also, as we discussed in the resistivity behavior section, we already entered into the quantum critical region around $x\geq0.3$ with a non-Fermi-liquid behavior. Our proposed energy levels suggest that the quantum critical behavior in CeTi$_{1-x}$V$_{x}$Ge$_3$ is related to the suppression of CEF splitting energies. However, given that our CEF fittings have quite large uncertainties in the higher doped region, the actual energy levels may not be as close as determined. The high-field magnetization trend in Fig.12(b) suggests that the $c$ axis remains the easy axis in the $x=0.4$ sample, and therefore it still has a $\ket{\pm 5/2}$ ground state with a certain magnitude of $\Delta_1$. Future work on inelastic neutron scattering will be interesting and give more conclusive evidences.

On a side note, we have also tried a second method to fit the magnetic susceptibility data by a summation of fractional contributions from CeTiGe$_3$ and CeVGe$_3$ together. The procedure is that first we find out the desired CEF parameters $B_{2}^{0}$ and $B_{4}^{0}$ for both pure compounds, and then fit the doped samples with a certain mixture of the magnetic susceptibilities from these two pure compounds. For example, for $x = 0.2$, we will add 20\% of CeVGe$_3$ and 80\% of CeTiGe$_3$ without changing $B_{2}^{0}$ and $B_{4}^{0}$, and then adjust other parameters. Note that this corresponds to a very different physical picture, which suggests that locally we will have a certain portion of Ce atoms in the $\ket{\pm 1/2}$ state, and the rest in the $\ket{\pm 5/2}$ state. This method is labeled as ``CEF method 2''(pink lines) in Figs.~\ref{fig:HoverM_all}(b) and \ref{fig:HoverM_all}(f). Although this method also gives a reasonable fit of the high-T susceptibility along the easy axis, it does not capture the curvature behavior (around $15-50$\,K in $x=0.1$, and above 100\,K in $x=0.9$) correctly in the hard axis as well as the previous method does. So we believe that the previous method gives a more accurate description where the substitution of V has a more global  and homogeneous impact on the physical properties.

\medskip
\section{Extended comparison with the pressure phase diagram}

\begin{figure*}[!htb]
\centering
\includegraphics[width=0.95\textwidth]{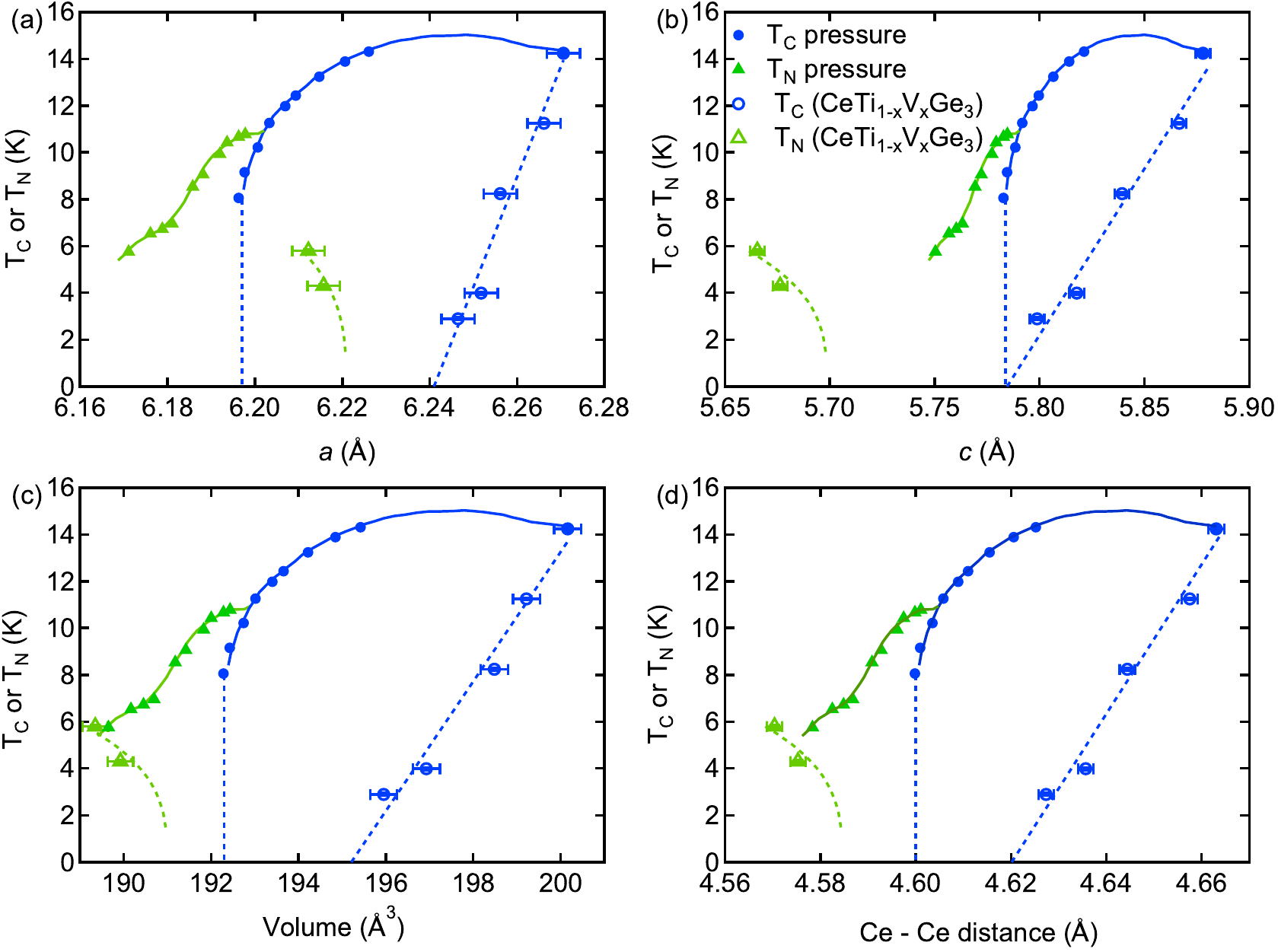}
\caption{The full comparison of the phase diagram of CeTiGe$_3$ under pressure with CeTi$_{1-x}$V$_x$Ge$_3$, with parameters of lattice constant (a) $a$, (b) $c$, (c) volume and (d) Ce-Ce distance. The critical temperature of the pressure study were obtained from Ref.~\cite{Kaluarachchi2018PRB}.}
\label{fig:combo}
\end{figure*}

The full phase transition temperatures of CeTiGe$_3$ under pressure and of CeTi$_{1-x}$V$_x$Ge$_3$, including the AFM region, are plotted together in Fig.~\ref{fig:combo}, as a function of lattice parameters $a$, $c$, lattice volume, and Ce-Ce distance. This phase diagrams were constructed in the same way as Fig.~\ref{fig:4Tc}, and the main difference is that we include the AFM region data and also change the legend for clarity.

By comparing the FM and AFM regions in both works, we can confirm that the lattice constant $a$ is not a good descriptor of the magnetic phases, since the AFM region in the substitution work overlap with the FM region of the pressure work. When the phase transitions are plotted as a function of lattice volume, or Ce-Ce distance, no clear trend or correlation can be identified. By contrast, as noted in the main text, under the parameter of lattice constant $c$, the suggested QCP in the substitution work is very close to the point where the FM order disappears in the pressure work, which is about $c = 5.78$\,\si{\angstrom}. However, the lattice constant $c$ does not describe the AFM region well since the two AFM regions are far apart. This might be due to the change of anisotropy in the substitution work. It is also unclear whether the AFM state in the pressure work is related to the AFM state in CeVGe$_3$ and if the observed change of anisotropy with V substitution also occurs under pressure.
%%%%%%%%%%%%%%%%%%%%%%%%%%%%%%%%%%%%%%%%%%%%%%%%%%%%%%%%%%%%%%%%%%%%

\bibliography{CeTiVGe3,biblio}

%apsrev4-2.bst 2019-01-14 (MD) hand-edited version of apsrev4-1.bst
%Control: key (0)
%Control: author (8) initials jnrlst
%Control: editor formatted (1) identically to author
%Control: production of article title (0) allowed
%Control: page (0) single
%Control: year (1) truncated
%Control: production of eprint (0) enabled
\begin{thebibliography}{65}%
\makeatletter
\providecommand \@ifxundefined [1]{%
 \@ifx{#1\undefined}
}%
\providecommand \@ifnum [1]{%
 \ifnum #1\expandafter \@firstoftwo
 \else \expandafter \@secondoftwo
 \fi
}%
\providecommand \@ifx [1]{%
 \ifx #1\expandafter \@firstoftwo
 \else \expandafter \@secondoftwo
 \fi
}%
\providecommand \natexlab [1]{#1}%
\providecommand \enquote  [1]{``#1''}%
\providecommand \bibnamefont  [1]{#1}%
\providecommand \bibfnamefont [1]{#1}%
\providecommand \citenamefont [1]{#1}%
\providecommand \href@noop [0]{\@secondoftwo}%
\providecommand \href [0]{\begingroup \@sanitize@url \@href}%
\providecommand \@href[1]{\@@startlink{#1}\@@href}%
\providecommand \@@href[1]{\endgroup#1\@@endlink}%
\providecommand \@sanitize@url [0]{\catcode `\\12\catcode `\$12\catcode
  `\&12\catcode `\#12\catcode `\^12\catcode `\_12\catcode `\%12\relax}%
\providecommand \@@startlink[1]{}%
\providecommand \@@endlink[0]{}%
\providecommand \url  [0]{\begingroup\@sanitize@url \@url }%
\providecommand \@url [1]{\endgroup\@href {#1}{\urlprefix }}%
\providecommand \urlprefix  [0]{URL }%
\providecommand \Eprint [0]{\href }%
\providecommand \doibase [0]{https://doi.org/}%
\providecommand \selectlanguage [0]{\@gobble}%
\providecommand \bibinfo  [0]{\@secondoftwo}%
\providecommand \bibfield  [0]{\@secondoftwo}%
\providecommand \translation [1]{[#1]}%
\providecommand \BibitemOpen [0]{}%
\providecommand \bibitemStop [0]{}%
\providecommand \bibitemNoStop [0]{.\EOS\space}%
\providecommand \EOS [0]{\spacefactor3000\relax}%
\providecommand \BibitemShut  [1]{\csname bibitem#1\endcsname}%
\let\auto@bib@innerbib\@empty
%</preamble>
\bibitem [{\citenamefont {Brando}\ \emph {et~al.}(2016)\citenamefont {Brando},
  \citenamefont {Belitz}, \citenamefont {Grosche},\ and\ \citenamefont
  {Kirkpatrick}}]{Brando2016RMP}%
  \BibitemOpen
  \bibfield  {author} {\bibinfo {author} {\bibfnamefont {M.}~\bibnamefont
  {Brando}}, \bibinfo {author} {\bibfnamefont {D.}~\bibnamefont {Belitz}},
  \bibinfo {author} {\bibfnamefont {F.~M.}\ \bibnamefont {Grosche}},\ and\
  \bibinfo {author} {\bibfnamefont {T.~R.}\ \bibnamefont {Kirkpatrick}},\
  }\bibfield  {title} {\bibinfo {title} {Metallic quantum ferromagnets},\
  }\href {https://doi.org/10.1103/RevModPhys.88.025006} {\bibfield  {journal}
  {\bibinfo  {journal} {Rev. Mod. Phys.}\ }\textbf {\bibinfo {volume} {88}},\
  \bibinfo {pages} {025006} (\bibinfo {year} {2016})}\BibitemShut {NoStop}%
\bibitem [{\citenamefont {Stewart}(2001)}]{Stewart2001RMP}%
  \BibitemOpen
  \bibfield  {author} {\bibinfo {author} {\bibfnamefont {G.~R.}\ \bibnamefont
  {Stewart}},\ }\bibfield  {title} {\bibinfo {title} {{Non-Fermi-liquid
  behavior in d- and f-electron metals}},\ }\href@noop {} {\bibfield  {journal}
  {\bibinfo  {journal} {{Rev. Mod. Phys.}}\ }\textbf {\bibinfo {volume}
  {{73}}},\ \bibinfo {pages} {797} (\bibinfo {year} {{2001}})}\BibitemShut
  {NoStop}%
\bibitem [{\citenamefont {L\"ohneysen}\ \emph {et~al.}(2007)\citenamefont
  {L\"ohneysen}, \citenamefont {Rosch}, \citenamefont {Vojta},\ and\
  \citenamefont {W\"olfle}}]{RevModPhys.79.1015}%
  \BibitemOpen
  \bibfield  {author} {\bibinfo {author} {\bibfnamefont {H.~v.}\ \bibnamefont
  {L\"ohneysen}}, \bibinfo {author} {\bibfnamefont {A.}~\bibnamefont {Rosch}},
  \bibinfo {author} {\bibfnamefont {M.}~\bibnamefont {Vojta}},\ and\ \bibinfo
  {author} {\bibfnamefont {P.}~\bibnamefont {W\"olfle}},\ }\bibfield  {title}
  {\bibinfo {title} {{Fermi-liquid instabilities at magnetic quantum phase
  transitions}},\ }\href {https://doi.org/10.1103/RevModPhys.79.1015}
  {\bibfield  {journal} {\bibinfo  {journal} {Rev. Mod. Phys.}\ }\textbf
  {\bibinfo {volume} {79}},\ \bibinfo {pages} {1015} (\bibinfo {year}
  {2007})}\BibitemShut {NoStop}%
\bibitem [{\citenamefont {Stewart}(2017)}]{unconvsc}%
  \BibitemOpen
  \bibfield  {author} {\bibinfo {author} {\bibfnamefont {G.~R.}\ \bibnamefont
  {Stewart}},\ }\bibfield  {title} {\bibinfo {title} {Unconventional
  superconductivity},\ }\href {https://doi.org/10.1080/00018732.2017.1331615}
  {\bibfield  {journal} {\bibinfo  {journal} {Advances in Physics}\ }\textbf
  {\bibinfo {volume} {66}},\ \bibinfo {pages} {75} (\bibinfo {year}
  {2017})}\BibitemShut {NoStop}%
\bibitem [{\citenamefont {Sefat}\ \emph
  {et~al.}(2009{\natexlab{a}})\citenamefont {Sefat}, \citenamefont {Singh},
  \citenamefont {VanBebber}, \citenamefont {Mozharivskyj}, \citenamefont
  {McGuire}, \citenamefont {Jin}, \citenamefont {Sales}, \citenamefont
  {Keppens},\ and\ \citenamefont {Mandrus}}]{Sefat2009PRB}%
  \BibitemOpen
  \bibfield  {author} {\bibinfo {author} {\bibfnamefont {A.~S.}\ \bibnamefont
  {Sefat}}, \bibinfo {author} {\bibfnamefont {D.~J.}\ \bibnamefont {Singh}},
  \bibinfo {author} {\bibfnamefont {L.~H.}\ \bibnamefont {VanBebber}}, \bibinfo
  {author} {\bibfnamefont {Y.}~\bibnamefont {Mozharivskyj}}, \bibinfo {author}
  {\bibfnamefont {M.~A.}\ \bibnamefont {McGuire}}, \bibinfo {author}
  {\bibfnamefont {R.}~\bibnamefont {Jin}}, \bibinfo {author} {\bibfnamefont
  {B.~C.}\ \bibnamefont {Sales}}, \bibinfo {author} {\bibfnamefont
  {V.}~\bibnamefont {Keppens}},\ and\ \bibinfo {author} {\bibfnamefont
  {D.}~\bibnamefont {Mandrus}},\ }\bibfield  {title} {\bibinfo {title}
  {{Absence of superconductivity in hole-doped
  ${\text{BaFe}}_{2\ensuremath{-}x}{\text{Cr}}_{x}{\text{As}}_{2}$ single
  crystals}},\ }\href {https://doi.org/10.1103/PhysRevB.79.224524} {\bibfield
  {journal} {\bibinfo  {journal} {Phys. Rev. B}\ }\textbf {\bibinfo {volume}
  {79}},\ \bibinfo {pages} {224524} (\bibinfo {year}
  {2009}{\natexlab{a}})}\BibitemShut {NoStop}%
\bibitem [{\citenamefont {Hu}\ and\ \citenamefont {Hu}(2010)}]{Hu2010JPCC}%
  \BibitemOpen
  \bibfield  {author} {\bibinfo {author} {\bibfnamefont {S.-J.}\ \bibnamefont
  {Hu}}\ and\ \bibinfo {author} {\bibfnamefont {X.}~\bibnamefont {Hu}},\
  }\bibfield  {title} {\bibinfo {title} {{Half-metallic antiferromagnet
  BaCrFeAs$_2$}},\ }\href {https://doi.org/10.1021/jp103328g} {\bibfield
  {journal} {\bibinfo  {journal} {The Journal of Physical Chemistry C}\
  }\textbf {\bibinfo {volume} {114}},\ \bibinfo {pages} {11614} (\bibinfo
  {year} {2010})},\ \Eprint
  {https://arxiv.org/abs/http://dx.doi.org/10.1021/jp103328g}
  {http://dx.doi.org/10.1021/jp103328g} \BibitemShut {NoStop}%
\bibitem [{\citenamefont {Filsinger}\ \emph {et~al.}(2017)\citenamefont
  {Filsinger}, \citenamefont {Schnelle}, \citenamefont {Adler}, \citenamefont
  {Fecher}, \citenamefont {Reehuis}, \citenamefont {Hoser}, \citenamefont
  {Hoffmann}, \citenamefont {Werner}, \citenamefont {Greenblatt},\ and\
  \citenamefont {Felser}}]{Filsinger2017PRB}%
  \BibitemOpen
  \bibfield  {author} {\bibinfo {author} {\bibfnamefont {K.~A.}\ \bibnamefont
  {Filsinger}}, \bibinfo {author} {\bibfnamefont {W.}~\bibnamefont {Schnelle}},
  \bibinfo {author} {\bibfnamefont {P.}~\bibnamefont {Adler}}, \bibinfo
  {author} {\bibfnamefont {G.~H.}\ \bibnamefont {Fecher}}, \bibinfo {author}
  {\bibfnamefont {M.}~\bibnamefont {Reehuis}}, \bibinfo {author} {\bibfnamefont
  {A.}~\bibnamefont {Hoser}}, \bibinfo {author} {\bibfnamefont {J.-U.}\
  \bibnamefont {Hoffmann}}, \bibinfo {author} {\bibfnamefont {P.}~\bibnamefont
  {Werner}}, \bibinfo {author} {\bibfnamefont {M.}~\bibnamefont {Greenblatt}},\
  and\ \bibinfo {author} {\bibfnamefont {C.}~\bibnamefont {Felser}},\
  }\bibfield  {title} {\bibinfo {title} {{Antiferromagnetic structure and
  electronic properties of ${\mathrm{BaCr}}_{2}{\mathrm{As}}_{2}$ and
  ${\mathrm{BaCrFeAs}}_{2}$}},\ }\href
  {https://doi.org/10.1103/PhysRevB.95.184414} {\bibfield  {journal} {\bibinfo
  {journal} {Phys. Rev. B}\ }\textbf {\bibinfo {volume} {95}},\ \bibinfo
  {pages} {184414} (\bibinfo {year} {2017})}\BibitemShut {NoStop}%
\bibitem [{\citenamefont {Xu}\ \emph {et~al.}(2012)\citenamefont {Xu},
  \citenamefont {Sun}, \citenamefont {Jiang}, \citenamefont {Xing},
  \citenamefont {Jiao}, \citenamefont {Yuan}, \citenamefont {Feng},
  \citenamefont {Xu},\ and\ \citenamefont {Cao}}]{Xu2012JPCS}%
  \BibitemOpen
  \bibfield  {author} {\bibinfo {author} {\bibfnamefont {S.-G.}\ \bibnamefont
  {Xu}}, \bibinfo {author} {\bibfnamefont {Y.-L.}\ \bibnamefont {Sun}},
  \bibinfo {author} {\bibfnamefont {S.}~\bibnamefont {Jiang}}, \bibinfo
  {author} {\bibfnamefont {H.}~\bibnamefont {Xing}}, \bibinfo {author}
  {\bibfnamefont {L.}~\bibnamefont {Jiao}}, \bibinfo {author} {\bibfnamefont
  {H.-Q.}\ \bibnamefont {Yuan}}, \bibinfo {author} {\bibfnamefont {C.-M.}\
  \bibnamefont {Feng}}, \bibinfo {author} {\bibfnamefont {Z.-A.}\ \bibnamefont
  {Xu}},\ and\ \bibinfo {author} {\bibfnamefont {G.-H.}\ \bibnamefont {Cao}},\
  }\bibfield  {title} {\bibinfo {title} {{Spin glass instead of
  superconductivity in Ba(Fe$_{1-x}$Cr$_{x/2}$Ni$_{x/2}$)$_2$As$_2$}},\ }\href
  {http://stacks.iop.org/1742-6596/400/i=3/a=032115} {\bibfield  {journal}
  {\bibinfo  {journal} {Journal of Physics: Conference Series}\ }\textbf
  {\bibinfo {volume} {400}},\ \bibinfo {pages} {032115} (\bibinfo {year}
  {2012})}\BibitemShut {NoStop}%
\bibitem [{\citenamefont {Sefat}\ \emph
  {et~al.}(2009{\natexlab{b}})\citenamefont {Sefat}, \citenamefont {Singh},
  \citenamefont {Jin}, \citenamefont {McGuire}, \citenamefont {Sales},\ and\
  \citenamefont {Mandrus}}]{Sefat2009PRBBaCo2As2}%
  \BibitemOpen
  \bibfield  {author} {\bibinfo {author} {\bibfnamefont {A.~S.}\ \bibnamefont
  {Sefat}}, \bibinfo {author} {\bibfnamefont {D.~J.}\ \bibnamefont {Singh}},
  \bibinfo {author} {\bibfnamefont {R.}~\bibnamefont {Jin}}, \bibinfo {author}
  {\bibfnamefont {M.~A.}\ \bibnamefont {McGuire}}, \bibinfo {author}
  {\bibfnamefont {B.~C.}\ \bibnamefont {Sales}},\ and\ \bibinfo {author}
  {\bibfnamefont {D.}~\bibnamefont {Mandrus}},\ }\bibfield  {title} {\bibinfo
  {title} {{Renormalized behavior and proximity of
  ${\text{BaCo}}_{2}{\text{As}}_{2}$ to a magnetic quantum critical point}},\
  }\href {https://doi.org/10.1103/PhysRevB.79.024512} {\bibfield  {journal}
  {\bibinfo  {journal} {Phys. Rev. B}\ }\textbf {\bibinfo {volume} {79}},\
  \bibinfo {pages} {024512} (\bibinfo {year} {2009}{\natexlab{b}})}\BibitemShut
  {NoStop}%
\bibitem [{\citenamefont {Pandey}\ \emph {et~al.}(2013)\citenamefont {Pandey},
  \citenamefont {Quirinale}, \citenamefont {Jayasekara}, \citenamefont
  {Sapkota}, \citenamefont {Kim}, \citenamefont {Dhaka}, \citenamefont {Lee},
  \citenamefont {Heitmann}, \citenamefont {Stephens}, \citenamefont
  {Ogloblichev}, \citenamefont {Kreyssig}, \citenamefont {McQueeney},
  \citenamefont {Goldman}, \citenamefont {Kaminski}, \citenamefont {Harmon},
  \citenamefont {Furukawa},\ and\ \citenamefont {Johnston}}]{Pandey2013PRB}%
  \BibitemOpen
  \bibfield  {author} {\bibinfo {author} {\bibfnamefont {A.}~\bibnamefont
  {Pandey}}, \bibinfo {author} {\bibfnamefont {D.~G.}\ \bibnamefont
  {Quirinale}}, \bibinfo {author} {\bibfnamefont {W.}~\bibnamefont
  {Jayasekara}}, \bibinfo {author} {\bibfnamefont {A.}~\bibnamefont {Sapkota}},
  \bibinfo {author} {\bibfnamefont {M.~G.}\ \bibnamefont {Kim}}, \bibinfo
  {author} {\bibfnamefont {R.~S.}\ \bibnamefont {Dhaka}}, \bibinfo {author}
  {\bibfnamefont {Y.}~\bibnamefont {Lee}}, \bibinfo {author} {\bibfnamefont
  {T.~W.}\ \bibnamefont {Heitmann}}, \bibinfo {author} {\bibfnamefont {P.~W.}\
  \bibnamefont {Stephens}}, \bibinfo {author} {\bibfnamefont {V.}~\bibnamefont
  {Ogloblichev}}, \bibinfo {author} {\bibfnamefont {A.}~\bibnamefont
  {Kreyssig}}, \bibinfo {author} {\bibfnamefont {R.~J.}\ \bibnamefont
  {McQueeney}}, \bibinfo {author} {\bibfnamefont {A.~I.}\ \bibnamefont
  {Goldman}}, \bibinfo {author} {\bibfnamefont {A.}~\bibnamefont {Kaminski}},
  \bibinfo {author} {\bibfnamefont {B.~N.}\ \bibnamefont {Harmon}}, \bibinfo
  {author} {\bibfnamefont {Y.}~\bibnamefont {Furukawa}},\ and\ \bibinfo
  {author} {\bibfnamefont {D.~C.}\ \bibnamefont {Johnston}},\ }\bibfield
  {title} {\bibinfo {title} {{Crystallographic, electronic, thermal, and
  magnetic properties of single-crystal SrCo$_2$As$_2$}},\ }\href
  {https://doi.org/10.1103/PhysRevB.88.014526} {\bibfield  {journal} {\bibinfo
  {journal} {Phys. Rev. B}\ }\textbf {\bibinfo {volume} {88}},\ \bibinfo
  {pages} {014526} (\bibinfo {year} {2013})}\BibitemShut {NoStop}%
\bibitem [{\citenamefont {Jayasekara}\ \emph {et~al.}(2013)\citenamefont
  {Jayasekara}, \citenamefont {Lee}, \citenamefont {Pandey}, \citenamefont
  {Tucker}, \citenamefont {Sapkota}, \citenamefont {Lamsal}, \citenamefont
  {Calder}, \citenamefont {Abernathy}, \citenamefont {Niedziela}, \citenamefont
  {Harmon}, \citenamefont {Kreyssig}, \citenamefont {Vaknin}, \citenamefont
  {Johnston}, \citenamefont {Goldman},\ and\ \citenamefont
  {McQueeney}}]{Jayasekara2013PRL}%
  \BibitemOpen
  \bibfield  {author} {\bibinfo {author} {\bibfnamefont {W.}~\bibnamefont
  {Jayasekara}}, \bibinfo {author} {\bibfnamefont {Y.}~\bibnamefont {Lee}},
  \bibinfo {author} {\bibfnamefont {A.}~\bibnamefont {Pandey}}, \bibinfo
  {author} {\bibfnamefont {G.~S.}\ \bibnamefont {Tucker}}, \bibinfo {author}
  {\bibfnamefont {A.}~\bibnamefont {Sapkota}}, \bibinfo {author} {\bibfnamefont
  {J.}~\bibnamefont {Lamsal}}, \bibinfo {author} {\bibfnamefont
  {S.}~\bibnamefont {Calder}}, \bibinfo {author} {\bibfnamefont {D.~L.}\
  \bibnamefont {Abernathy}}, \bibinfo {author} {\bibfnamefont {J.~L.}\
  \bibnamefont {Niedziela}}, \bibinfo {author} {\bibfnamefont {B.~N.}\
  \bibnamefont {Harmon}}, \bibinfo {author} {\bibfnamefont {A.}~\bibnamefont
  {Kreyssig}}, \bibinfo {author} {\bibfnamefont {D.}~\bibnamefont {Vaknin}},
  \bibinfo {author} {\bibfnamefont {D.~C.}\ \bibnamefont {Johnston}}, \bibinfo
  {author} {\bibfnamefont {A.~I.}\ \bibnamefont {Goldman}},\ and\ \bibinfo
  {author} {\bibfnamefont {R.~J.}\ \bibnamefont {McQueeney}},\ }\bibfield
  {title} {\bibinfo {title} {{Stripe Antiferromagnetic Spin Fluctuations in
  ${\mathrm{SrCo}}_{2}{\mathrm{As}}_{2}$}},\ }\href
  {https://doi.org/10.1103/PhysRevLett.111.157001} {\bibfield  {journal}
  {\bibinfo  {journal} {Phys. Rev. Lett.}\ }\textbf {\bibinfo {volume} {111}},\
  \bibinfo {pages} {157001} (\bibinfo {year} {2013})}\BibitemShut {NoStop}%
\bibitem [{\citenamefont {Guterding}\ \emph {et~al.}(2017)\citenamefont
  {Guterding}, \citenamefont {Jeschke}, \citenamefont {Mazin}, \citenamefont
  {Glasbrenner}, \citenamefont {Bascones},\ and\ \citenamefont
  {Valent\'{\i}}}]{Guterding2017PRL}%
  \BibitemOpen
  \bibfield  {author} {\bibinfo {author} {\bibfnamefont {D.}~\bibnamefont
  {Guterding}}, \bibinfo {author} {\bibfnamefont {H.~O.}\ \bibnamefont
  {Jeschke}}, \bibinfo {author} {\bibfnamefont {I.~I.}\ \bibnamefont {Mazin}},
  \bibinfo {author} {\bibfnamefont {J.~K.}\ \bibnamefont {Glasbrenner}},
  \bibinfo {author} {\bibfnamefont {E.}~\bibnamefont {Bascones}},\ and\
  \bibinfo {author} {\bibfnamefont {R.}~\bibnamefont {Valent\'{\i}}},\
  }\bibfield  {title} {\bibinfo {title} {{Nontrivial Role of Interlayer Cation
  States in Iron-Based Superconductors}},\ }\href
  {https://doi.org/10.1103/PhysRevLett.118.017204} {\bibfield  {journal}
  {\bibinfo  {journal} {Phys. Rev. Lett.}\ }\textbf {\bibinfo {volume} {118}},\
  \bibinfo {pages} {017204} (\bibinfo {year} {2017})}\BibitemShut {NoStop}%
\bibitem [{\citenamefont {Wiecki}\ \emph
  {et~al.}(2015{\natexlab{a}})\citenamefont {Wiecki}, \citenamefont
  {Ogloblichev}, \citenamefont {Pandey}, \citenamefont {Johnston},\ and\
  \citenamefont {Furukawa}}]{Wiecki2015PRB}%
  \BibitemOpen
  \bibfield  {author} {\bibinfo {author} {\bibfnamefont {P.}~\bibnamefont
  {Wiecki}}, \bibinfo {author} {\bibfnamefont {V.}~\bibnamefont {Ogloblichev}},
  \bibinfo {author} {\bibfnamefont {A.}~\bibnamefont {Pandey}}, \bibinfo
  {author} {\bibfnamefont {D.~C.}\ \bibnamefont {Johnston}},\ and\ \bibinfo
  {author} {\bibfnamefont {Y.}~\bibnamefont {Furukawa}},\ }\bibfield  {title}
  {\bibinfo {title} {{Coexistence of antiferromagnetic and ferromagnetic spin
  correlations in ${\mathrm{SrCo}}_{2}{\mathrm{As}}_{2}$ revealed by
  $^{59}\mathrm{Co}$ and $^{75}\mathrm{As}$ NMR}},\ }\href
  {https://doi.org/10.1103/PhysRevB.91.220406} {\bibfield  {journal} {\bibinfo
  {journal} {Phys. Rev. B}\ }\textbf {\bibinfo {volume} {91}},\ \bibinfo
  {pages} {220406(R)} (\bibinfo {year} {2015}{\natexlab{a}})}\BibitemShut
  {NoStop}%
\bibitem [{\citenamefont {Wiecki}\ \emph
  {et~al.}(2015{\natexlab{b}})\citenamefont {Wiecki}, \citenamefont {Roy},
  \citenamefont {Johnston}, \citenamefont {Bud'ko}, \citenamefont {Canfield},\
  and\ \citenamefont {Furukawa}}]{Wiecki2015PRL}%
  \BibitemOpen
  \bibfield  {author} {\bibinfo {author} {\bibfnamefont {P.}~\bibnamefont
  {Wiecki}}, \bibinfo {author} {\bibfnamefont {B.}~\bibnamefont {Roy}},
  \bibinfo {author} {\bibfnamefont {D.~C.}\ \bibnamefont {Johnston}}, \bibinfo
  {author} {\bibfnamefont {S.~L.}\ \bibnamefont {Bud'ko}}, \bibinfo {author}
  {\bibfnamefont {P.~C.}\ \bibnamefont {Canfield}},\ and\ \bibinfo {author}
  {\bibfnamefont {Y.}~\bibnamefont {Furukawa}},\ }\bibfield  {title} {\bibinfo
  {title} {{Competing Magnetic Fluctuations in Iron Pnictide Superconductors:
  Role of Ferromagnetic Spin Correlations Revealed by NMR}},\ }\href
  {https://doi.org/10.1103/PhysRevLett.115.137001} {\bibfield  {journal}
  {\bibinfo  {journal} {Phys. Rev. Lett.}\ }\textbf {\bibinfo {volume} {115}},\
  \bibinfo {pages} {137001} (\bibinfo {year} {2015}{\natexlab{b}})}\BibitemShut
  {NoStop}%
\bibitem [{\citenamefont {Jesche}\ \emph {et~al.}(2017)\citenamefont {Jesche},
  \citenamefont {Ball\'e}, \citenamefont {Kliemt}, \citenamefont {Geibel},
  \citenamefont {Brando},\ and\ \citenamefont {Krellner}}]{Jesche2017PSSB}%
  \BibitemOpen
  \bibfield  {author} {\bibinfo {author} {\bibfnamefont {A.}~\bibnamefont
  {Jesche}}, \bibinfo {author} {\bibfnamefont {T.}~\bibnamefont {Ball\'e}},
  \bibinfo {author} {\bibfnamefont {K.}~\bibnamefont {Kliemt}}, \bibinfo
  {author} {\bibfnamefont {C.}~\bibnamefont {Geibel}}, \bibinfo {author}
  {\bibfnamefont {M.}~\bibnamefont {Brando}},\ and\ \bibinfo {author}
  {\bibfnamefont {C.}~\bibnamefont {Krellner}},\ }\bibfield  {title} {\bibinfo
  {title} {{Avoided ferromagnetic quantum critical point: Antiferromagnetic
  ground state in substituted CeFePO}},\ }\href
  {https://doi.org/10.1002/pssb.201600169} {\bibfield  {journal} {\bibinfo
  {journal} {physica status solidi (b)}\ }\textbf {\bibinfo {volume} {254}},\
  \bibinfo {pages} {1600169} (\bibinfo {year} {2017})}\BibitemShut {NoStop}%
\bibitem [{\citenamefont {Lausberg}\ \emph {et~al.}(2012)\citenamefont
  {Lausberg}, \citenamefont {Spehling}, \citenamefont {Steppke}, \citenamefont
  {Jesche}, \citenamefont {Luetkens}, \citenamefont {Amato}, \citenamefont
  {Baines}, \citenamefont {Krellner}, \citenamefont {Brando}, \citenamefont
  {Geibel}, \citenamefont {Klauss},\ and\ \citenamefont
  {Steglich}}]{Lausberg2012PRL}%
  \BibitemOpen
  \bibfield  {author} {\bibinfo {author} {\bibfnamefont {S.}~\bibnamefont
  {Lausberg}}, \bibinfo {author} {\bibfnamefont {J.}~\bibnamefont {Spehling}},
  \bibinfo {author} {\bibfnamefont {A.}~\bibnamefont {Steppke}}, \bibinfo
  {author} {\bibfnamefont {A.}~\bibnamefont {Jesche}}, \bibinfo {author}
  {\bibfnamefont {H.}~\bibnamefont {Luetkens}}, \bibinfo {author}
  {\bibfnamefont {A.}~\bibnamefont {Amato}}, \bibinfo {author} {\bibfnamefont
  {C.}~\bibnamefont {Baines}}, \bibinfo {author} {\bibfnamefont
  {C.}~\bibnamefont {Krellner}}, \bibinfo {author} {\bibfnamefont
  {M.}~\bibnamefont {Brando}}, \bibinfo {author} {\bibfnamefont
  {C.}~\bibnamefont {Geibel}}, \bibinfo {author} {\bibfnamefont {H.~H.}\
  \bibnamefont {Klauss}},\ and\ \bibinfo {author} {\bibfnamefont
  {F.}~\bibnamefont {Steglich}},\ }\bibfield  {title} {\bibinfo {title}
  {{Avoided Ferromagnetic Quantum Critical Point: Unusual Short-Range Ordered
  State in CeFePO}},\ }\href {https://doi.org/10.1103/PhysRevLett.109.216402}
  {\bibfield  {journal} {\bibinfo  {journal} {{Phys. Rev. Lett.}}\ }\textbf
  {\bibinfo {volume} {{109}}},\ \bibinfo {pages} {216402} (\bibinfo {year}
  {{2012}})}\BibitemShut {NoStop}%
\bibitem [{\citenamefont {Kopp}\ \emph {et~al.}(2007)\citenamefont {Kopp},
  \citenamefont {Ghosal},\ and\ \citenamefont {Chakravarty}}]{Kopp2007PNAS}%
  \BibitemOpen
  \bibfield  {author} {\bibinfo {author} {\bibfnamefont {A.}~\bibnamefont
  {Kopp}}, \bibinfo {author} {\bibfnamefont {A.}~\bibnamefont {Ghosal}},\ and\
  \bibinfo {author} {\bibfnamefont {S.}~\bibnamefont {Chakravarty}},\
  }\bibfield  {title} {\bibinfo {title} {Competing ferromagnetism in
  high-temperature copper oxide superconductors},\ }\href
  {https://doi.org/10.1073/pnas.0701265104} {\bibfield  {journal} {\bibinfo
  {journal} {Proc. Natl. Acad. Sci. U.S.A.}\ }\textbf {\bibinfo {volume}
  {104}},\ \bibinfo {pages} {6123} (\bibinfo {year} {2007})},\ \Eprint
  {https://arxiv.org/abs/https://www.pnas.org/content/104/15/6123.full.pdf}
  {https://www.pnas.org/content/104/15/6123.full.pdf} \BibitemShut {NoStop}%
\bibitem [{\citenamefont {Kurashima}\ \emph {et~al.}(2018)\citenamefont
  {Kurashima}, \citenamefont {Adachi}, \citenamefont {Suzuki}, \citenamefont
  {Fukunaga}, \citenamefont {Kawamata}, \citenamefont {Noji}, \citenamefont
  {Miyasaka}, \citenamefont {Watanabe}, \citenamefont {Miyazaki}, \citenamefont
  {Koda}, \citenamefont {Kadono},\ and\ \citenamefont
  {Koike}}]{Kurashima2018PRL}%
  \BibitemOpen
  \bibfield  {author} {\bibinfo {author} {\bibfnamefont {K.}~\bibnamefont
  {Kurashima}}, \bibinfo {author} {\bibfnamefont {T.}~\bibnamefont {Adachi}},
  \bibinfo {author} {\bibfnamefont {K.~M.}\ \bibnamefont {Suzuki}}, \bibinfo
  {author} {\bibfnamefont {Y.}~\bibnamefont {Fukunaga}}, \bibinfo {author}
  {\bibfnamefont {T.}~\bibnamefont {Kawamata}}, \bibinfo {author}
  {\bibfnamefont {T.}~\bibnamefont {Noji}}, \bibinfo {author} {\bibfnamefont
  {H.}~\bibnamefont {Miyasaka}}, \bibinfo {author} {\bibfnamefont
  {I.}~\bibnamefont {Watanabe}}, \bibinfo {author} {\bibfnamefont
  {M.}~\bibnamefont {Miyazaki}}, \bibinfo {author} {\bibfnamefont
  {A.}~\bibnamefont {Koda}}, \bibinfo {author} {\bibfnamefont {R.}~\bibnamefont
  {Kadono}},\ and\ \bibinfo {author} {\bibfnamefont {Y.}~\bibnamefont
  {Koike}},\ }\bibfield  {title} {\bibinfo {title} {{Development of
  Ferromagnetic Fluctuations in Heavily Overdoped
  $(\mathrm{Bi},\mathrm{Pb}{)}_{2}{\mathrm{Sr}}_{2}{\mathrm{CuO}}_{6+\ensuremath{\delta}}$
  Copper Oxides}},\ }\href {https://doi.org/10.1103/PhysRevLett.121.057002}
  {\bibfield  {journal} {\bibinfo  {journal} {Phys. Rev. Lett.}\ }\textbf
  {\bibinfo {volume} {121}},\ \bibinfo {pages} {057002} (\bibinfo {year}
  {2018})}\BibitemShut {NoStop}%
\bibitem [{\citenamefont {Ishikawa}\ \emph {et~al.}(2001)\citenamefont
  {Ishikawa}, \citenamefont {Takeda}, \citenamefont {Ahmet}, \citenamefont
  {Ishikawa}, \citenamefont {Takeda}, \citenamefont {Karaki},\ and\
  \citenamefont {Ishimoto}}]{Ishikawa2001JPCM}%
  \BibitemOpen
  \bibfield  {author} {\bibinfo {author} {\bibfnamefont {M.}~\bibnamefont
  {Ishikawa}}, \bibinfo {author} {\bibfnamefont {N.}~\bibnamefont {Takeda}},
  \bibinfo {author} {\bibfnamefont {P.}~\bibnamefont {Ahmet}}, \bibinfo
  {author} {\bibfnamefont {M.}~\bibnamefont {Ishikawa}}, \bibinfo {author}
  {\bibfnamefont {N.}~\bibnamefont {Takeda}}, \bibinfo {author} {\bibfnamefont
  {Y.}~\bibnamefont {Karaki}},\ and\ \bibinfo {author} {\bibfnamefont
  {H.}~\bibnamefont {Ishimoto}},\ }\bibfield  {title} {\bibinfo {title}
  {{Ferromagnetic interaction and superconductivity of CeCu$_2$Si$_2$}},\
  }\href {http://stacks.iop.org/0953-8984/13/i=1/a=104} {\bibfield  {journal}
  {\bibinfo  {journal} {J. Phys.: Condens. Matter}\ }\textbf {\bibinfo {volume}
  {13}},\ \bibinfo {pages} {L25} (\bibinfo {year} {2001})}\BibitemShut
  {NoStop}%
\bibitem [{\citenamefont {Ishikawa}\ \emph {et~al.}(2002)\citenamefont
  {Ishikawa}, \citenamefont {Takeda}, \citenamefont {Ahmet}, \citenamefont
  {Karaki}, \citenamefont {Ishimoto}, \citenamefont {Huo},\ and\ \citenamefont
  {Sakurai}}]{Ishikawa2002JPCS}%
  \BibitemOpen
  \bibfield  {author} {\bibinfo {author} {\bibfnamefont {M.}~\bibnamefont
  {Ishikawa}}, \bibinfo {author} {\bibfnamefont {N.}~\bibnamefont {Takeda}},
  \bibinfo {author} {\bibfnamefont {P.}~\bibnamefont {Ahmet}}, \bibinfo
  {author} {\bibfnamefont {Y.}~\bibnamefont {Karaki}}, \bibinfo {author}
  {\bibfnamefont {H.}~\bibnamefont {Ishimoto}}, \bibinfo {author}
  {\bibfnamefont {D.}~\bibnamefont {Huo}},\ and\ \bibinfo {author}
  {\bibfnamefont {J.}~\bibnamefont {Sakurai}},\ }\bibfield  {title} {\bibinfo
  {title} {{Ferromagnetic interaction and superconductivity of
  CeCu$_2$Si$_2$}},\ }\href
  {http://www.sciencedirect.com/science/article/pii/S0022369702001208}
  {\bibfield  {journal} {\bibinfo  {journal} {J. Phys. Chem. Solids}\ }\textbf
  {\bibinfo {volume} {63}},\ \bibinfo {pages} {1165} (\bibinfo {year}
  {2002})}\BibitemShut {NoStop}%
\bibitem [{\citenamefont {Ran}\ \emph {et~al.}(2019)\citenamefont {Ran},
  \citenamefont {Eckberg}, \citenamefont {Ding}, \citenamefont {Furukawa},
  \citenamefont {Metz}, \citenamefont {Saha}, \citenamefont {Liu},
  \citenamefont {Zic}, \citenamefont {Kim}, \citenamefont {Paglione},\ and\
  \citenamefont {Butch}}]{Ran2019Science}%
  \BibitemOpen
  \bibfield  {author} {\bibinfo {author} {\bibfnamefont {S.}~\bibnamefont
  {Ran}}, \bibinfo {author} {\bibfnamefont {C.}~\bibnamefont {Eckberg}},
  \bibinfo {author} {\bibfnamefont {Q.-P.}\ \bibnamefont {Ding}}, \bibinfo
  {author} {\bibfnamefont {Y.}~\bibnamefont {Furukawa}}, \bibinfo {author}
  {\bibfnamefont {T.}~\bibnamefont {Metz}}, \bibinfo {author} {\bibfnamefont
  {S.~R.}\ \bibnamefont {Saha}}, \bibinfo {author} {\bibfnamefont {I.-L.}\
  \bibnamefont {Liu}}, \bibinfo {author} {\bibfnamefont {M.}~\bibnamefont
  {Zic}}, \bibinfo {author} {\bibfnamefont {H.}~\bibnamefont {Kim}}, \bibinfo
  {author} {\bibfnamefont {J.}~\bibnamefont {Paglione}},\ and\ \bibinfo
  {author} {\bibfnamefont {N.~P.}\ \bibnamefont {Butch}},\ }\bibfield  {title}
  {\bibinfo {title} {Nearly ferromagnetic spin-triplet superconductivity},\
  }\href {https://doi.org/10.1126/science.aav8645} {\bibfield  {journal}
  {\bibinfo  {journal} {Science}\ }\textbf {\bibinfo {volume} {365}},\ \bibinfo
  {pages} {684} (\bibinfo {year} {2019})}\BibitemShut {NoStop}%
\bibitem [{\citenamefont {Sundar}\ \emph {et~al.}(2019)\citenamefont {Sundar},
  \citenamefont {Gheidi}, \citenamefont {Akintola}, \citenamefont {C\^ot\'e},
  \citenamefont {Dunsiger}, \citenamefont {Ran}, \citenamefont {Butch},
  \citenamefont {Saha}, \citenamefont {Paglione},\ and\ \citenamefont
  {Sonier}}]{Sundar2019PRB}%
  \BibitemOpen
  \bibfield  {author} {\bibinfo {author} {\bibfnamefont {S.}~\bibnamefont
  {Sundar}}, \bibinfo {author} {\bibfnamefont {S.}~\bibnamefont {Gheidi}},
  \bibinfo {author} {\bibfnamefont {K.}~\bibnamefont {Akintola}}, \bibinfo
  {author} {\bibfnamefont {A.~M.}\ \bibnamefont {C\^ot\'e}}, \bibinfo {author}
  {\bibfnamefont {S.~R.}\ \bibnamefont {Dunsiger}}, \bibinfo {author}
  {\bibfnamefont {S.}~\bibnamefont {Ran}}, \bibinfo {author} {\bibfnamefont
  {N.~P.}\ \bibnamefont {Butch}}, \bibinfo {author} {\bibfnamefont {S.~R.}\
  \bibnamefont {Saha}}, \bibinfo {author} {\bibfnamefont {J.}~\bibnamefont
  {Paglione}},\ and\ \bibinfo {author} {\bibfnamefont {J.~E.}\ \bibnamefont
  {Sonier}},\ }\bibfield  {title} {\bibinfo {title} {{Coexistence of
  ferromagnetic fluctuations and superconductivity in the actinide
  superconductor ${\mathrm{UTe}}_{2}$}},\ }\href
  {https://doi.org/10.1103/PhysRevB.100.140502} {\bibfield  {journal} {\bibinfo
   {journal} {Phys. Rev. B}\ }\textbf {\bibinfo {volume} {100}},\ \bibinfo
  {pages} {140502(R)} (\bibinfo {year} {2019})}\BibitemShut {NoStop}%
\bibitem [{\citenamefont {Duan}\ \emph {et~al.}(2020)\citenamefont {Duan},
  \citenamefont {Sasmal}, \citenamefont {Maple}, \citenamefont {Podlesnyak},
  \citenamefont {Zhu}, \citenamefont {Si},\ and\ \citenamefont
  {Dai}}]{Duan32020PRL}%
  \BibitemOpen
  \bibfield  {author} {\bibinfo {author} {\bibfnamefont {C.}~\bibnamefont
  {Duan}}, \bibinfo {author} {\bibfnamefont {K.}~\bibnamefont {Sasmal}},
  \bibinfo {author} {\bibfnamefont {M.~B.}\ \bibnamefont {Maple}}, \bibinfo
  {author} {\bibfnamefont {A.}~\bibnamefont {Podlesnyak}}, \bibinfo {author}
  {\bibfnamefont {J.-X.}\ \bibnamefont {Zhu}}, \bibinfo {author} {\bibfnamefont
  {Q.}~\bibnamefont {Si}},\ and\ \bibinfo {author} {\bibfnamefont
  {P.}~\bibnamefont {Dai}},\ }\bibfield  {title} {\bibinfo {title}
  {{Incommensurate Spin Fluctuations in the Spin-Triplet Superconductor
  Candidate ${\mathrm{UTe}}_{2}$}},\ }\href
  {https://doi.org/10.1103/PhysRevLett.125.237003} {\bibfield  {journal}
  {\bibinfo  {journal} {Phys. Rev. Lett.}\ }\textbf {\bibinfo {volume} {125}},\
  \bibinfo {pages} {237003} (\bibinfo {year} {2020})}\BibitemShut {NoStop}%
\bibitem [{\citenamefont {Kaluarachchi}\ \emph {et~al.}(2018)\citenamefont
  {Kaluarachchi}, \citenamefont {Taufour}, \citenamefont {Bud'ko},\ and\
  \citenamefont {Canfield}}]{Kaluarachchi2018PRB}%
  \BibitemOpen
  \bibfield  {author} {\bibinfo {author} {\bibfnamefont {U.~S.}\ \bibnamefont
  {Kaluarachchi}}, \bibinfo {author} {\bibfnamefont {V.}~\bibnamefont
  {Taufour}}, \bibinfo {author} {\bibfnamefont {S.~L.}\ \bibnamefont
  {Bud'ko}},\ and\ \bibinfo {author} {\bibfnamefont {P.~C.}\ \bibnamefont
  {Canfield}},\ }\bibfield  {title} {\bibinfo {title} {{Quantum tricritical
  point in the temperature-pressure-magnetic field phase diagram of
  ${\mathrm{CeTiGe}}_{3}$}},\ }\href
  {https://doi.org/10.1103/PhysRevB.97.045139} {\bibfield  {journal} {\bibinfo
  {journal} {Phys. Rev. B}\ }\textbf {\bibinfo {volume} {97}},\ \bibinfo
  {pages} {045139} (\bibinfo {year} {2018})}\BibitemShut {NoStop}%
\bibitem [{\citenamefont {Friedemann}\ \emph {et~al.}(2018)\citenamefont
  {Friedemann}, \citenamefont {Duncan}, \citenamefont {Hirschberger},
  \citenamefont {Bauer}, \citenamefont {Küchler}, \citenamefont {Neubauer},
  \citenamefont {Brando}, \citenamefont {Pfleiderer},\ and\ \citenamefont
  {Grosche}}]{NbFe2}%
  \BibitemOpen
  \bibfield  {author} {\bibinfo {author} {\bibfnamefont {S.}~\bibnamefont
  {Friedemann}}, \bibinfo {author} {\bibfnamefont {W.~J.}\ \bibnamefont
  {Duncan}}, \bibinfo {author} {\bibfnamefont {M.}~\bibnamefont
  {Hirschberger}}, \bibinfo {author} {\bibfnamefont {T.~W.}\ \bibnamefont
  {Bauer}}, \bibinfo {author} {\bibfnamefont {R.}~\bibnamefont {Küchler}},
  \bibinfo {author} {\bibfnamefont {A.}~\bibnamefont {Neubauer}}, \bibinfo
  {author} {\bibfnamefont {M.}~\bibnamefont {Brando}}, \bibinfo {author}
  {\bibfnamefont {C.}~\bibnamefont {Pfleiderer}},\ and\ \bibinfo {author}
  {\bibfnamefont {F.}~\bibnamefont {Grosche}},\ }\bibfield  {title} {\bibinfo
  {title} {{Quantum tricritical points in NbFe$_2$}},\ }\href
  {https://doi.org/10.1038/nphys4242} {\bibfield  {journal} {\bibinfo
  {journal} {Nature Physics}\ }\textbf {\bibinfo {volume} {14}},\ \bibinfo
  {pages} {62} (\bibinfo {year} {2018})}\BibitemShut {NoStop}%
\bibitem [{\citenamefont {Kittler}\ \emph {et~al.}(2013)\citenamefont
  {Kittler}, \citenamefont {Fritsch}, \citenamefont {Weber}, \citenamefont
  {Fischer}, \citenamefont {Lamago}, \citenamefont {Andr{\'e}},\ and\
  \citenamefont {von L{\"o}hneysen}}]{Kittler:2013bc}%
  \BibitemOpen
  \bibfield  {author} {\bibinfo {author} {\bibfnamefont {W.}~\bibnamefont
  {Kittler}}, \bibinfo {author} {\bibfnamefont {V.}~\bibnamefont {Fritsch}},
  \bibinfo {author} {\bibfnamefont {F.}~\bibnamefont {Weber}}, \bibinfo
  {author} {\bibfnamefont {G.}~\bibnamefont {Fischer}}, \bibinfo {author}
  {\bibfnamefont {D.}~\bibnamefont {Lamago}}, \bibinfo {author} {\bibfnamefont
  {G.}~\bibnamefont {Andr{\'e}}},\ and\ \bibinfo {author} {\bibfnamefont
  {H.}~\bibnamefont {von L{\"o}hneysen}},\ }\bibfield  {title} {\bibinfo
  {title} {{Suppression of ferromagnetism of CeTiGe$_3$ by V substitution}},\
  }\href {https://doi.org/10.1103/PhysRevB.88.165123} {\bibfield  {journal}
  {\bibinfo  {journal} {Physical Review B}\ }\textbf {\bibinfo {volume} {88}},\
  \bibinfo {pages} {165123} (\bibinfo {year} {2013})}\BibitemShut {NoStop}%
\bibitem [{\citenamefont {Khan}\ \emph {et~al.}(2016)\citenamefont {Khan},
  \citenamefont {Yang}, \citenamefont {Wang}, \citenamefont {Mao},
  \citenamefont {Du}, \citenamefont {Xu}, \citenamefont {Zhou}, \citenamefont
  {Zhang}, \citenamefont {Chen},\ and\ \citenamefont {Fang}}]{Khan:vx}%
  \BibitemOpen
  \bibfield  {author} {\bibinfo {author} {\bibfnamefont {R.}~\bibnamefont
  {Khan}}, \bibinfo {author} {\bibfnamefont {J.}~\bibnamefont {Yang}}, \bibinfo
  {author} {\bibfnamefont {H.}~\bibnamefont {Wang}}, \bibinfo {author}
  {\bibfnamefont {Q.}~\bibnamefont {Mao}}, \bibinfo {author} {\bibfnamefont
  {J.}~\bibnamefont {Du}}, \bibinfo {author} {\bibfnamefont {B.}~\bibnamefont
  {Xu}}, \bibinfo {author} {\bibfnamefont {Y.}~\bibnamefont {Zhou}}, \bibinfo
  {author} {\bibfnamefont {Y.}~\bibnamefont {Zhang}}, \bibinfo {author}
  {\bibfnamefont {B.}~\bibnamefont {Chen}},\ and\ \bibinfo {author}
  {\bibfnamefont {M.}~\bibnamefont {Fang}},\ }\bibfield  {title} {\bibinfo
  {title} {{Ferromagnetic quantum critical behavior in heavy-fermion compounds
  CeTi$_{1-x}$Ni$_x$Ge$_3$}},\ }\href
  {https://doi.org/10.1088/2053-1591/3/10/106101} {\bibfield  {journal}
  {\bibinfo  {journal} {Materials Research Express}\ }\textbf {\bibinfo
  {volume} {3}},\ \bibinfo {pages} {106101} (\bibinfo {year}
  {2016})}\BibitemShut {NoStop}%
\bibitem [{\citenamefont {Das}\ \emph {et~al.}(2015)\citenamefont {Das},
  \citenamefont {Bhattacharyya}, \citenamefont {Anand}, \citenamefont
  {Hillier}, \citenamefont {Taylor}, \citenamefont {Gruner}, \citenamefont
  {Geibel}, \citenamefont {Adroja},\ and\ \citenamefont {Hossain}}]{Das:wy}%
  \BibitemOpen
  \bibfield  {author} {\bibinfo {author} {\bibfnamefont {D.}~\bibnamefont
  {Das}}, \bibinfo {author} {\bibfnamefont {A.}~\bibnamefont {Bhattacharyya}},
  \bibinfo {author} {\bibfnamefont {V.~K.}\ \bibnamefont {Anand}}, \bibinfo
  {author} {\bibfnamefont {A.~D.}\ \bibnamefont {Hillier}}, \bibinfo {author}
  {\bibfnamefont {J.~W.}\ \bibnamefont {Taylor}}, \bibinfo {author}
  {\bibfnamefont {T.}~\bibnamefont {Gruner}}, \bibinfo {author} {\bibfnamefont
  {C.}~\bibnamefont {Geibel}}, \bibinfo {author} {\bibfnamefont {D.~T.}\
  \bibnamefont {Adroja}},\ and\ \bibinfo {author} {\bibfnamefont
  {Z.}~\bibnamefont {Hossain}},\ }\bibfield  {title} {\bibinfo {title} {{Muon
  spin relaxation study on itinerant ferromagnet CeCrGe$_3$ and the effect of
  Ti substitution on magnetism of CeCrGe$_3$}},\ }\href
  {https://doi.org/10.1088/0953-8984/27/1/016004} {\bibfield  {journal}
  {\bibinfo  {journal} {Journal of Physics: Condensed Matter}\ }\textbf
  {\bibinfo {volume} {27}},\ \bibinfo {pages} {016004} (\bibinfo {year}
  {2015})}\BibitemShut {NoStop}%
\bibitem [{\citenamefont {Bie}\ and\ \citenamefont {Mar}(2009)}]{Bie:uh}%
  \BibitemOpen
  \bibfield  {author} {\bibinfo {author} {\bibfnamefont {H.}~\bibnamefont
  {Bie}}\ and\ \bibinfo {author} {\bibfnamefont {A.}~\bibnamefont {Mar}},\
  }\bibfield  {title} {\bibinfo {title} {{Structure and magnetic properties of
  hexagonal perovskite-type rare-earth vanadium germanides REVGe$_3$ (RE =
  La–Nd)}},\ }\href {https://doi.org/10.1039/B908781H} {\bibfield  {journal}
  {\bibinfo  {journal} {J. Mater. Chem.}\ }\textbf {\bibinfo {volume} {19}},\
  \bibinfo {pages} {6225} (\bibinfo {year} {2009})}\BibitemShut {NoStop}%
\bibitem [{\citenamefont {Inamdar}\ \emph {et~al.}(2014)\citenamefont
  {Inamdar}, \citenamefont {Thamizhavel},\ and\ \citenamefont
  {Dhar}}]{Inamdar:2014io}%
  \BibitemOpen
  \bibfield  {author} {\bibinfo {author} {\bibfnamefont {M.}~\bibnamefont
  {Inamdar}}, \bibinfo {author} {\bibfnamefont {A.}~\bibnamefont
  {Thamizhavel}},\ and\ \bibinfo {author} {\bibfnamefont {S.~K.}\ \bibnamefont
  {Dhar}},\ }\bibfield  {title} {\bibinfo {title} {{Anisotropic magnetic
  behavior of single crystalline CeTiGe$_3$ and CeVGe$_3$}},\ }\href
  {https://doi.org/10.1088/0953-8984/26/32/326003} {\bibfield  {journal}
  {\bibinfo  {journal} {Journal of Physics: Condensed Matter}\ }\textbf
  {\bibinfo {volume} {26}},\ \bibinfo {pages} {326003} (\bibinfo {year}
  {2014})}\BibitemShut {NoStop}%
\bibitem [{\citenamefont {Manfrinetti}\ \emph {et~al.}(2005)\citenamefont
  {Manfrinetti}, \citenamefont {Dhar}, \citenamefont {Kulkarni},\ and\
  \citenamefont {Morozkin}}]{MANFRINETTI2005444}%
  \BibitemOpen
  \bibfield  {author} {\bibinfo {author} {\bibfnamefont {P.}~\bibnamefont
  {Manfrinetti}}, \bibinfo {author} {\bibfnamefont {S.}~\bibnamefont {Dhar}},
  \bibinfo {author} {\bibfnamefont {R.}~\bibnamefont {Kulkarni}},\ and\
  \bibinfo {author} {\bibfnamefont {A.}~\bibnamefont {Morozkin}},\ }\bibfield
  {title} {\bibinfo {title} {{Crystal structure and the magnetic properties of
  CeTiGe$_3$}},\ }\href
  {https://doi.org/https://doi.org/10.1016/j.ssc.2005.05.026} {\bibfield
  {journal} {\bibinfo  {journal} {Solid State Communications}\ }\textbf
  {\bibinfo {volume} {135}},\ \bibinfo {pages} {444} (\bibinfo {year}
  {2005})}\BibitemShut {NoStop}%
\bibitem [{\citenamefont {Fritsch}\ \emph {et~al.}(2015)\citenamefont
  {Fritsch}, \citenamefont {Stockert}, \citenamefont {Huang}, \citenamefont
  {Bagrets}, \citenamefont {Kittler}, \citenamefont {Taubenheim}, \citenamefont
  {Pilawa}, \citenamefont {Woitschach}, \citenamefont {Huesges}, \citenamefont
  {Lucas}, \citenamefont {Schneidewind}, \citenamefont {Grube},\ and\
  \citenamefont {Löhneysen}}]{Fritsch2015}%
  \BibitemOpen
  \bibfield  {author} {\bibinfo {author} {\bibfnamefont {V.}~\bibnamefont
  {Fritsch}}, \bibinfo {author} {\bibfnamefont {O.}~\bibnamefont {Stockert}},
  \bibinfo {author} {\bibfnamefont {C.-L.}\ \bibnamefont {Huang}}, \bibinfo
  {author} {\bibfnamefont {N.}~\bibnamefont {Bagrets}}, \bibinfo {author}
  {\bibfnamefont {W.}~\bibnamefont {Kittler}}, \bibinfo {author} {\bibfnamefont
  {C.}~\bibnamefont {Taubenheim}}, \bibinfo {author} {\bibfnamefont
  {B.}~\bibnamefont {Pilawa}}, \bibinfo {author} {\bibfnamefont
  {S.}~\bibnamefont {Woitschach}}, \bibinfo {author} {\bibfnamefont
  {Z.}~\bibnamefont {Huesges}}, \bibinfo {author} {\bibfnamefont
  {S.}~\bibnamefont {Lucas}}, \bibinfo {author} {\bibfnamefont
  {A.}~\bibnamefont {Schneidewind}}, \bibinfo {author} {\bibfnamefont
  {K.}~\bibnamefont {Grube}},\ and\ \bibinfo {author} {\bibfnamefont {H.~v.}\
  \bibnamefont {Löhneysen}},\ }\bibfield  {title} {\bibinfo {title} {{Role of
  the tuning parameter at magnetic quantum phase transitions}},\ }\href
  {https://doi.org/10.1140/epjst/e2015-02443-6} {\bibfield  {journal} {\bibinfo
   {journal} {Eur. Phys. J. Spec. Top.}\ }\textbf {\bibinfo {volume} {224}},\
  \bibinfo {pages} {997} (\bibinfo {year} {2015})}\BibitemShut {NoStop}%
\bibitem [{\citenamefont {Majumder}\ \emph {et~al.}(2019)\citenamefont
  {Majumder}, \citenamefont {Kittler}, \citenamefont {Fritsch}, \citenamefont
  {L\"ohneysen}, \citenamefont {Yasuoka},\ and\ \citenamefont
  {Baenitz}}]{Majumder:2018vt}%
  \BibitemOpen
  \bibfield  {author} {\bibinfo {author} {\bibfnamefont {M.}~\bibnamefont
  {Majumder}}, \bibinfo {author} {\bibfnamefont {W.}~\bibnamefont {Kittler}},
  \bibinfo {author} {\bibfnamefont {V.}~\bibnamefont {Fritsch}}, \bibinfo
  {author} {\bibfnamefont {H.~v.}\ \bibnamefont {L\"ohneysen}}, \bibinfo
  {author} {\bibfnamefont {H.}~\bibnamefont {Yasuoka}},\ and\ \bibinfo {author}
  {\bibfnamefont {M.}~\bibnamefont {Baenitz}},\ }\bibfield  {title} {\bibinfo
  {title} {{Competing magnetic correlations across the ferromagnetic quantum
  critical point in the Kondo system
  ${\mathrm{CeTi}}_{1\ensuremath{-}x}{\mathrm{V}}_{x}{\mathrm{Ge}}_{3}$:
  $^{51}\mathrm{V}$ NMR as a local probe}},\ }\href
  {https://doi.org/10.1103/PhysRevB.100.134432} {\bibfield  {journal} {\bibinfo
   {journal} {Phys. Rev. B}\ }\textbf {\bibinfo {volume} {100}},\ \bibinfo
  {pages} {134432} (\bibinfo {year} {2019})}\BibitemShut {NoStop}%
\bibitem [{\citenamefont {Canfield}\ and\ \citenamefont
  {Fisk}(1992)}]{Canfield:uq}%
  \BibitemOpen
  \bibfield  {author} {\bibinfo {author} {\bibfnamefont {P.~C.}\ \bibnamefont
  {Canfield}}\ and\ \bibinfo {author} {\bibfnamefont {Z.}~\bibnamefont
  {Fisk}},\ }\bibfield  {title} {\bibinfo {title} {{Growth of single crystals
  from metallic fluxes}},\ }\href {https://doi.org/10.1080/13642819208215073}
  {\bibfield  {journal} {\bibinfo  {journal} {Philosophical Magazine B}\
  }\textbf {\bibinfo {volume} {65}},\ \bibinfo {pages} {1117} (\bibinfo {year}
  {1992})}\BibitemShut {NoStop}%
\bibitem [{\citenamefont {Jin}\ \emph {et~al.}(2021)\citenamefont {Jin},
  \citenamefont {Badger}, \citenamefont {Klavins}, \citenamefont {Zhao},\ and\
  \citenamefont {Taufour}}]{JIN2021158354}%
  \BibitemOpen
  \bibfield  {author} {\bibinfo {author} {\bibfnamefont {H.}~\bibnamefont
  {Jin}}, \bibinfo {author} {\bibfnamefont {J.}~\bibnamefont {Badger}},
  \bibinfo {author} {\bibfnamefont {P.}~\bibnamefont {Klavins}}, \bibinfo
  {author} {\bibfnamefont {J.-T.}\ \bibnamefont {Zhao}},\ and\ \bibinfo
  {author} {\bibfnamefont {V.}~\bibnamefont {Taufour}},\ }\bibfield  {title}
  {\bibinfo {title} {{Stabilization of CeGe$_{3}$ with Ti and O featuring
  tetravalent Ce ions: (Ce$_{0.85}$Ti$_{0.15}$)Ge$_{3}$O$_{0.5}$}},\ }\href
  {https://doi.org/https://doi.org/10.1016/j.jallcom.2020.158354} {\bibfield
  {journal} {\bibinfo  {journal} {Journal of Alloys and Compounds}\ }\textbf
  {\bibinfo {volume} {863}},\ \bibinfo {pages} {158354} (\bibinfo {year}
  {2021})}\BibitemShut {NoStop}%
\bibitem [{\citenamefont {Fukuoka}\ and\ \citenamefont
  {Yamanaka}(2004)}]{Fukuoka2004CL}%
  \BibitemOpen
  \bibfield  {author} {\bibinfo {author} {\bibfnamefont {H.}~\bibnamefont
  {Fukuoka}}\ and\ \bibinfo {author} {\bibfnamefont {S.}~\bibnamefont
  {Yamanaka}},\ }\bibfield  {title} {\bibinfo {title} {{High-Pressure Synthesis
  and Properties of a Cerium Germanide CeGe$_3$ with the Cubic Cu$_3$Au Type
  Structure}},\ }\href {https://doi.org/10.1246/cl.2004.1334} {\bibfield
  {journal} {\bibinfo  {journal} {Chem. Lett.}\ }\textbf {\bibinfo {volume}
  {33}},\ \bibinfo {pages} {1334} (\bibinfo {year} {2004})}\BibitemShut
  {NoStop}%
\bibitem [{\citenamefont {Bud{\textquotesingle}ko}\ \emph
  {et~al.}(2014)\citenamefont {Bud{\textquotesingle}ko}, \citenamefont
  {Hodovanets}, \citenamefont {Panchula}, \citenamefont {Prozorov},\ and\
  \citenamefont {Canfield}}]{Bud_ko_2014}%
  \BibitemOpen
  \bibfield  {author} {\bibinfo {author} {\bibfnamefont {S.~L.}\ \bibnamefont
  {Bud{\textquotesingle}ko}}, \bibinfo {author} {\bibfnamefont
  {H.}~\bibnamefont {Hodovanets}}, \bibinfo {author} {\bibfnamefont
  {A.}~\bibnamefont {Panchula}}, \bibinfo {author} {\bibfnamefont
  {R.}~\bibnamefont {Prozorov}},\ and\ \bibinfo {author} {\bibfnamefont
  {P.~C.}\ \bibnamefont {Canfield}},\ }\bibfield  {title} {\bibinfo {title}
  {{Physical properties of CeGe$_{2-x}$ (x= 0.24) single crystals}},\ }\href
  {https://doi.org/10.1088/0953-8984/26/14/146005} {\bibfield  {journal}
  {\bibinfo  {journal} {Journal of Physics: Condensed Matter}\ }\textbf
  {\bibinfo {volume} {26}},\ \bibinfo {pages} {146005} (\bibinfo {year}
  {2014})}\BibitemShut {NoStop}%
\bibitem [{\citenamefont {Bruker}(2016)}]{bruker2016apex3}%
  \BibitemOpen
  \bibfield  {author} {\bibinfo {author} {\bibfnamefont {A.}~\bibnamefont
  {Bruker}},\ }\href@noop {} {\emph {\bibinfo {title} {APEX3 Package, APEX3,
  SAINT and SADABS}}} (\bibinfo {year} {2016})\BibitemShut {NoStop}%
\bibitem [{\citenamefont {Sheldrick}(2008)}]{Sheldrick:sc5010}%
  \BibitemOpen
  \bibfield  {author} {\bibinfo {author} {\bibfnamefont {G.~M.}\ \bibnamefont
  {Sheldrick}},\ }\bibfield  {title} {\bibinfo {title} {{A short history of
  {\it SHELX}}},\ }\href {https://doi.org/10.1107/S0108767307043930} {\bibfield
   {journal} {\bibinfo  {journal} {Acta Crystallographica Section A}\ }\textbf
  {\bibinfo {volume} {64}},\ \bibinfo {pages} {112} (\bibinfo {year}
  {2008})}\BibitemShut {NoStop}%
\bibitem [{\citenamefont {Dolomanov}\ \emph {et~al.}(2009)\citenamefont
  {Dolomanov}, \citenamefont {Bourhis}, \citenamefont {Gildea}, \citenamefont
  {Howard},\ and\ \citenamefont {Puschmann}}]{Dolomanov:kk5042}%
  \BibitemOpen
  \bibfield  {author} {\bibinfo {author} {\bibfnamefont {O.~V.}\ \bibnamefont
  {Dolomanov}}, \bibinfo {author} {\bibfnamefont {L.~J.}\ \bibnamefont
  {Bourhis}}, \bibinfo {author} {\bibfnamefont {R.~J.}\ \bibnamefont {Gildea}},
  \bibinfo {author} {\bibfnamefont {J.~A.~K.}\ \bibnamefont {Howard}},\ and\
  \bibinfo {author} {\bibfnamefont {H.}~\bibnamefont {Puschmann}},\ }\bibfield
  {title} {\bibinfo {title} {{{\it OLEX2}: a complete structure solution,
  refinement and analysis program}},\ }\href
  {https://doi.org/10.1107/S0021889808042726} {\bibfield  {journal} {\bibinfo
  {journal} {Journal of Applied Crystallography}\ }\textbf {\bibinfo {volume}
  {42}},\ \bibinfo {pages} {339} (\bibinfo {year} {2009})}\BibitemShut
  {NoStop}%
\bibitem [{\citenamefont {Gruner}\ and\ \citenamefont
  {Zawadowski}(1974)}]{Gruner_1974}%
  \BibitemOpen
  \bibfield  {author} {\bibinfo {author} {\bibfnamefont {G.}~\bibnamefont
  {Gruner}}\ and\ \bibinfo {author} {\bibfnamefont {A.}~\bibnamefont
  {Zawadowski}},\ }\bibfield  {title} {\bibinfo {title} {Magnetic impurities in
  non-magnetic metals},\ }\href {https://doi.org/10.1088/0034-4885/37/12/001}
  {\bibfield  {journal} {\bibinfo  {journal} {Reports on Progress in Physics}\
  }\textbf {\bibinfo {volume} {37}},\ \bibinfo {pages} {1497} (\bibinfo {year}
  {1974})}\BibitemShut {NoStop}%
\bibitem [{\citenamefont {Krishna-murthy}\ \emph {et~al.}(1975)\citenamefont
  {Krishna-murthy}, \citenamefont {Wilson},\ and\ \citenamefont
  {Wilkins}}]{PhysRevLett.35.1101}%
  \BibitemOpen
  \bibfield  {author} {\bibinfo {author} {\bibfnamefont {H.~R.}\ \bibnamefont
  {Krishna-murthy}}, \bibinfo {author} {\bibfnamefont {K.~G.}\ \bibnamefont
  {Wilson}},\ and\ \bibinfo {author} {\bibfnamefont {J.~W.}\ \bibnamefont
  {Wilkins}},\ }\bibfield  {title} {\bibinfo {title} {{Temperature-Dependent
  Susceptibility of the Symmetric Anderson Model: Connection to the Kondo
  Model}},\ }\href {https://doi.org/10.1103/PhysRevLett.35.1101} {\bibfield
  {journal} {\bibinfo  {journal} {Phys. Rev. Lett.}\ }\textbf {\bibinfo
  {volume} {35}},\ \bibinfo {pages} {1101} (\bibinfo {year}
  {1975})}\BibitemShut {NoStop}%
\bibitem [{\citenamefont {Hafner}\ \emph {et~al.}(2019)\citenamefont {Hafner},
  \citenamefont {Rai}, \citenamefont {Banda}, \citenamefont {Kliemt},
  \citenamefont {Krellner}, \citenamefont {Sichelschmidt}, \citenamefont
  {Morosan}, \citenamefont {Geibel},\ and\ \citenamefont {Brando}}]{KLFM}%
  \BibitemOpen
  \bibfield  {author} {\bibinfo {author} {\bibfnamefont {D.}~\bibnamefont
  {Hafner}}, \bibinfo {author} {\bibfnamefont {B.~K.}\ \bibnamefont {Rai}},
  \bibinfo {author} {\bibfnamefont {J.}~\bibnamefont {Banda}}, \bibinfo
  {author} {\bibfnamefont {K.}~\bibnamefont {Kliemt}}, \bibinfo {author}
  {\bibfnamefont {C.}~\bibnamefont {Krellner}}, \bibinfo {author}
  {\bibfnamefont {J.}~\bibnamefont {Sichelschmidt}}, \bibinfo {author}
  {\bibfnamefont {E.}~\bibnamefont {Morosan}}, \bibinfo {author} {\bibfnamefont
  {C.}~\bibnamefont {Geibel}},\ and\ \bibinfo {author} {\bibfnamefont
  {M.}~\bibnamefont {Brando}},\ }\bibfield  {title} {\bibinfo {title}
  {{Kondo-lattice ferromagnets and their peculiar order along the magnetically
  hard axis determined by the crystalline electric field}},\ }\href
  {https://doi.org/10.1103/PhysRevB.99.201109} {\bibfield  {journal} {\bibinfo
  {journal} {Phys. Rev. B}\ }\textbf {\bibinfo {volume} {99}},\ \bibinfo
  {pages} {201109} (\bibinfo {year} {2019})}\BibitemShut {NoStop}%
\bibitem [{\citenamefont {Sidorov}\ \emph {et~al.}(2003)\citenamefont
  {Sidorov}, \citenamefont {Bauer}, \citenamefont {Frederick}, \citenamefont
  {Jeffries}, \citenamefont {Nakatsuji}, \citenamefont {Moreno}, \citenamefont
  {Thompson}, \citenamefont {Maple},\ and\ \citenamefont
  {Fisk}}]{Sidorov:2003et}%
  \BibitemOpen
  \bibfield  {author} {\bibinfo {author} {\bibfnamefont {V.~A.}\ \bibnamefont
  {Sidorov}}, \bibinfo {author} {\bibfnamefont {E.~D.}\ \bibnamefont {Bauer}},
  \bibinfo {author} {\bibfnamefont {N.~A.}\ \bibnamefont {Frederick}}, \bibinfo
  {author} {\bibfnamefont {J.~R.}\ \bibnamefont {Jeffries}}, \bibinfo {author}
  {\bibfnamefont {S.}~\bibnamefont {Nakatsuji}}, \bibinfo {author}
  {\bibfnamefont {N.~O.}\ \bibnamefont {Moreno}}, \bibinfo {author}
  {\bibfnamefont {J.~D.}\ \bibnamefont {Thompson}}, \bibinfo {author}
  {\bibfnamefont {M.~B.}\ \bibnamefont {Maple}},\ and\ \bibinfo {author}
  {\bibfnamefont {Z.}~\bibnamefont {Fisk}},\ }\bibfield  {title} {\bibinfo
  {title} {{Magnetic phase diagram of the ferromagnetic Kondo-lattice compound
  ${\mathrm{CeAgSb}}_{2}$ up to 80 kbar}},\ }\href
  {https://doi.org/10.1103/PhysRevB.67.224419} {\bibfield  {journal} {\bibinfo
  {journal} {Phys. Rev. B}\ }\textbf {\bibinfo {volume} {67}},\ \bibinfo
  {pages} {224419} (\bibinfo {year} {2003})}\BibitemShut {NoStop}%
\bibitem [{\citenamefont {Andersen}(1980)}]{f:wd}%
  \BibitemOpen
  \bibfield  {author} {\bibinfo {author} {\bibfnamefont {N.~H.}\ \bibnamefont
  {Andersen}},\ }\href@noop {} {\emph {\bibinfo {title} {{Crystalline Field and
  Structural Effects in f-Electron Systems, edited by J. E. Crow, R. P. Guertin
  and T. W. Mihalisin}}}}\ (\bibinfo  {publisher} {Plenum Press, New York},\
  \bibinfo {year} {1980})\ p.\ \bibinfo {pages} {373}\BibitemShut {NoStop}%
\bibitem [{\citenamefont {Yamamoto}\ and\ \citenamefont
  {Si}(2010)}]{Yamamoto15704}%
  \BibitemOpen
  \bibfield  {author} {\bibinfo {author} {\bibfnamefont {S.~J.}\ \bibnamefont
  {Yamamoto}}\ and\ \bibinfo {author} {\bibfnamefont {Q.}~\bibnamefont {Si}},\
  }\bibfield  {title} {\bibinfo {title} {{Metallic ferromagnetism in the Kondo
  lattice}},\ }\href {https://doi.org/10.1073/pnas.1009498107} {\bibfield
  {journal} {\bibinfo  {journal} {Proceedings of the National Academy of
  Sciences}\ }\textbf {\bibinfo {volume} {107}},\ \bibinfo {pages} {15704}
  (\bibinfo {year} {2010})}\BibitemShut {NoStop}%
\bibitem [{\citenamefont {Smith}\ \emph {et~al.}(2008)\citenamefont {Smith},
  \citenamefont {Sutherland}, \citenamefont {Lonzarich}, \citenamefont
  {Saxena}, \citenamefont {Kimura}, \citenamefont {Takashima}, \citenamefont
  {Nohara},\ and\ \citenamefont {Takagi}}]{Smith2008}%
  \BibitemOpen
  \bibfield  {author} {\bibinfo {author} {\bibfnamefont {R.~P.}\ \bibnamefont
  {Smith}}, \bibinfo {author} {\bibfnamefont {M.}~\bibnamefont {Sutherland}},
  \bibinfo {author} {\bibfnamefont {G.~G.}\ \bibnamefont {Lonzarich}}, \bibinfo
  {author} {\bibfnamefont {S.~S.}\ \bibnamefont {Saxena}}, \bibinfo {author}
  {\bibfnamefont {N.}~\bibnamefont {Kimura}}, \bibinfo {author} {\bibfnamefont
  {S.}~\bibnamefont {Takashima}}, \bibinfo {author} {\bibfnamefont
  {M.}~\bibnamefont {Nohara}},\ and\ \bibinfo {author} {\bibfnamefont
  {H.}~\bibnamefont {Takagi}},\ }\bibfield  {title} {\bibinfo {title} {Marginal
  breakdown of the fermi-liquid state on the border of metallic
  ferromagnetism},\ }\href {https://doi.org/10.1038/nature07401} {\bibfield
  {journal} {\bibinfo  {journal} {Nature}\ }\textbf {\bibinfo {volume} {455}},\
  \bibinfo {pages} {1220} (\bibinfo {year} {2008})}\BibitemShut {NoStop}%
\bibitem [{\citenamefont {Jobiliong}\ \emph {et~al.}(2005)\citenamefont
  {Jobiliong}, \citenamefont {Brooks}, \citenamefont {Choi}, \citenamefont
  {Lee},\ and\ \citenamefont {Fisk}}]{CeAgSb2}%
  \BibitemOpen
  \bibfield  {author} {\bibinfo {author} {\bibfnamefont {E.}~\bibnamefont
  {Jobiliong}}, \bibinfo {author} {\bibfnamefont {J.~S.}\ \bibnamefont
  {Brooks}}, \bibinfo {author} {\bibfnamefont {E.~S.}\ \bibnamefont {Choi}},
  \bibinfo {author} {\bibfnamefont {H.}~\bibnamefont {Lee}},\ and\ \bibinfo
  {author} {\bibfnamefont {Z.}~\bibnamefont {Fisk}},\ }\bibfield  {title}
  {\bibinfo {title} {{Magnetization and electrical-transport investigation of
  the dense Kondo system $\mathrm{Ce}\mathrm{Ag}{\mathrm{Sb}}_{2}$}},\ }\href
  {https://doi.org/10.1103/PhysRevB.72.104428} {\bibfield  {journal} {\bibinfo
  {journal} {Phys. Rev. B}\ }\textbf {\bibinfo {volume} {72}},\ \bibinfo
  {pages} {104428} (\bibinfo {year} {2005})}\BibitemShut {NoStop}%
\bibitem [{\citenamefont {Palstra}\ \emph {et~al.}(1986)\citenamefont
  {Palstra}, \citenamefont {Menovsky},\ and\ \citenamefont {Mydosh}}]{URu2Si2}%
  \BibitemOpen
  \bibfield  {author} {\bibinfo {author} {\bibfnamefont {T.~T.~M.}\
  \bibnamefont {Palstra}}, \bibinfo {author} {\bibfnamefont {A.~A.}\
  \bibnamefont {Menovsky}},\ and\ \bibinfo {author} {\bibfnamefont {J.~A.}\
  \bibnamefont {Mydosh}},\ }\bibfield  {title} {\bibinfo {title} {{Anisotropic
  electrical resistivity of the magnetic heavy-fermion superconductor
  ${\mathrm{URu}}_{2}{\mathrm{Si}}_{2}$}},\ }\href
  {https://doi.org/10.1103/PhysRevB.33.6527} {\bibfield  {journal} {\bibinfo
  {journal} {Phys. Rev. B}\ }\textbf {\bibinfo {volume} {33}},\ \bibinfo
  {pages} {6527} (\bibinfo {year} {1986})}\BibitemShut {NoStop}%
\bibitem [{\citenamefont {Ahilan}\ \emph {et~al.}(2008)\citenamefont {Ahilan},
  \citenamefont {Balasubramaniam}, \citenamefont {Ning}, \citenamefont {Imai},
  \citenamefont {Sefat}, \citenamefont {Jin}, \citenamefont {McGuire},
  \citenamefont {Sales},\ and\ \citenamefont {Mandrus}}]{BaFeCoAs2}%
  \BibitemOpen
  \bibfield  {author} {\bibinfo {author} {\bibfnamefont {K.}~\bibnamefont
  {Ahilan}}, \bibinfo {author} {\bibfnamefont {J.}~\bibnamefont
  {Balasubramaniam}}, \bibinfo {author} {\bibfnamefont {F.~L.}\ \bibnamefont
  {Ning}}, \bibinfo {author} {\bibfnamefont {T.}~\bibnamefont {Imai}}, \bibinfo
  {author} {\bibfnamefont {A.~S.}\ \bibnamefont {Sefat}}, \bibinfo {author}
  {\bibfnamefont {R.}~\bibnamefont {Jin}}, \bibinfo {author} {\bibfnamefont
  {M.~A.}\ \bibnamefont {McGuire}}, \bibinfo {author} {\bibfnamefont {B.~C.}\
  \bibnamefont {Sales}},\ and\ \bibinfo {author} {\bibfnamefont
  {D.}~\bibnamefont {Mandrus}},\ }\bibfield  {title} {\bibinfo {title}
  {{Pressure effects on the electron-doped high Tc superconductor
  BaFe$_{2-x}$Co$_x$As$_2$}},\ }\href
  {https://doi.org/10.1088/0953-8984/20/47/472201} {\bibfield  {journal}
  {\bibinfo  {journal} {Journal of Physics: Condensed Matter}\ }\textbf
  {\bibinfo {volume} {20}},\ \bibinfo {pages} {472201} (\bibinfo {year}
  {2008})}\BibitemShut {NoStop}%
\bibitem [{\citenamefont {Tran}\ \emph {et~al.}(2009)\citenamefont {Tran},
  \citenamefont {Khan}, \citenamefont {Wisniewski},\ and\ \citenamefont
  {Bauer}}]{Mo3Sb7}%
  \BibitemOpen
  \bibfield  {author} {\bibinfo {author} {\bibfnamefont {V.~H.}\ \bibnamefont
  {Tran}}, \bibinfo {author} {\bibfnamefont {R.~T.}\ \bibnamefont {Khan}},
  \bibinfo {author} {\bibfnamefont {P.}~\bibnamefont {Wisniewski}},\ and\
  \bibinfo {author} {\bibfnamefont {E.}~\bibnamefont {Bauer}},\ }\href
  {https://doi.org/10.48550/ARXIV.0907.5530} {\bibinfo {title}
  {{Pressure-induced spin-density-wave transition in superconducting
  Mo$_3$Sb$_7$ (arXiv)}}} (\bibinfo {year} {2009})\BibitemShut {NoStop}%
\bibitem [{\citenamefont {Sang}\ \emph {et~al.}(2014)\citenamefont {Sang},
  \citenamefont {Belitz},\ and\ \citenamefont {Kirkpatrick}}]{disorderFMQPT}%
  \BibitemOpen
  \bibfield  {author} {\bibinfo {author} {\bibfnamefont {Y.}~\bibnamefont
  {Sang}}, \bibinfo {author} {\bibfnamefont {D.}~\bibnamefont {Belitz}},\ and\
  \bibinfo {author} {\bibfnamefont {T.~R.}\ \bibnamefont {Kirkpatrick}},\
  }\bibfield  {title} {\bibinfo {title} {{Disorder Dependence of the
  Ferromagnetic Quantum Phase Transition}},\ }\href
  {https://doi.org/10.1103/PhysRevLett.113.207201} {\bibfield  {journal}
  {\bibinfo  {journal} {Phys. Rev. Lett.}\ }\textbf {\bibinfo {volume} {113}},\
  \bibinfo {pages} {207201} (\bibinfo {year} {2014})}\BibitemShut {NoStop}%
\bibitem [{\citenamefont {Ruderman}\ and\ \citenamefont
  {Kittel}(1954)}]{PhysRev.96.99}%
  \BibitemOpen
  \bibfield  {author} {\bibinfo {author} {\bibfnamefont {M.~A.}\ \bibnamefont
  {Ruderman}}\ and\ \bibinfo {author} {\bibfnamefont {C.}~\bibnamefont
  {Kittel}},\ }\bibfield  {title} {\bibinfo {title} {{Indirect Exchange
  Coupling of Nuclear Magnetic Moments by Conduction Electrons}},\ }\href
  {https://doi.org/10.1103/PhysRev.96.99} {\bibfield  {journal} {\bibinfo
  {journal} {Phys. Rev.}\ }\textbf {\bibinfo {volume} {96}},\ \bibinfo {pages}
  {99} (\bibinfo {year} {1954})}\BibitemShut {NoStop}%
\bibitem [{\citenamefont {Kasuya}(1956)}]{10.1143/PTP.16.45}%
  \BibitemOpen
  \bibfield  {author} {\bibinfo {author} {\bibfnamefont {T.}~\bibnamefont
  {Kasuya}},\ }\bibfield  {title} {\bibinfo {title} {{A Theory of Metallic
  Ferro- and Antiferromagnetism on Zener's Model}},\ }\href
  {https://doi.org/10.1143/PTP.16.45} {\bibfield  {journal} {\bibinfo
  {journal} {Progress of Theoretical Physics}\ }\textbf {\bibinfo {volume}
  {16}},\ \bibinfo {pages} {45} (\bibinfo {year} {1956})}\BibitemShut {NoStop}%
\bibitem [{\citenamefont {Doniach}(1977)}]{DONIACH1977231}%
  \BibitemOpen
  \bibfield  {author} {\bibinfo {author} {\bibfnamefont {S.}~\bibnamefont
  {Doniach}},\ }\bibfield  {title} {\bibinfo {title} {{The Kondo lattice and
  weak antiferromagnetism}},\ }\href
  {https://doi.org/https://doi.org/10.1016/0378-4363(77)90190-5} {\bibfield
  {journal} {\bibinfo  {journal} {Physica B+C}\ }\textbf {\bibinfo {volume}
  {91}},\ \bibinfo {pages} {231} (\bibinfo {year} {1977})}\BibitemShut
  {NoStop}%
\bibitem [{\citenamefont {Iglesias}\ \emph {et~al.}(1997)\citenamefont
  {Iglesias}, \citenamefont {Lacroix},\ and\ \citenamefont
  {Coqblin}}]{PhysRevB.56.11820}%
  \BibitemOpen
  \bibfield  {author} {\bibinfo {author} {\bibfnamefont {J.~R.}\ \bibnamefont
  {Iglesias}}, \bibinfo {author} {\bibfnamefont {C.}~\bibnamefont {Lacroix}},\
  and\ \bibinfo {author} {\bibfnamefont {B.}~\bibnamefont {Coqblin}},\
  }\bibfield  {title} {\bibinfo {title} {{Revisited Doniach diagram: Influence
  of short-range antiferromagnetic correlations in the Kondo lattice}},\ }\href
  {https://doi.org/10.1103/PhysRevB.56.11820} {\bibfield  {journal} {\bibinfo
  {journal} {Phys. Rev. B}\ }\textbf {\bibinfo {volume} {56}},\ \bibinfo
  {pages} {11820} (\bibinfo {year} {1997})}\BibitemShut {NoStop}%
\bibitem [{\citenamefont {Badger}\ \emph {et~al.}(2022)\citenamefont {Badger},
  \citenamefont {Miyawaki}, \citenamefont {Brubaker}, \citenamefont {Klavins},
  \citenamefont {Zieve}, \citenamefont {Matsuda},\ and\ \citenamefont
  {Taufour}}]{KondoCEF_Jackson}%
  \BibitemOpen
  \bibfield  {author} {\bibinfo {author} {\bibfnamefont {J.~R.}\ \bibnamefont
  {Badger}}, \bibinfo {author} {\bibfnamefont {R.}~\bibnamefont {Miyawaki}},
  \bibinfo {author} {\bibfnamefont {Z.~E.}\ \bibnamefont {Brubaker}}, \bibinfo
  {author} {\bibfnamefont {P.}~\bibnamefont {Klavins}}, \bibinfo {author}
  {\bibfnamefont {R.}~\bibnamefont {Zieve}}, \bibinfo {author} {\bibfnamefont
  {T.~D.}\ \bibnamefont {Matsuda}},\ and\ \bibinfo {author} {\bibfnamefont
  {V.}~\bibnamefont {Taufour}},\ }\bibfield  {title} {\bibinfo {title}
  {{Separation of Kondo lattice coherence from crystal electric field in
  ${\mathrm{CeIn}}_{3}$ with Nd substitutions}},\ }\href
  {https://doi.org/10.1103/PhysRevB.105.075125} {\bibfield  {journal} {\bibinfo
   {journal} {Phys. Rev. B}\ }\textbf {\bibinfo {volume} {105}},\ \bibinfo
  {pages} {075125} (\bibinfo {year} {2022})}\BibitemShut {NoStop}%
\bibitem [{\citenamefont {Ajeesh}\ \emph {et~al.}(2017)\citenamefont {Ajeesh},
  \citenamefont {Shang}, \citenamefont {Jiang}, \citenamefont {Xie},
  \citenamefont {dos Reis}, \citenamefont {Smidman}, \citenamefont {Geibel},
  \citenamefont {Yuan},\ and\ \citenamefont {Nicklas}}]{CePd2As2}%
  \BibitemOpen
  \bibfield  {author} {\bibinfo {author} {\bibfnamefont {M.~O.}\ \bibnamefont
  {Ajeesh}}, \bibinfo {author} {\bibfnamefont {T.}~\bibnamefont {Shang}},
  \bibinfo {author} {\bibfnamefont {W.~B.}\ \bibnamefont {Jiang}}, \bibinfo
  {author} {\bibfnamefont {W.}~\bibnamefont {Xie}}, \bibinfo {author}
  {\bibfnamefont {R.~D.}\ \bibnamefont {dos Reis}}, \bibinfo {author}
  {\bibfnamefont {M.}~\bibnamefont {Smidman}}, \bibinfo {author} {\bibfnamefont
  {C.}~\bibnamefont {Geibel}}, \bibinfo {author} {\bibfnamefont {H.~Q.}\
  \bibnamefont {Yuan}},\ and\ \bibinfo {author} {\bibfnamefont
  {M.}~\bibnamefont {Nicklas}},\ }\bibfield  {title} {\bibinfo {title}
  {{{Ising-type Magnetic Anisotropy in CePd$_2$As$_2$}}},\ }\href
  {https://doi.org/10.1038/s41598-017-07595-w} {\bibfield  {journal} {\bibinfo
  {journal} {Scientific Reports}\ }\textbf {\bibinfo {volume} {7}},\ \bibinfo
  {pages} {7338} (\bibinfo {year} {2017})}\BibitemShut {NoStop}%
\bibitem [{\citenamefont {Stevens}(1952)}]{Stevens_1952}%
  \BibitemOpen
  \bibfield  {author} {\bibinfo {author} {\bibfnamefont {K.~W.~H.}\
  \bibnamefont {Stevens}},\ }\bibfield  {title} {\bibinfo {title} {Matrix
  elements and operator equivalents connected with the magnetic properties of
  rare earth ions},\ }\href {https://doi.org/10.1088/0370-1298/65/3/308}
  {\bibfield  {journal} {\bibinfo  {journal} {Proceedings of the Physical
  Society. Section A}\ }\textbf {\bibinfo {volume} {65}},\ \bibinfo {pages}
  {209} (\bibinfo {year} {1952})}\BibitemShut {NoStop}%
\bibitem [{\citenamefont {Hutchings}(1964)}]{HUTCHINGS1964227}%
  \BibitemOpen
  \bibfield  {author} {\bibinfo {author} {\bibfnamefont {M.}~\bibnamefont
  {Hutchings}},\ }\bibfield  {title} {\bibinfo {title} {Point-charge
  calculations of energy levels of magnetic ions in crystalline electric
  fields},\ }\href {https://doi.org/10.1016/S0081-1947(08)60517-2} {\bibfield
  {journal} {\bibinfo  {journal} {Solid State Physics}\ }\textbf {\bibinfo
  {volume} {16}},\ \bibinfo {pages} {227} (\bibinfo {year} {1964})}\BibitemShut
  {NoStop}%
\bibitem [{\citenamefont {Banda}\ \emph {et~al.}(2018)\citenamefont {Banda},
  \citenamefont {Rai}, \citenamefont {Rosner}, \citenamefont {Morosan},
  \citenamefont {Geibel},\ and\ \citenamefont {Brando}}]{PhysRevB.98.195120}%
  \BibitemOpen
  \bibfield  {author} {\bibinfo {author} {\bibfnamefont {J.}~\bibnamefont
  {Banda}}, \bibinfo {author} {\bibfnamefont {B.~K.}\ \bibnamefont {Rai}},
  \bibinfo {author} {\bibfnamefont {H.}~\bibnamefont {Rosner}}, \bibinfo
  {author} {\bibfnamefont {E.}~\bibnamefont {Morosan}}, \bibinfo {author}
  {\bibfnamefont {C.}~\bibnamefont {Geibel}},\ and\ \bibinfo {author}
  {\bibfnamefont {M.}~\bibnamefont {Brando}},\ }\bibfield  {title} {\bibinfo
  {title} {{Crystalline electric field of Ce in trigonal symmetry:
  ${\mathrm{CeIr}}_{3}{\mathrm{Ge}}_{7}$ as a model case}},\ }\href
  {https://doi.org/10.1103/PhysRevB.98.195120} {\bibfield  {journal} {\bibinfo
  {journal} {Phys. Rev. B}\ }\textbf {\bibinfo {volume} {98}},\ \bibinfo
  {pages} {195120} (\bibinfo {year} {2018})}\BibitemShut {NoStop}%
\bibitem [{\citenamefont {Bowden}\ \emph {et~al.}(1971)\citenamefont {Bowden},
  \citenamefont {Bunbury},\ and\ \citenamefont {McCausland}}]{Bowden_1971}%
  \BibitemOpen
  \bibfield  {author} {\bibinfo {author} {\bibfnamefont {G.~J.}\ \bibnamefont
  {Bowden}}, \bibinfo {author} {\bibfnamefont {D.~S.~P.}\ \bibnamefont
  {Bunbury}},\ and\ \bibinfo {author} {\bibfnamefont {M.~A.~H.}\ \bibnamefont
  {McCausland}},\ }\bibfield  {title} {\bibinfo {title} {Crystal fields and
  magnetic anisotropy in the molecular field approximation. i. general
  considerations},\ }\href {https://doi.org/10.1088/0022-3719/4/13/035}
  {\bibfield  {journal} {\bibinfo  {journal} {Journal of Physics C: Solid State
  Physics}\ }\textbf {\bibinfo {volume} {4}},\ \bibinfo {pages} {1840}
  (\bibinfo {year} {1971})}\BibitemShut {NoStop}%
\bibitem [{\citenamefont {Wang}(1971)}]{WANG1971383}%
  \BibitemOpen
  \bibfield  {author} {\bibinfo {author} {\bibfnamefont {Y.-L.}\ \bibnamefont
  {Wang}},\ }\bibfield  {title} {\bibinfo {title} {Crystal-field effects of
  paramagnetic curie temperature},\ }\href
  {https://doi.org/https://doi.org/10.1016/0375-9601(71)90750-X} {\bibfield
  {journal} {\bibinfo  {journal} {Physics Letters A}\ }\textbf {\bibinfo
  {volume} {35}},\ \bibinfo {pages} {383} (\bibinfo {year} {1971})}\BibitemShut
  {NoStop}%
\bibitem [{\citenamefont {Kabeya}\ \emph {et~al.}(2022)\citenamefont {Kabeya},
  \citenamefont {Takahara}, \citenamefont {Arisumi}, \citenamefont {Kimura},
  \citenamefont {Araki}, \citenamefont {Katoh},\ and\ \citenamefont
  {Ochiai}}]{Ce2Pd2Pb}%
  \BibitemOpen
  \bibfield  {author} {\bibinfo {author} {\bibfnamefont {N.}~\bibnamefont
  {Kabeya}}, \bibinfo {author} {\bibfnamefont {S.}~\bibnamefont {Takahara}},
  \bibinfo {author} {\bibfnamefont {T.}~\bibnamefont {Arisumi}}, \bibinfo
  {author} {\bibfnamefont {S.}~\bibnamefont {Kimura}}, \bibinfo {author}
  {\bibfnamefont {K.}~\bibnamefont {Araki}}, \bibinfo {author} {\bibfnamefont
  {K.}~\bibnamefont {Katoh}},\ and\ \bibinfo {author} {\bibfnamefont
  {A.}~\bibnamefont {Ochiai}},\ }\bibfield  {title} {\bibinfo {title}
  {{Eigenstate analysis of the crystal electric field at low-symmetry sites:
  Application for an orthogonal site in the tetragonal crystal
  ${\mathrm{Ce}}_{2}{\mathrm{Pd}}_{2}\mathrm{Pb}$}},\ }\href
  {https://doi.org/10.1103/PhysRevB.105.014419} {\bibfield  {journal} {\bibinfo
   {journal} {Phys. Rev. B}\ }\textbf {\bibinfo {volume} {105}},\ \bibinfo
  {pages} {014419} (\bibinfo {year} {2022})}\BibitemShut {NoStop}%
\bibitem [{\citenamefont {Shu}\ \emph {et~al.}(2021)\citenamefont {Shu},
  \citenamefont {Adroja}, \citenamefont {Hillier}, \citenamefont {Zhang},
  \citenamefont {Chen}, \citenamefont {Shen}, \citenamefont {Orlandi},
  \citenamefont {Walker}, \citenamefont {Liu}, \citenamefont {Cao},
  \citenamefont {Steglich}, \citenamefont {Yuan},\ and\ \citenamefont
  {Smidman}}]{CeRh6Ge4}%
  \BibitemOpen
  \bibfield  {author} {\bibinfo {author} {\bibfnamefont {J.~W.}\ \bibnamefont
  {Shu}}, \bibinfo {author} {\bibfnamefont {D.~T.}\ \bibnamefont {Adroja}},
  \bibinfo {author} {\bibfnamefont {A.~D.}\ \bibnamefont {Hillier}}, \bibinfo
  {author} {\bibfnamefont {Y.~J.}\ \bibnamefont {Zhang}}, \bibinfo {author}
  {\bibfnamefont {Y.~X.}\ \bibnamefont {Chen}}, \bibinfo {author}
  {\bibfnamefont {B.}~\bibnamefont {Shen}}, \bibinfo {author} {\bibfnamefont
  {F.}~\bibnamefont {Orlandi}}, \bibinfo {author} {\bibfnamefont {H.~C.}\
  \bibnamefont {Walker}}, \bibinfo {author} {\bibfnamefont {Y.}~\bibnamefont
  {Liu}}, \bibinfo {author} {\bibfnamefont {C.}~\bibnamefont {Cao}}, \bibinfo
  {author} {\bibfnamefont {F.}~\bibnamefont {Steglich}}, \bibinfo {author}
  {\bibfnamefont {H.~Q.}\ \bibnamefont {Yuan}},\ and\ \bibinfo {author}
  {\bibfnamefont {M.}~\bibnamefont {Smidman}},\ }\bibfield  {title} {\bibinfo
  {title} {{Magnetic order and crystalline electric field excitations of the
  quantum critical heavy-fermion ferromagnet
  $\mathrm{Ce}{\mathrm{Rh}}_{6}{\mathrm{Ge}}_{4}$}},\ }\href
  {https://doi.org/10.1103/PhysRevB.104.L140411} {\bibfield  {journal}
  {\bibinfo  {journal} {Phys. Rev. B}\ }\textbf {\bibinfo {volume} {104}},\
  \bibinfo {pages} {L140411} (\bibinfo {year} {2021})}\BibitemShut {NoStop}%
\end{thebibliography}%

\end{document}